# Predictive modeling of football injuries

Stylianos Kampakis

**Supervisors**

Professor Philip Treleaven, Dr Ioannis Kosmidis

A dissertation submitted in partial fulfillment
of the requirements for the degree of
**Doctor of Philosophy**
of
**University College London.**

Department of Computer Science
University College London

April 2016

# Abstract

The goal of this thesis is to investigate the potential of predictive modelling for football injuries. This work was conducted in close collaboration with Tottenham Hotspurs FC (THFC), the PGA European tour and the participation of Wolverhampton Wanderers (WW).

Three investigations were conducted:

1. **Predicting the recovery time of football injuries using the UEFA injury recordings**: The UEFA recordings is a common standard for recording injuries in professional football. For this investigation, three datasets of UEFA injury recordings were available: one from THFC, one from WW and one that was constructed by merging both. Poisson, negative binomial and ordinal regression were used to model the recovery time after an injury and assess the significance of various injury-related covariates. Then, different machine learning algorithms (support vector machines, Gaussian processes, neural networks, random forests, naïve Bayes and k-nearest neighbours) were used in order to build a predictive model. The performance of the machine learning models is then improved by using feature selection conducted through correlation-based subset feature selection and random forests.

2. **Predicting injuries in professional football using exposure records**: The relationship between exposure (in training hours and match hours) in professional football athletes and injury incidence was studied. A common problem in football is understanding how the training schedule of an athlete can affect the chance of him getting injured. The task was to predict the number of days a player can train before he gets injured. The dataset consisted of the exposure records of professional footballers in Tottenham Hotspur Football Club from the season 2012-2013. The problem was approached by a Gaussian process model equipped with a dynamic time warping kernel that allowed the calculation of the similarity of exposure records of different lengths.

3. **Predicting intrinsic injury incidence using in-training GPS measurements**: A significant percentage of football injuries can be attributed to overtraining and fatigue. GPS data collected during training sessions might provide indicators of fatigue, or might be used to detect very intense training sessions which can lead to overtraining. This research used GPS data gathered during training sessions of the first team of THFC, in order to predict whether an injury would take place during a week. The data consisted of 69 variables in total. Two different binary classification approaches were followed and a variety of algorithms were applied (supervised principal component analysis, random forests, naïve Bayes, support vector machines, Gaussian process, neural networks, ridge logistic regression and k-nearest neighbours). Supervised principal component analysis shows the best results, while it also allows the extraction of components that reduce the total number of variables to 3 or 4 components which correlate with injury incidence.

The first investigation contributes the following to the field:

- It provides models based on the UEFA injury recordings, a standard used by many clubs, which makes it easier to replicate and apply the results.

- It investigates which variables seem to be more highly related to the prediction of recovery after an injury.
- It provides a comparison of models for predicting the time to return to play after injury.

The second investigation contributes the following to the field:

- It provides a model that can be used to predict the time when the first injury of the season will take place.
- It provides a kernel that can be utilized by a Gaussian process in order to measure the similarity of training and match schedules, even if the time series involved are of different lengths.

The third investigation contributes the following to the field:

- It provides a model to predict injury on a given week based on GPS data gathered from training sessions.
- It provides components, extracted through supervised principal component analysis, that correlate with injury incidence and can be used to summarize the large number of GPS variables in a parsimonious way.

# Acknowledgments

First of all, I would like to thank my primary supervisor Philip Treleaven. His feedback when advising on this thesis, and his help in the collaboration with the football clubs was immense.

I would also like to thank my second supervisor Dr. Ioannis Kosmidis. His feedback on the scientific aspects of the thesis helped clarify many points, and improve the methodology throughout the thesis. Also, his help was invaluable when preparing the papers that led to the publications related to this thesis.

The medical team of THFC provided the data to conduct this thesis, but also provided useful feedback on the direction of this research and its potential applicability. Therefore, I would like to thank all the people in the team that helped me with their suggestions and support, but first and foremost I would like to thank Wayne Diesel, the head medic of THFC. His instructions and feedback where really valuable in making this work more solid.

I would also like to thank my parents, family, friends and Tünde Tolvaj for their support and encouragement.

Finally, I would like to thank Neil Jain from Wolverthampton Wanderers. The discussions we had helped me improve my understanding of sports science and football.

# Contents

# List of Figures

# List of Tables

# 1 Introduction

*This chapter presents an overview of the thesis. First, it provides an overview of sports analytics, offering a definition and some discussion on the current state of the field. Then, it discusses the main topic of this thesis which is the predictive modeling of football injuries. Finally, it presents the structure of this thesis.*

## 1.1 Introduction and definitions

The field of sports analytics has been introduced to the popular audience through the book Moneyball (Lewis, 2004) which was also turned into film (Miller, 2011).

The use of machine learning and statistical techniques in sports is becoming increasingly popular (Cochran, 2010). Regarding academic developments the American Statistical Association Journal of Quantitative Analysis in sports was founded in 2004, while in 2006, MIT's Sloan School of Management founded the annual Sports Analytics Conference (Sloan, 2006). The field has also attracted interest in the industry with prominent examples including Nate Silver's PECOTA algorithm (Silver, 2003) for predicting outcomes in baseball, Accenture's partnership with the Australian rugby union[1], as well as SAP's partnership with the Germany national football team[2].

A common challenge in sports is injury occurrence and management (please consult Chapter 2 for literature for these claims) including:

- Injuries have a personal impact on the athlete, because this results in stress, both physical and psychological.
- Injuries have an impact on the team, because of a player's absence, which can limit the options of the team and the strategic advantage over the opposition.
- Injuries have a financial impact on the club that has finance the player's rehabilitation.

However, in spite of the importance of injuries in football and other sports, there is still little research to predict and prevent injuries.

## 1.2 Research objectives

This thesis researched the construction of predictive models for injuries in professional football. This broad goal can be split down into the following sub-goals:

- Provide a set of models, benchmarks and insights for scientists who will work in the future in the area of injury prediction.
- Provide a quantitative analysis of the relationships between injuries and relevant variables. Many variables (e.g. exposure records) are being collected based on common sense or informed opinion, but there is not a quantitative model to explain their relationship to injuries.
- Construct models which can be applied, and possibly extended, by the professional staff of a football team.

---

[1] http://www.lifehacker.com.au/2013/07/business-analytics-lessons-from-the-wallabieslions-rugby-series/
[2] http://www.news-sap.com/sap-dfb-turn-big-data-smart-data-world-cup-brazil/

## 1.3 Research methods

This thesis made use of algorithms and methods developed in the fields of statistics and machine learning.

The main statistical tool that was used was the generalized linear model (Nelder & Wedderburn, 1972). The generalized linear model enjoys various benefits such as that its coefficients can have a clear interpretable meaning and that significance testing can be used to compare models. The models used include Poisson regression, negative binomial regression and ordinal regression.

Machine learning has been used for various purposes in sports such as cycling (Ofoghi, Zeleznikow, MacMahon, & Dwyer, 2013), swimming (Meżyk & Unold, 2011) and football (Joseph, Fenton, & Neil, 2006; Min, Kim, Choe, Eom, & McKay, 2008). Machine learning algorithms were used in this thesis in order to construct predictive models.

In particular, in the first investigation (*predicting the recovery time after a football injury using the UEFA injury recordings*) the generalized linear model was fitted to the data in order to get a better understanding of the data and statistical hypothesis testing was used in order to assess which predictor variables were important.

This was particularly useful, not only for getting a better understanding of the data, but also for selecting a subset of features which improved the performance of the predictive models. Negative binomial regression was used in order to deal with some problems relating to overdispersion that were observed with Poisson regression. Ordinal regression was used because the response variable could also be treated as an ordered categorical variable.

Various machine learning algorithms (random forests, neural networks, naïve Bayes, support vector machines, k-nearest neighbours and Gaussian processes) were used in order to build a predictive model. The results of the predictive models were subsequently improved by doing feature selection based on genetic algorithms and random forests.

The second investigation (*predicting injuries in professional football using exposure records*) used a Gaussian process for the predictive model. An important challenge in the second investigation was that the input variables, which consisted of the training and match records of the footballers, were time series of different lengths. A covariance kernel that was based on dynamic time warping was proposed in order to tackle this challenge.

The third investigation (*predicting intrinsic injury incidence using in-training GPS measurements*) used GPS variables collected during the training in order to predict injuries. The problem was to discriminate between weeks where a particular athlete had been injured and weeks where the same athlete was not injured. In the first approach, only injured players were kept, while in the second approach all players were used.

The main challenge of this investigation was the large number of variables used (69 in total). Various methods were used such as supervised principal component analysis (supervised PCA), ridge logistic regression, support vector machines, Gaussian processes, neural networks, naïve Bayes, random forests and k-nearest neighbours. Supervised PCA proved to be the most effective method, as measured by the kappa statistic, and it led to the extraction of components that correlate with injury and can be used to summarize the dataset.

## 1.4  Structure of this thesis

The remainder thesis is structured as follows.

Chapter 2 provides a general background on the nature of sports analytics and related fields before moving on to sports injuries. It first presents a literature review on the current state of academic research in the field of sports analytics and the current state of commercial solutions. Then, it moves on to presenting the current work in sports analytics for injuries, demonstrating the gap in the research, and thus establishing the motivation behind this thesis.

Chapter 3 provides an introduction to some of the issues when working with data in football how they relate to the research in this thesis.

Chapter 4 outlines the statistical and machine learning methods that were used throughout this study.

Chapter 5 describes the first investigation (*predicting the recovery time after a football injury using the UEFA injury recordings*). This chapter starts with a more detailed look on the datasets which consisted of injuries from the clubs of THFC and WW. Then it moves on to the statistical analysis that was conducted using the generalized linear model. The insights of the analysis are discussed and then are used in the next step which is the creation of a predictive model. A variety of algorithms (support vector machines, Gaussian processes, neural networks, random forests, naïve Bayes and k-nearest neighbours) are compared and then the results are improved even further using feature selection.

Chapter 6 describes the second investigation (*using exposure records for predicting injuries in professional football*). The chapter starts with an explanation of the problem and some of the challenges of working with exposure records. It then moves on to a statistical investigation of the dataset which consisted of 35 exposure records from the professional team of THFC. A Gaussian process model equipped with a dynamic time warping kernel is then built for predicting the first injury of the season. The chapter then discusses some of the limitations of this study as well as the insights gained from the successful application of the model.

Chapter 7 describes the third investigation (*predicting intrinsic injury incidence using in-training GPS measurements*). The dataset consisted of the records of the training GPS records of 29 professional footballers from THFC. Two different approaches were followed in order to pose the problem as a binary classification task. Different methods were used: supervised principal component analysis, random forests, naïve Bayes, support vector machines, Gaussian process, neural networks, ridge logistic regression and k-nearest neighbours. The components derived by the supervised PCA are used in order to reduce the dimensionality and summarize the dataset in a way that is easier to digest by a non-technical audience.

Chapter 8, concludes by providing a summary of the contributions of this thesis and some suggestions for future work.

## 1.5 Contributions

Sports analytics is a new discipline and the study of injuries within it even more so. The domain is still at its infancy where standard practices have yet to be established. Current research work focuses on the following:

- Data collection protocols.
- Formalizing problems in sports in a mathematical way.
- Making the models work alongside the reality of everyday practice.

This thesis touches upon all those points.

Regarding the first point, the lack of standards in data collection can create inconsistencies in categorical variables (e.g. different names referring to the same entity), and missing or erroneous values. Chapter 3 discusses issues related to data collection in detail, since these issues were met in every one of the investigations.

Regarding the second point, it is not always clear how the data can be used to answer questions meaningful for a football club. There are many different questions that can be asked regarding the prediction of injuries, depending on the data that is available and the goal:

- The problem can be specified as a classification problem (is someone going to get on a given week?). This is for example the approach taken in Chapter 7.
- It can also become a regression problem (how many days will pass before a player gets injured). This is the approach taken in Chapters 5 and 6.
- It can be attacked on the individual, or the aggregate level (individual injuries vs team injuries).

Regarding the third point, the models developed need not only answer questions meaningful for a club, but they also need to work alongside the realities of the sport. Football clubs have an established structure and way of working, which might make difficult the adoption of new techniques, and can also influence what research questions can be posed.

For example, models that are based on easily available data, are easier to use than models that are based on expensive or time-consuming tests. Also, models that answer questions of problems solved by following common medical practices are less interesting than models that attack problems where decisions are made based on intuition and educated guesses with no formal methodology to guide the medical staff.

This thesis contributes to that point, by answering research questions that have practical relevance for the professional, like the ones stated above.

Hopefully, this thesis will set new standards for sports analytics and the use of predictive modelling for football injuries.

This thesis led to the following publications:

**Publication 1:**

Stylianos Kampakis (2013), Comparison of machine learning methods for predicting the recovery time of professional football players after an undiagnosed injury, *European Conference on Machine Learning and Principles and Practice of Knowledge Discovery in Databases (ECML-PKDD 2013), Workshop in Sports Analytics*

**Publication 2:**

Stylianos Kampakis (2014), Predictive modelling for joint and ligament football injuries, *13$^{th}$ International Conference on Sports Rehabilitation and Traumatology*

**Publication 3:**

Stylianos Kampakis, Ioannis Kosmidis, Wayne Diesel, Ed Leng (2015), A supervised PCA logistic regression model for predicting fatigue-related injuries using training GPS data, Mathsports International 2015

**Publication 4:**

(In press) Stylianos Kampakis, Ioannis Kosmidis (2015), Prediction of Injuries in Professional Football Using Gaussian Processes with Dynamic Time Warping Kernel, Statistical Analysis and Data Mining: Sports Analytics Special Issue

**Publication 5:**

(under review) Stylianos Kampakis (2016), A cost sensitive logistic regression model for predicting injuries based on exposure records, Data Mining and Knowledge Discovery (special issue in sports analytics)

## 1.6 Reference style
This thesis follows the reference style of the American Psychological Association, 6$^{th}$ edition.

# 2 Background and literature review

*This chapter starts by presenting some general background on sports analytics and introduces other scientific fields related to sports analytics. It then discusses the current state of research in sports analytics before moving on to the problem of football injuries and discussing how it relates to this thesis.*

## 2.1 Current state of research in sports analytics

### 2.1.1 General overview of academic publications

Even though the use of the term "sports analytics" is a relatively recent development, there has been research in the last few decades which, by the current terms, constitutes sports analytics. Wright (2009) reports that operations research in sports has a history of more than 50 years. Coleman (2012) discusses that even though sports analytics are constantly rising in popularity, the field is still fragmented.

Wright (2009) reports that many authors do not pursue further research once they publish it, and also that the field is considered too incoherent to support dedicated academic programs. Also, many results might not get published because the corresponding sports professionals who collaborated for a research might want to use the results in order to get an advantage over the competition.

Another problem is that there are still few conferences and journals that are a natural target for sports analytics research. A large part of research that can be termed as "sports analytics" has been published in journals that are not focused on sports analytics. Some examples include economics journals (e.g. the Journal of Applied Econometrics has published related research) or journals in computational intelligence (e.g. IEEE Transactions on Fuzzy Systems).

Coleman (2012) provides a list of journals, papers and institutions in order to identify the main contributors of sports analytics research. From Coleman's research the following points stand out:

- From the list of 40 top-cited journals that cite sports analytics research there are only two journals that are dedicated to sports: Journal of Sports Economics and Journal of Quantitative Analysis in Sports. From these two, the latter is the one that could be described as a "pure" sports analytics journal.

- There are 648 institutions that have produced sports analytics papers. Out of these, only 183 have produced more than two articles, 78 have contributed more than five and 22 institutions have contributed more than 10. The contributions by an institution are usually performed by few researchers.

However, in spite of the fragmentation of the field, there is a clear trend of rising popularity. From that perspective it is clear that sports analytics is a prosperous field for research. The absence of clear structure in the field provides lots of opportunities for exploring it through a PhD thesis.

However, while many of the papers that are published each year are stand-alone, it is not very difficult to see that there are some trends on which research is more active. Some of these trends are described in turn:

**Predicting wins/losses**: A significant portion of research is dedicated to predicting outcomes of games or tournaments. There are a few ways to pose this problem such as:

- Predicting which team is going to win a particular game: Weissbock et al. (2013) built a predictive model for hockey by using a neural network using as features team-level summary statistics and performance metrics. Zimmerman et al. (2013) did similar work for basketball by comparing various machine learning classifiers (naïve Bayes, random forests, neural networks and C4.5).
- Goals/points scored for each team: Groll and Abedieh (2013) used the goals scored by a football team as the response variable for a Poisson regression model and applied it for predicting the results of EURO 2012.
- Predicting the outcome of a specific tournament: Leitner et al. (2010) built a statistical model for predicting the winner of EURO 2008 by building the winning probabilities of each team using the betting odds and then performing pairwise comparisons.
- Sinha et al. (2013) examined the use of data mined from Twitter for predicting results of games from the National Football League.

**Analyzing performance**: Many researches concentrate on modeling performance. There are different approaches to this problem. Some examples include:

- Explaining the performance of a team or a player compared to other teams or players: Jarvandi et al. (2013) used a semi-Markov decision process in order to measure the impact of a player in a football team.
- Deriving metrics for performance of a team or a player: Stöckl (2011) created a new metric for shot quality at golf. Blackburn (2013) derived a new way to assess performance in women's tennis.
- Understanding the variables that can affect performance and outcomes either at the team or the individual level: Torin et al. (2013) used logistic regression to identify the most important factors of a successful field goal in a National Football League game.

**Analyzing movements and strategy**: Some research concentrates on breaking down important patterns in sports on an individual or collective level. Some examples include:

- Maheswaran et al. (2014) analyzed the structure of a basketball rebound.
- Haase and Brefeld (2013) derived a method for discovering similar positional streams in football.
- Annis (2006) studied the optimal strategy for basketball in the end-game.
- Bialkowski et al. (2013) used team occupancy maps and centroids for the identification of team activities in hockey.

### 2.1.2 Current state of commercial solutions

Sports constitute significant economic activity in many countries. For example, the US sports industry's value is estimated at $470 billion (Plunkett Research, Ltd., 2010). It is natural then that many commercial systems for sports analysis have been developed, with lots of research taking place outside of academia.

The main bulk of sports analytics software currently in the market concentrates on video capturing and analysis. This software does not offer advanced analytical capabilities. Rather, it concentrates on collecting data on the players and the team and presenting it in a way that is easy to understand and analyze by sports coaches, usually by the aid of visualizations and user-friendly interfaces. The most prominent example of sports analytics software currently in the market is Prozone[3]. Other examples include Elite sports analysis[4], Kinovea[5] and Quintic[6].

Obviously, the current sports analytics software available in the market can help the coaching staff make more informed decisions, by providing many statistics that might have been unavailable otherwise, in a way that is easy to digest, in the form of graphs and tables. However, the currently available software ignores more advanced capabilities that can be offered by machine learning techniques and can help towards the decision making process of an athlete or a team.

## 2.2 Injuries and sports analytics

Injuries are commonplace in all levels of football with many injuries occurring within the professional game.

While sports analytics is a field of increasing popularity the problem of predicting football injuries has largely evaded the sports analytics community. This section will present the current literature on injuries, before explaining some of the issues with it and moving on to the motivation behind this thesis.

### 2.2.1 General research on football injuries

Parry & Drust (2006) identified injuries as the main factor that prevents elite players from not being able to train and player during the football season. They also report that injuries at the first team are responsible for 49% of match unavailability and 60% of training unavailability.

Junge & Dvorak (2004) report identified that elite football players get injured on average once per year with 10-35 injuries occurring per 1000 game hours (Dvorak & Junge, 2000). The causes for injuries can vary (Junge & Dvorak, 2000; Dvorak & Junge, 2000; Junge & Dvorak, 2004) and are distinguished in intrinsic and extrinsic factors (Dvorak & Junge, 2000), with intrinsic factors being those that involve only the athlete's body and physiology, while extrinsic factors being the ones that involve external influencers, such as contact with other players.

The cost of the medical treatment per injury in football has been estimated by Dvorak and Junge (2000) at $150. Other studies place the cost of medical treatment at $188 (de Loes, 1990). However, these calculations do not take into account collateral costs that can occur as a result of an injury, nor insurance costs.

For example, the injury of a star player can affect the quality of a game, making it less spectacular and driving fans away from the stadium. The injury of a new transfer could effectively reduce the player's value. The absence of a player during important games could affect the team's chances of winning, which could also result in reduced profits. Therefore, the cost of injury can be higher than the cost of simply treating it and rehabilitating the player.

---

[3] http://www.prozonesports.com/
[4] http://www.elitesportsanalysis.com/
[5] http://www.kinovea.org/
[6] http://www.quintic.com

Most football injuries are traumatic. Around 29% of the traumatic injuries are caused by foul play (Hawkins & Fuller, 1996). A percentage of the injuries (9%-34%) can be attributed to overuse (Nielsen & Yde, 1989; Arnason, Gudmundsson, Dahl, & Jóhannsson, 1996). Most of the injuries take place during play, with most acute injuries occurring due to body contact with another player (Peterson L., Junge, Chomiak, Graf-Baumann, & Dvorak, 2000).

Many studies have studied which factors can influence injuries. For example, Dallinga et al. (2012) studied which screening tools can be used to predict injuries in the lower extremities in team sports. They found that the screening tools that concentrated on muscle and joint functionality can be predictive.

Psychological factors can affect the proneness of a player towards injury (Junge, 2000), something that might also hold for football (Junge, Dvorak, & Rösch, 2000). Johnson and Ivarson (2011) tried to identify psychological factors that can predict injuries in young soccer players and discovered a structure that can explain 23% of injury occurrence. Similar results hold for senior football players. Johnson and Ivarsson (2010) report that psychological factors could explain 14.6% of injury occurrence.

The majority of injuries happen at the lower extremities (Inklar, 1994), with the exception of goalkeepers who have more upper extremities injuries (Dvorak & Junge, 2000). The most common injuries happen at the knees and ankles and the ligaments of the thigh and calf (Inklar, 1994; Fried & Lloyd, 1992; Tucker, 1997).

### 2.2.2 Quantifying football injuries: tests and risk factors

There are various ways in the literature that have been used to quantify injuries in football. Some of these tests track the general fitness of the athlete, while other studies have concentrated on specific risk factors.

A test for assessing the general functioning of an athlete is the Functional Movement Screen$^{TM}$ (FMS) test (Kiesel, Plisky, & Voight, 2007). The FMS tests measures weaknesses or imbalances by using 7 tests: deep squat, hurdle step, in-line lunge, shoulder mobility, active straight leg raise, trunk stability push-up, and rotary stability. Each test is used in order to assess the functionality of a different body part, detecting imbalances which could potentially lead to injury.

Another test that can be used to assess the fitness of an athlete is the Yo-Yo Intermittent Recovery test (Bangsbo, 1994). This test is used to evaluate the athlete's maximum oxygen uptake under repeated intense exercise.

Another test is the single hop test, which consists of a family of various tests, such as hop tests for distance or time. The hop test can be particularly useful for assessing the stability of the knee joint (Fitzgerald, Lephart, Hwang, & Wainner, 2001). The eccentric-to-concentric power output of the knee joint has also been found to be predictive of injuries in elite soccer players (Dauty, Potiron-Josse, & Rochcongar, 2003).

A general risk factor for leg injuries is imbalance, either on the same leg or between legs. For example adductor-to-abductor ratio has been found to be a risk factor for leg injuries (Tyler, Nicholas, Campbell, Donellan, & McHugh, 2002). Hamstring strength imbalances have also been correlated to injury incidence (Croisier, Ganteaume, Binet, Genty, & Ferret, 2008).

Other tests with predictive quality for injuries include the straight-leg raise test (or Lasègue test) (Daniëlle , Devillé, Dzaferagić, Bezemer, & Bouter, 2000), assessments of ankle dorsiflexion (Malliaras, Cook, & Kent, 2006) and hip abduction (Arnason, et al., 2004) test. Ankle dorsiflexion can be assessed through various ways, such as by assessing the distance-to-wall technique or a goniometer (Konor, Morton, Eckerson, & Grindstaff, 2012).

Besides objective physical tests, it is also possible to use subjective measures to predict injury. The session-rating of perceived exertion (RPE) is another tool that can be used for understanding the proneness of athletes to injury (Haddad, Padulo, & Chamari, 2014). The RPE asks the athlete to rate the subjective feeling of the intensity of a training session. Even though the scale is subjective, it can provide good results. RPE has already been utilized in Canadian football (not soccer) with Clarke et al. (2013) reporting results of high accuracy. Similarly good results have been reported for the use of this instrument in football (Impellizzeri, Rampinini, Coutts, Sassi, & Marcora, 2004).

Many studies have been performed on specific risk factors for various kinds of injuries, where sometimes the researchers propose new tests and measures designed for specific injuries or areas.

Henderson, Barnes and Portas (2010) used a logistic regression model in order to identify risk factors for injury such as active hip flexion range of movement, the age or the lean mass. They concluded that older, more powerful and less flexible soccer players are at greater risk of sustaining a hamstring injury. Worell and Perrin (1992) studied strength, flexibility, warm-up and fatigue as risk factors for hamstring injury, outlying their importance on all of them. Gabbe et al. (2006) studied risk factors for hamstring injuries in Australian football players, identifying old age and prior injuries as the most significant factors.

Witvrouw et al. (2003) studied flexibility as a risk factor for muscle injury in professional male football athletes. Crow et al. (2010) identified that hip adductor muscle strength can be an early indicator of groin pain. Similar results with regards to the functioning of the hip joint and groin pain were uncovered by Verrall et al. (2007). Willems et al. (2005) studied the intrinsic factors that influence the incidence of ankle sprains concluding that a combination of factors (running speed, cardiorespiratory endurance, balance, dorsiflexion strength, coordination, muscle reaction, and dorsiflexion range of motion) place an athlete at risk. De Noronha et al. (2006) tried to find risk factors that can be used as predictors for sprained ankles, concluding that dorsiflexion range was the most important factor.

### 2.3   The issue with current research

A common problem that many of the methods outlined in the previous section face is that they take a simple approach to the problem. The body is a complicated system with inter-related components. The aforementioned researches focus on one or more risk factors or tests, ignoring the multitude of other factors that can lead to injury. This means that the physician is left with the task of combining the results of multitude of tests and measurements in order to assess the risk that a player faces.

For that purpose, studies, as well as commercial solutions, have tried using methodologies that can handle more complicated relationships between variables. For example, Senanayke et al. (2014) used a hybrid intelligent system, utilizing fuzzy clustering, an SVM and a neurofuzzy inference system, for monitoring the progress of anterior cruciate ligament rehabilitation. Lu et al. (2013) used a neural network model for the prediction of cartilage stress. Machine learning methods have also been applied for modeling the degeneration of the knee cartilage (Wu & Krishnan, 2011) and for the classification of knees that have been ruptured in the anterior cruciate ligament (Hashemi, Arabalibeik, & Farahmand, 2014).

However, the literature on this area is very scarce. Furthermore, the aforementioned research concentrates only on a few specific types of injuries. In addition, understanding the risk factors of an injury is less useful than actually predicting when an injury can take place. A few companies, like Kitman labs[7] and Top Sports Lab[8], have tried to fill this gap by providing commercial solutions for predicting injuries. However, no academic research, at least to the author's knowledge, has been done on predictive injury modeling in football.

Predictive modelling in football (or sports) injuries would greatly help the practitioner and the field of sports science. First, by disentangling the multitude of relationships between factors that contribute to injury. Secondly, by offering a direct answer to injury-related outcomes, instead of letting the medical practitioner carry this burden.

---

[7] http://kitmanlabs.com/
[8] http://www.topsportslab.com/en/

# 3  Data in football and datasets in the current research

*This section discusses the special challenges met when handling football data. Football data suffer from lack of standards and other issues that can interfere with the smooth development of statistical models. The chapter also offers a brief description of the datasets that were used and how they relate to the research questions in this thesis.*

## 3.1  The challenges of handling football medical data

Football is a complicated sport and this complexity reflects on all aspects of the game: from the pitch, to the training and to the physiotherapist's room. With few exceptions, there is lack of universally accepted standards on how to record aspects of the game, which extends to the recording of data related to injuries.

A main challenge regarding the recording of injury data in football is that they are handled by different parties with different priorities. The training staff might be solely interested in the progress of an athlete and they might require only the storage of specific metrics that can be used to measure performance and progress. The medical staff might not have an interested in the details behind the training, but rather on the total minutes that athletes are exposed to training or match. Some parties, like the medics might care about the medical record of the athlete over the last few years, while for the coach this might not be as relevant. The definitions behind injuries can vary in different studies (Junge & Dvorak, 2000). The problems that different definitions can cause has been stated in the past by Fuller et al. (2006).

All these requirements have an impact on both how data are recorded and handled. From the perspective of a data analyst, the data ideally should be recorded as detailed as possible, in periodic intervals over time, following specific standards. However, many parties within the team simply care about isolated events (e.g. injuries or matches), ignoring valuable information about activities that take place before the event happened. For example, the medical staff might record an injury but might ignore the specific training protocol that a player followed until the injury took place.

This problem is indicated very clearly by the fact that there is not even a common definition of what an injury is. Many studies treat as an injury only an event where the player lost days from training or playing (Hawkins & Fuller, 1999; Hawkins, Hulse, & Wilkinson, 2001; Andersen, Floerenes, Árnason, & Bahr, 2004; Árnason, Sigurdsson, & Gudmundsson, 2004).

Others follow the "medical assistance" definition where an injury is considered to have taken place if it requires medical assistance (Hawkins & Fuller, 1996; Fuller, Smith, Junge, & Dvorak, 2004; Junge, Dvorak, & Graf-Baumann , 2004).

In addition, some other studies record an injury irrespective of whether it requires medical attention or if it causes a player to stay of the game or training ("tissue injury definition") (Peterson L., Junge, Chomiak, Graf-Baumann, & Dvorak, 2000; Junge , Dvorak, Graf-Baumann, & Peterson, 2004) while some others use a combination of all these definitions.

Finally, some kinds of information lack a set of standards, so the staff has to resort to using free text or other unstandardized means of recording information. For example, the medical professional might require additional detail in specifying a knee injury, therefore creating additional sub-classifications according to his/her experience. However, other professional within the team (and in different teams) might have different experiences, which can lead them to using slightly different descriptions for the exact same injury pattern.

### 3.1.1 The challenges for data analysis

The lack of standardization in the collection and handling of data across the club and within the sport in general can pose unique challenges to the data analyst.

First of all, the data analyst must be able to detect and deal with any inconsistencies in the data which might affect the analysis. Some of these can be:

- **Lack of periodicity**: In data that should be collected periodically (e.g. every day), there can be many missing values. The analyst has to impute these values, while making sure that the missing values do not represent events that could affect the analysis. For example, if a player was injured or ill and missed a training session but didn't report it, might have gone down unrecorded, while a naïve imputation strategy might treat this as a recovery day.

- **Inconsistencies between datasets**: Different departments within a team can handle datasets with overlapping information. For example, the training staff might record information on who is attending training sessions and when, while the medical staff is recording the data that an injury has taken place and the date when the player got back into play or training. However, a player could recover with one department being aware of that, and some other department being unaware. Inconsistencies like these, which come as a result of miscommunication between the different departments of a team and lack of centralization of the data, need to be resolved, since they can severely affect the results of the analysis.

- **Inconsistencies in data entry for categorical variables**: In many cases, the datasets are compiled by more than one individual. For example, the injuries are recorded by various medics within the club, depending on who is responsible for the current athlete at that time. As it was mentioned previously, the medical staff can sometimes choose slightly different terminology to describe the same injury, depending on their experience or personal preference of terminology. This, however, poses a problem for the analyst and the statistical algorithms which might have no information to detect that two categories can refer to the exact same thing. For that purpose, the data has to be thoroughly examined in order to detect similar problems and the analyst has to consult with the medical experts in order to integrate the variables.

- **Non-statistically relevant field jargon**: There are some cases where the professionals can record variables down in a way that is highly relevant to their profession, but it might not be relevant for statistical analysis. For example, injuries can be classified in severity categories. An example discretization used by teams (discussed in Chapter 5) is breaking down injuries as "transient" (recovery time < 7 days), "mild" (recovery time <28 days), "moderate" (recovery time < 84 days), "severe" (recovery time $\geq$ 84 days). This categorization can be relevant for medical or training purposes, but otherwise, there is nothing to suggest that this discretization of the recovery time is the best response for using in a predictive model. This problem has been stated outside of statistics as well. For example, Pollock et al. (2014) have stated that the current grading system for muscle injury classification provides little prognostic information and they suggest the adoption of a new system.

- **Missing data**: In some cases, data can be missing for various reasons. For example, database failures, or omissions on the part of the professionals can lead to missing values. Some other times, missing values come as a result of constraints imposed by the profession itself. For example, if a player goes out on loan to some other team, it might not be feasible for the club to know the exact training protocol that the player was previously following.

### 3.1.2 Handling typical data challenges

In order to make sure that the analysis is useful, a set of criteria has to be followed.

First, the models have to answer a practical problem. A statistical analysis can be entirely practically irrelevant if the results for whatever reason are not useful for the football professional. Some examples include:

- Results that might seem self-evident to anyone who has even some minimum experience in the field.

- Results that might depend on "ideal" data collection procedures. Data collection strategy in football can be very complicated, with many different professionals and divisions involved, and it is normal to expect errors in data entry. Any model or analysis should take into account the fact that the data can have a variety of issues, such as missing values, which will have to be dealt with.

- Results that can seem significant from a statistical perspective but irrelevant from a football perspective. For example, achieving X error on a predictive model might not be really practically applicable, even if it is a statistically significant result.

Therefore, a set of principles have to be set and followed when analyzing football medical data. The results have to be practically relevant, which can mean one of the following things:

- The results are directly applicable. This can be the case, for example, for predictive modeling.

- The results are not directly applicable, but quantify relationships which, until now, were based on experience and other qualitative measures.

- The results indicate the existence of information or patterns in a dataset, that, even if the models are not applicable, can indicate future research directions which could lead to practically applicable models.

These principles were followed in this thesis when running all investigations.

## 3.2 Datasets overview and problem explanation

The research in this thesis was conducted with datasets that were provided by Tottenham Hotspur Football Club (THFC) and one dataset from Wolverthampton Wanderers (WW). Both THFC and WW provided a dataset of injuries for the season 2012-2013. THFC also provided a dataset with exposure records, for the same season, plus a dataset with GPS variables recorded from training from the season 2014-2015.

Figure 3.1 shows the injuries that took place in the season 2012-2013 in THFC. The x-axis represents the number of days since the 8$^{th}$ of July which was the beginning of the pre-season period for the 2012-2013 season. Each line represents a player of the first team. The red dots indicate days where the players were injured.

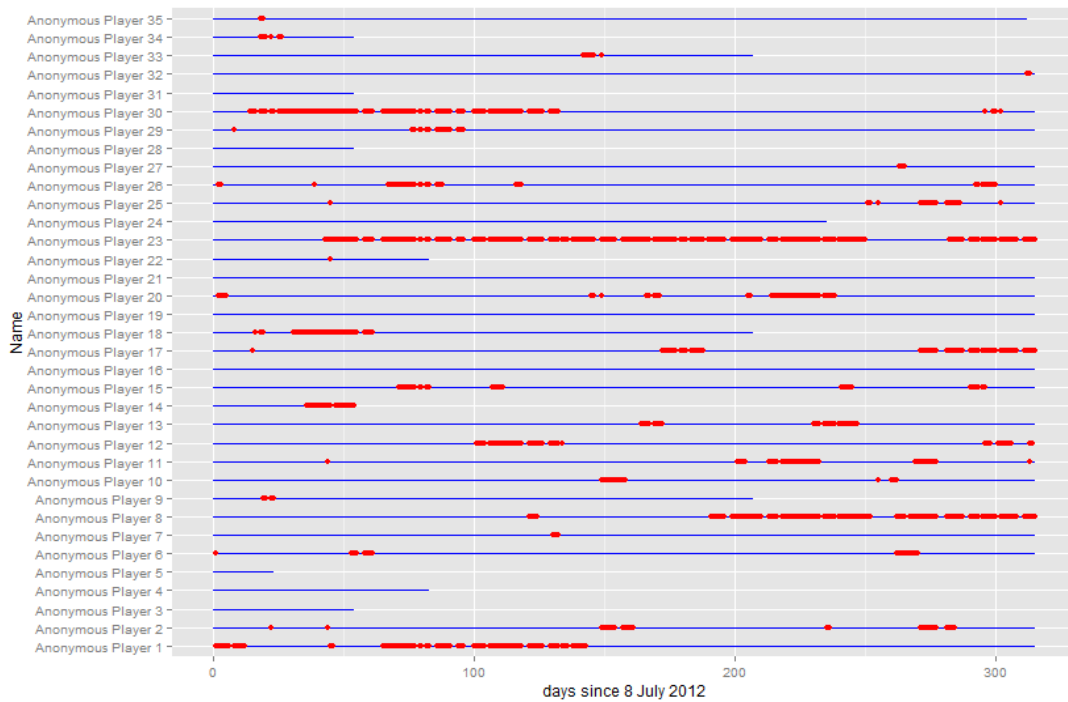

**Figure 3.1. Injuries for the first team of THFC in the season 2012-2013.**

A few things become evident. First, injuries are very common. Secondly, injuries can have large variability in their severity. Some injuries force a player to stay out of play for 1 or 2 days, while other injuries take months.

A question that can be asked is when will a player get injured?

Having a model for that would help the team in two ways. First, the team would understand how to structure the training in a better way, assessing each individual athlete's needs in a better way. Secondly, it would help the team plan in advance the schedule based on the expected number of injuries for each week or month. For example, it is clear from Figure 3.1 that there was a large incidence of injuries towards the second half of the season. If the team is not prepared for this, then it might end up going into important games handicapped. Having an appropriate model can only improve the chances of a the team to end up in a higher place in a tournament.

This problem was the focus of the second and the third investigations. In the second investigation training and exposure records from the professional squad of THFC from season 2012-2013 were used. The exposure records contained the number of minutes a player trained or played each day. The goal of this investigation was to predict when the first injury of the season would take place for each player, concentrating only on overuse and fatigue injuries.

In the third investigation, GPS data were used, collected from the training sessions of THFC from season 2014-2015. The GPS system collects a large number of variables, which can be useful indicators of fatigue and overuse. This dataset was used in order to predict the week of injury, focusing once again on overuse and fatigue injuries.

A second issue being raised is what happens during the injury and more specifically, when will a player return to play or training after the injury took place? There is some uncertainty related to that outcome. If the team had an idea of when players can return back from an injury they can devise a better strategy for the upcoming weeks.

This was the focus of the first investigation. Data from both THFC and WW were used that had been collected according to the UEFA injury standard. The data provide, mainly, extrinsic information about an injury.

Some of the challenges that were described in the previous section were present in this datasets as well. Noisy data, definition problems, and missing values showed up in all of these datasets, which required preprocessing before they become usable. Extensive preprocessing was done in collaboration with the professional staff of THFC, in order to ensure that any corrections were medically accurate. Also, in many occasions the medical records were checked against records of other departments in the club, in order to ensure consistency. Example inconsistencies include a player to have been recorded as injured when at the same time he has been recorded as actively training, or conflicting diagnoses about an injury.

# 4 Techniques and methods

*This section outlines the techniques and methods that were used in the research outlined in this thesis. An overview of each method is presented in this chapter. A justification behind the use of a method in an investigation lies in the relevant investigation chapter.*

## 4.1 Machine learning and statistical methods

### 4.1.1 Gaussian processes

A Gaussian process is a collection of random variables, any finite number of which have a joint Gaussian distribution (Rasmussen & Williams, 2006). A Gaussian process constitutes a distribution over functions. It is fully specified by a mean function $\boldsymbol{m(x)} = E[f(\boldsymbol{x})]$ and a covariance function $\boldsymbol{\Sigma(x, x')} = K(\boldsymbol{x, x'}) = E[(f(\boldsymbol{x}) - \boldsymbol{m(x)})(f(\boldsymbol{x'}) - \boldsymbol{m(x')})]$, where $f(\boldsymbol{x})$ is a real process.

Let $\boldsymbol{f}^*$ be the test outputs, $\boldsymbol{f}$ a vector that contains the response variable for the training output, let $X$ be the training input, $X_*$ the test input and $I$ the identity matrix. The dataset responses follow the distribution defined by (1):

$$\begin{bmatrix} \boldsymbol{f} \\ \boldsymbol{f}^* \end{bmatrix} \sim N \left( \boldsymbol{0}, \begin{bmatrix} K(X,X) + \varepsilon I & K(X, X_*) \\ K(X_*, X) & K(X_*, X_*) \end{bmatrix} \right) \qquad (1)$$

The term $K(X, X_*)$ is defined as the $n \times n_*$ matrix of the covariances evaluated at all pairs of the training inputs $x$ and test inputs $x_*$. The term $\varepsilon I$ represents the variance of Gaussian noise $\varepsilon$ added to each subject.

The prediction for instance $i$ is defined by equations (2)-(4):

$$f_i^* | X_i, X, \mathbf{f} \sim N(\hat{\mu}_i, \hat{\sigma}_i^2) \qquad (2)$$

where

$$\hat{\mu}_i = K(X_*, X)[K(X, X) + \varepsilon I]^{-1} \mathbf{f} \qquad (3)$$

and

$$\hat{\sigma}_i^2 = K(X_*, X_*) - K(X_*, X) K(X, X + \varepsilon I)^{-1} K(X, X_*) \qquad (4)$$

The covariance of a Gaussian process is specified by a positive semi-definite covariance kernel $K(\boldsymbol{x, x'})$. A kernel $\boldsymbol{K}$ is defined to be positive semi-definite if (5) holds:

$$\sum_{i=1}^{n} \sum_{j=1}^{n} \boldsymbol{K}(x_i, x_j) c_i c_j \geq 0, \qquad (5)$$

for any choice of real numbers $c_1, c_2, \ldots, c_n$.

Any positive semi-definite kernel can be used as a covariance function. Examples of covariance kernels include:

**Constant covariance kernel**

The constant covariance function defined in (6) and it represents the trivial case where the covariance is constant:

$$k(\boldsymbol{x}, \boldsymbol{x}') = C \tag{6}$$

**Radial basis function (RBF) kernel**

The radial basis function kernel defined in (7):

$$k(\boldsymbol{x}, \boldsymbol{x}') = \exp(-\sigma(\boldsymbol{x} - \boldsymbol{x}')) \tag{7}$$

where $\sigma$ is the lengthscale parameter. The radial basis kernel is a common choice of kernel, being the default in most cases (Duvenaud, 2014). One of the advantages of this kernel is that it is universal (Michelli, Yuesheng, & Haizhang, 2006), and so it can approximate any continuous function on a compact subset of the input space.

A problem with this type of kernel is that it cannot handle non-linearities very well. In that case, the lengthscale tends to become smaller as more data is added (Duvenaud, 2014).

**Polynomial kernel**

The polynomial covariance function defined in (8):

$$k(\boldsymbol{x}, \boldsymbol{x}') = (\sigma(\boldsymbol{x}^T \boldsymbol{x}'))^d \tag{8}$$

where once again $\sigma$ is the lengthscale parameter. The polynomial kernel is another popular kernel. In this case, a particularly interesting fact is that the degree of the kernel defines the highest order interaction of the variables. So, for example, a polynomial kernel of degree two, models all interactions up to the second degree.

Prior knowledge can be used in order to choose the best suited covariance function for a problem.

Training in Gaussian processes tries to solve the problem of identifying the best mean and covariance functions, along with the best hyperparameters (if hyperparameters are used) for these functions.

A common optimization method for this purpose is the maximization of the marginal log-likelihood, for which any gradient-based algorithm can be used. Some other approaches include grid search, Bayesian inference through Monte Carlo and multiple kernel learning (Murphy, 2012).

Gaussian processes, as a probabilistic method, have a number of advantages. First, it is possible to use prior knowledge in Gaussian processes through the careful choice of a covariance kernel. Secondly, Gaussian processes specify their output in probabilities and confidence intervals. Finally, Gaussian processes can produce smooth solutions and can work well in problem where this is an underlying assumption. An example of a Gaussian process is shown in Figure 4.1.

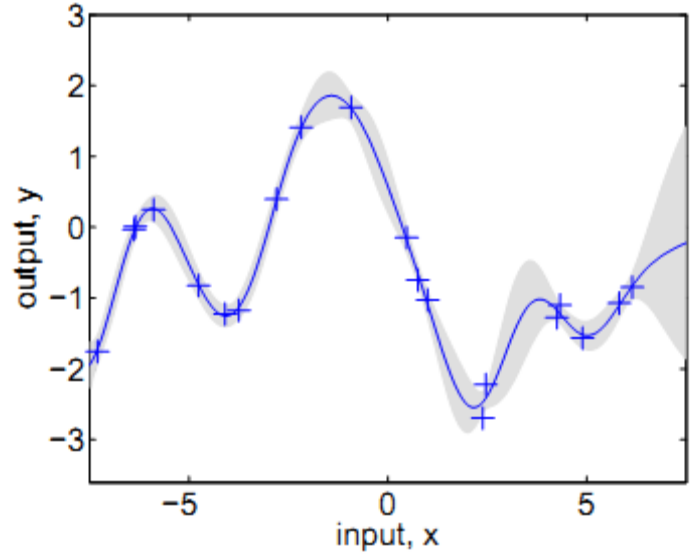

**Figure 4.1. Example of Gaussian process regression (Rasmussen & Williams, 2006). The crosses are the data points, the blue line is the prediction and the grey area is the 95% confidence interval. The confidence is distinctly greater in areas where there are more points.**

The main disadvantage of Gaussian processes is that they require inverting the covariance matrix $\Sigma$. This operation has a memory complexity of $O(n^2)$ and a computational complexity of $O(n^3)$ (Rasmussen & Williams, 2006) and it limits the applicability of Gaussian processes to big datasets.

Gaussian processes can be extended to classification, and ordinal regression (Chu & Ghahramani, 2005).

### 4.1.2 Support vector machines

Support vector machines (SVMs) are binary linear classifiers that can also be used for multi-class classification problems and regression. Their original formulation belongs to Vapnik and Cortes (1995).

SVMs solve the problem of finding the maximum margin hyperplane between two different classes of points. As a linear classifier, the decision function of an SVM is defined by formula (9):

$$\hat{y} = sign\left(\sum_i w_i x_i\right) \quad (9)$$

With $w_i$ being the weight that corresponds to the input $x_i$ and $\hat{y}$ being the result of classification (or regression).

The intuition behind support vector machines comes from the fact that linear classifiers can converge to different decision boundaries depending on their initialization settings and parameters. For example, all hyperplanes reproduced in Figure 4.2 are valid hyperplanes.

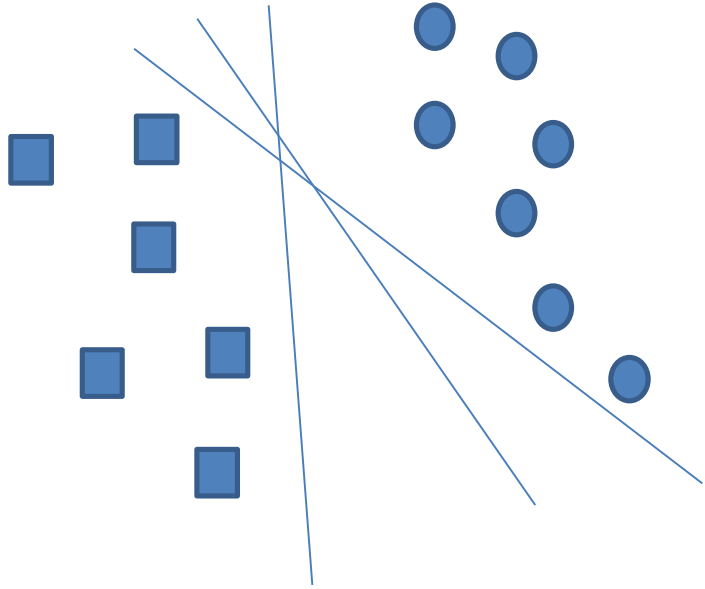

**Figure 4.2. Hyperplanes separating two clouds of points.**

However, not all hyperplanes seem equally plausible. The intuition behind support vector machines is that the optimal hyperplane is the one that has the maximum distance from both classes. This is called the "maximum margin hyperplane".

The objective function for support vector machines can be defined by (10):

$$J = C \sum_{i=1}^{N} L_e(y_i, \hat{y}_i) + \frac{1}{2} \|w\|^2 \qquad (10)$$

The term $L_e$ symbolizes some loss function[9]. The term $\|w\|^2$ is the norm of the weight vector and the term $C$ is acts as a regularization term.

Two important parts of SVMs is the kernel trick and the sparsity of the solutions. As part of the optimization objective of the SVM, $\hat{y}(x)$ can be reformulated by (11):

$$\hat{y}(x) = \hat{w}_0 + \sum_i a_i x_i^T x \qquad (11)$$

with $a_i$ being Lagrange multiplier for point $x_i$. The dot product can be replaced by a kernel function giving rise to equation (12):

$$\hat{y}(x) = \hat{w}_0 + \sum_i a_i k(x_i, x) \qquad (12)$$

The kernel is used in order to induce a non-linear transformation of the input space to a feature space. Since the original formulation of the SVM supports only the solution of linearly separable problems, the kernel trick, allows the transformation of non-linearly separable spaces to separable ones.

---

[9] This symbolism is useful because different loss functions are used for classification and regression.

The Langrange multipliers $a_i$ are what enforces the sparsity in the solution. In every problem there are usually only a few input instances that have $a_i > 0$. These are the instances that define the maximum-margin hyperplane, since the other instances are not taken into account. These instances are called support vectors and lend SVMs their name. A depiction of support vectors, along with the maximum-margin hyperplane is shown in Figure 4.3.

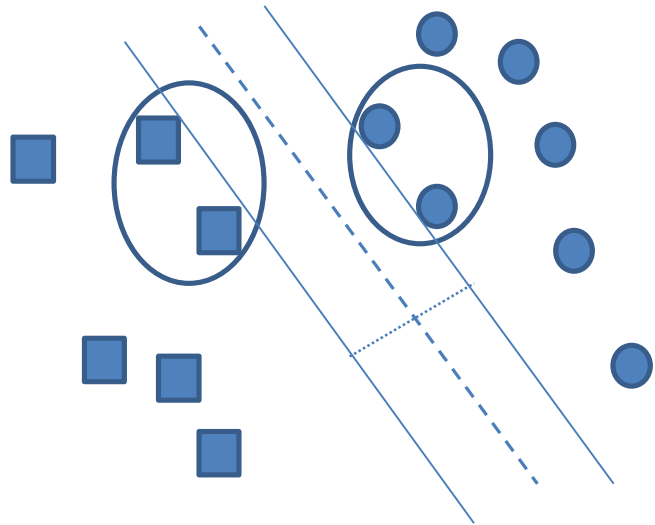

**Figure 4.3. Depiction of the support vectors. The fat dashed line is the maximum-margin hyperplane. The circled points are the support vectors.**

There are extensions for support vector machines for regression and multi-class classification, as well as ordinal regression (Chu & Keerthi, 2012).

So, in summary SVMs can be characterized by three main points (Murphy, 2012): the kernel trick, the maximum-margin hyperplane principle and sparsity. SVMs have been successfully applied to many problems (Wang, 2005). Their main advantages are their theoretical grounding in computational learning theory and their ability to find sparse solutions that generalize well. Also, the optimization problem for SVMs is convex, and so it has a single global optimum.

### 4.1.3 Neural networks

"Neural networks" is a term that describes a whole family of techniques and methods for both supervised and unsupervised learning. However, in machine learning, the term "neural networks" usually refers to what is the multi-layer perceptron (MLP) model. A multi-layer perceptron is a classification or regression model. It consists of different layers of nodes (or "neurons"), with connections among the layers. An example of an MLP is shown in Figure 4.4.

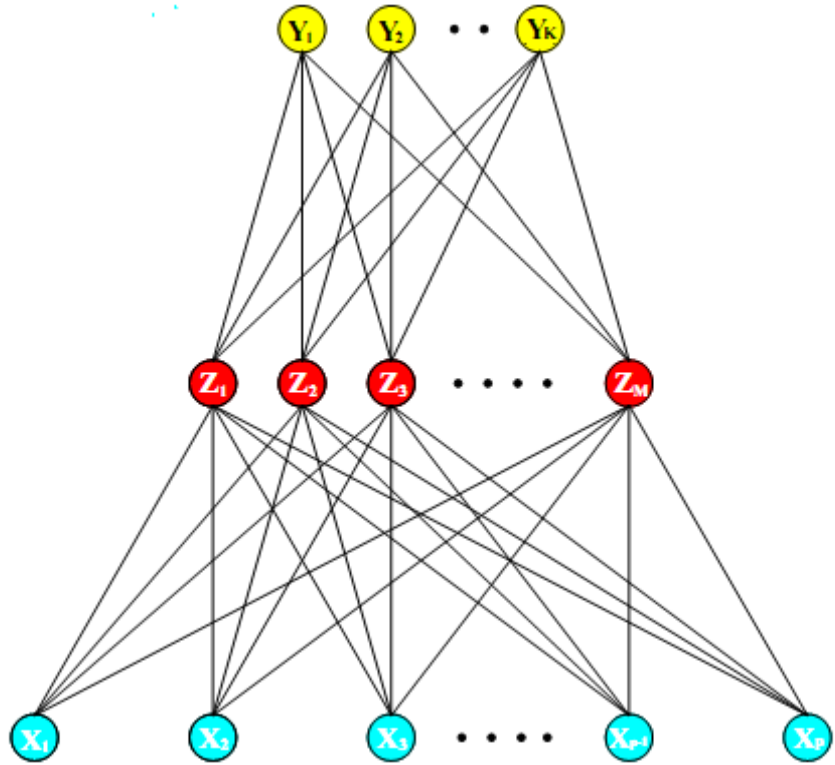

**Figure 4.4. Feedforward neural network example (Hastie, Tibshirani, & Friedman, 2009).**

The first layer is the input layer which consists of the values for each of the features of an instance of the dataset. The last layer is called the output layer and produces the network's prediction. The middle is comprised of one or more layers of neurons called "hidden layers".

The activation for a single neuron is calculated as the sum of its inputs, as can be seen in (13):

$$g = \sum_i w_i x_i \qquad (13)$$

with $g$ being the activation, and $w_i$ the weight corresponding to the input $x_i$.

For the hidden layers (and sometimes the output layer) this sum is passed through a non-linear function, called the "activation function". Probably, the most common activation function is the sigmoid shown in formula (14):

$$\sigma(x) = \frac{1}{(1 + e^{-x})} \qquad (14)$$

If the activation function chosen is a radial basis function, then the network is called a radial basis network.

MLPs are universal approximators. An MLP with a single hidden layer can approximate with arbitrarily small error any bounded continuous function, while an MLP with two hidden layers can approximate to an arbitrary accuracy any function (Cybenko, 1989).

The optimization goal is usually the root mean squared error (RMSE) between the predictions and the target values. Neural networks are usually trained through backpropagation, or other gradient descent methods such as Levenberg-Marquadt or conjugate gradient algorithms (Hagan & Menhaj, 1994). However, there have been successful applications of other optimization techniques, such as genetic algorithms or particle swarm optimization, in neural networks (Heaton, 2008). Neural networks can also be trained through stochastic gradient descent, which makes them really suitable for online applications.

The study of neural networks is an active area of research for decades and there is a multitude of successful models and applications in various fields (Paliwal & Kumar, 2009). There are many extensions of this model such as Kohonnen's self organizing maps (Kohonen, 1982) and deep neural networks (Hinton, Deng, Dong, & Dahl, 2012).

Neural networks, however, possess some disadvantages. First, their solutions are not easily interpretable.

Secondly, training a neural network can be as much an art, as it can be a science. Neural network training suffers form local minima and the choice of a learning algorithm and the choice of parameters can have a great effect on the final result. Therefore, training a neural network can be a computationally expensive procedure. However, their ability to handle non-linear interactions can be useful for particular problems.

It is common to use L2 regularization for neural networks, where, in this context, it is usually referred to as "weight decay" (Moody, 1995). Regularization in discussed in more detail in section 4.1.8.

### 4.1.4 Decision trees

The term "decision trees" describes a family of algorithms that concentrate on the idea of partitioning the space in different regions, and then applying a model (which is usually a constant) to each region (Hastie, Tibshirani, & Friedman, 2009). There are various algorithms for decision trees such as CART (Breiman L. , Friedman, Olshen, & Stone, 1984), ID3 (Quinlan, 1986) and C4.5 (Quinlan, 1993). There are decision tree algorithms for both regression and classification.

An example of a decision tree is shown in Figure 4.5. This example was taken from (Witten, Frank, & Hall, 2011) and considers the classification problem of a dataset with three categorical features (outlook, humidity and windy) and the binary response variable "rain today".

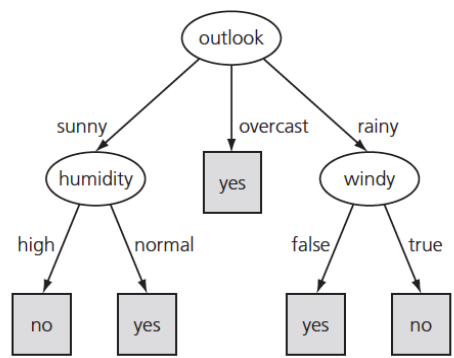

**Figure 4.5. Decision tree taken from (Witten, Frank, & Hall, 2011).**

The following pseudocode provides a simple implementation of a classification tree.

1. *Start at the root node.*
2. *Choose the feature F with the lowest impurity measure. Let K be the number of categories in this feature.*
3. *Create K different child nodes, one for each category in F.*
4. *Repeat the second step until all instances have been classified*

Common impurity measures are the misclassification rate, the Gini index, or the information gain (Witten, Frank, & Hall, 2011).

One of the main benefits of decision trees are that the output is easy to read and interpret. Another advantage is the fact that they can naturally handle categorical attributes and do not require dummification. However, numeric attributes require discretization.

**4.1.5 Random forests**

The random forest is a classification method developed by Leo Breiman (Breiman L., 2001). It is an ensemble learning method that combines two techniques: the random subspace method and bagged trees.

Bagging is an abbreviation of "bootstrapped aggregation". Bagging is an ensemble learning method where a set of *K* different classifiers are trained on *K* different sampled (sampling with replacement) versions $S_i$ of the original dataset *D*. Bagging is particularly effective on high variance-low bias algorithms (Hastie, Tibshirani, & Friedman, 2009). Decision trees are a common algorithm for classification and regression. There are different types of trees such as ID3 (Quinlan, 1986), CART (Breiman L., Friedman, Olshen, & Stone, 1984) and C4.5 (Quinlan, 1993).

The random subspace method is equivalent to bagging, but is only applied to the features of the dataset.

Random forests combine these methods into one. So, the final algorithm is summarized in the following pseudocode:

1. *Create K different datasets from the original dataset D, using sampling with replacement*
2. *For each sample $S_i$ choose a number of n features*
3. *Grow a decision tree for each $S_i$*
4. *When predicting a class $\hat{C}$, aggregate the K predictions $P_k$ of each tree. Assign $\hat{C} = majority\ vote(P_k)$*

A common setting for $n$ (the number of features chosen for building each tree) is equal to the square root of the number of features (Breiman L., 2001) and this is the approach that was followed in this thesis as well. An interesting property of random forests is that the out-of-bag error (OOB) is a very good estimate of the n-fold cross validation error (Hastie, Tibshirani, & Friedman, 2009). The OOB is calculated by taking, for each training instance, the proportion of trees in the ensemble that misclassified it.

Random forests embed feature selection and can also be used for assessing the importance of each feature. A way to do that is to calculate the average performance metric of a feature across all trees in the ensemble. For classification, this can be the Gini index or some other metric. For regression, this can be the root mean squared or other appropriate criterion for regression. This ability of random forests makes them a particularly suited technique in problems with many features.

### 4.1.6  Naïve Bayes

Naïve Bayes is a probabilistic classification method. Naïve Bayes makes the assumption that the input features are independent given the class label (conditional independence assumption). Figure 4.6 shows a graphical model representation of naïve Bayes.

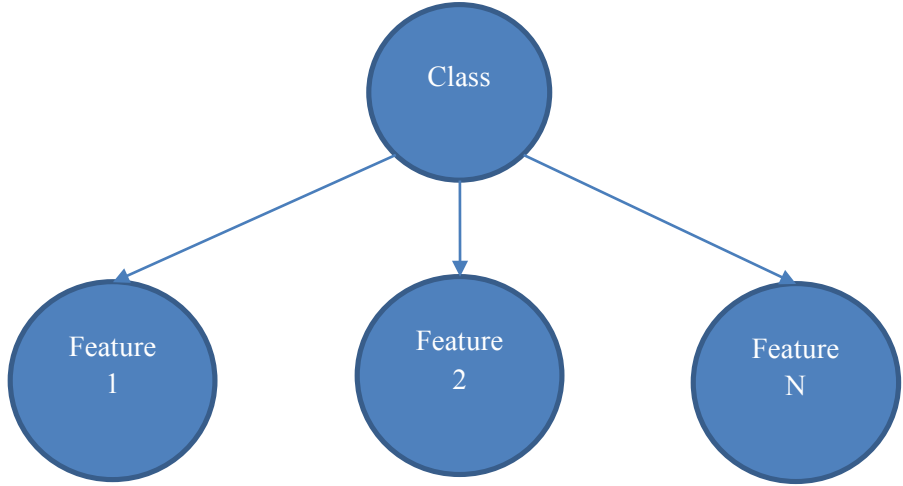

**Figure 4.6. Graphical model for Naïve Bayes.**

Let a binary classification problem, where the class is defined as $C \in \{0, 1\}$. The probability of the class taking a particular value given the input features is defined by (15):

$$p(C|F_1, \ldots, F_n) = \frac{p(C)p(F_1, \ldots, F_n|C)}{p(F_1, \ldots, F_n)} \qquad (15)$$

The problem with this equation is the term $p(C)p(F_1, \ldots, F_n|C)$. If we wanted to be precise and include all the interactions between all input features this term is re-written (using the definition of conditional probability) as seen in (16):

$$p(C)p(F_1, \ldots, F_n|C) = p(C)p(F_1|C)p(F_2|C, F_1) \ldots p(F_n|C, F_1, F_2, F_3, \ldots, F_{n-1}) \qquad (16)$$

The naïve Bayes assumption simplifies this to the much more simplified expression shown in formula (17):

$$p(C)p(F_1, \ldots, F_n|C) = p(C)p(F_1|C)p(F_2|C) \ldots p(F_n|C) \qquad (17)$$

So, the final form of the naïve Bayes classifier is defined by (18):

$$\hat{C} = argmax\ p(C)p(F_1|C)p(F_2|C) \ldots p(F_n|C) \qquad (18)$$

In case that the features are numerical a common strategy is to fit a Gaussian distribution over the data and use this to get estimates of $p(F_n|C)$.

Naïve Bayes is very fast algorithm to train, since all that is required are the frequencies for each feature. The simplicity of the naïve Bayes assumption makes it a particularly good model as a benchmark for more complicated models. Even though the naïve Bayes assumption might seem restrictive, it is the optimal classifier under both quadratic and zero-one loss when the assumption is correct. Also, the classifier can be optimal under zero-one loss even when the assumption is violated (Domingos & Pazzani, 1997).

### 4.1.7 K-nearest neighbours

K-nearest neighbours (k-NN) is a non-parametric classification method. It is a "lazy" method in that there is no real training stage. The classifier simply stores all instances and when a new instance appears, the classifier gets the K closest instances, based on some distance metric (such as Euclidean distance) and the outcome of the prediction is decided by majority voting.

In formal terms this is written as in (19) (Murphy, 2012):

$$p(y = c|\boldsymbol{x}, D, K) = \frac{1}{K} \sum_{i \in N_K(\boldsymbol{x},D)} I(y_i = c) \tag{19}$$

$$I(e) = \begin{cases} 1 \ if \ e \ is \ true \\ 0 \ if \ e \ is \ not \ true \end{cases}$$

where $D$ is the dataset, $\boldsymbol{x}$ the input vector and $I$ the indicator function.

For regression, the voting can be replaced by some other method, such as taking an average of the points.

K-NN is a low bias-high variance classifier (Hastie, Tibshirani, & Friedman, 2009). This can make it particularly prone to overfitting, but its simplicity can make it a relatively good benchmark for more complicated methods. Also, k-NN enjoys some consistency guarantees. As the amount of data tends to infinity, the error is no worse than twice the Bayes error rate (Cover & Hart, 1967). However, the algorithm is very prone to the curse of dimensionality, so it might not be a very good choice for datasets with lots of features (Murphy, 2012).

### 4.1.8 The generalized linear model and regularizers

Let $y$ be a response variable, $\boldsymbol{x}$ the covariates and $\boldsymbol{b}$ a vector of coefficients. The standard model for regression is written as in (20):

$$y = \boldsymbol{x}^T \boldsymbol{b} + \varepsilon \tag{20}$$
$$\varepsilon \sim N(0, \sigma^2)$$

However, this relationship forces the response to vary linearly with the predictors, which be a restrictive modeling assumption in many cases.

The generalized linear model extends this relationship by introducing the notion of the link function $g$ defined in (21):

$$E(y) = \mu = g^{-1}(\boldsymbol{x}^T \boldsymbol{b}) \tag{21}$$

The variable $y$ can follow any distribution from the exponential family and the function $g$ is called the link function.

A common measure-of-fit used in the generalized linear model is the deviance defined as in formula (22):

$$D = -2\left(\log\left(p(y|\hat{\theta}_0)\right) - \log\left(p(y|\hat{\theta}_s)\right)\right) \tag{22}$$

where $\hat{\theta}_0$ are the parameters of the model, and $\hat{\theta}_s$ are the parameters of the saturated model, with a parameter for every observation.

In place of the standard residuals used in linear regression, there are different kinds of residuals that can be used such as deviance residuals. For the sum of deviance residuals, relationship (23) holds:

$$D^2 = \sum_{1}^{n} d_i^2 \tag{23}$$

where $d_i$ is the deviance residual, whose form depends on the distribution of the response.

Some models used in these thesis include:

**Logistic regression**

Setting the link function to the one shown in (24):

$$g(\mu) = \log \frac{F(\mu)}{1 - F(\mu)} \\ \text{where } F(x) = \frac{1}{1 + e^{-x}} \tag{24}$$

leads to the logistic regression model. The link function provides the log odds of a class A versus class B but can easily be converted to the probability of class A shown in (25):

$$P(class = A) = \frac{e^x}{1 + e^x} \tag{25}$$

**Ordinal regression**

It is possible to extend the previous model to handle response variables with more than 2 ordered categories. Let the response be composed of $r$ categories. We define the sequence of cumulative logits as shown in (26):

$$L_1 = \log\left(\frac{p_1}{p_2 + p_3 + \cdots + p_r}\right)$$
$$L_2 = \log\left(\frac{p_1 + p_2}{p_3 + p_4 + \cdots + p_r}\right) \quad (26)$$
$$\cdots$$
$$L_{r-1} = \log\left(\frac{p_1 + p_2 + \cdots + p_{r-1}}{p_r}\right)$$

Then, the final model is defined as shown in (27):

$$\boldsymbol{L} = \boldsymbol{a} + \boldsymbol{\beta}^T \boldsymbol{x} \quad (27)$$

where **L** is a vector containing the cumulative logits for each category. This is equivalent to running multiple logistic regression models.

**Poisson regression and Negative binomial regression**

Another common choice for a link function is the log function, which is used in Poisson regression. Poisson regression can be applied when the range of the response variable lies in the range of the positive integers.

A core assumption of the Poisson distribution is that the mean is equal to the variance (overdispersion or underdispersion). This assumption can be checked before running the analysis by getting the mean and the variance of the response variable but also it can be tested after the analysis by using residual plots of the fitted values against the true values. In a plot that satisfies the assumption the variance of the points stays approximately the same across the whole range. If this assumption does not hold, a possible remedy is to use negative binomial regression which allows the variance to be different to the mean.

Also, when estimating the standard errors of the coefficients under heteroscedasticity, alternative methods can be used, such as White's heteroscedasticity-consistent estimator. The variance of the estimate of the coefficient $\hat{\beta}$ is given by (28):

$$v[\hat{\beta}] = (X'X)^{-1}(X' diag(\hat{\varepsilon}_1, \ldots, \hat{\varepsilon}_n)X)(X'X)^{-1} \quad (28)$$

where $\hat{\varepsilon}$ are the fitted residuals and $X$ is an $n \cdot k$ design matrix where $n$ is the total number of datapoints and $k$ the total number of covariates.

**Optimization and regularization**

The usual optimization goal for these models is the minimization of the sum of squares defined by (29):

$$\min_{\boldsymbol{b}} \sum_{i=1}^{n} (y - \boldsymbol{x}^T \boldsymbol{b})^2 \quad (29)$$

It is possible to improve the performance of a model on predictive tasks, by imposing a penalty on the size of the weights. This technique is called regularization. A specific kind of regularization ridge regression (or L2 penalty or else weight decay in neural networks) defined by (30):

$$\min_{\boldsymbol{b}} \sum_{i=1}^{n}(y - \boldsymbol{x}^T\boldsymbol{b})^2 + \lambda ||\boldsymbol{b}||^2 \tag{30}$$

where the parameter $\lambda \geq 0$ controls the amount of the penalty. Another choice is the LASSO (or L1) penalty defined by (31):

$$\min_{\beta} \sum_{i=1}^{n}(y - \boldsymbol{x}^T\boldsymbol{b})^2 + \lambda ||\boldsymbol{b}|| \tag{31}$$

The LASSO leads to sparse solutions, but it does not deal well with highly correlated variables. LASSO tends to select some of the variables arbitrarily and ridge regression has been shown to have better performance than LASSO in this context (Murphy, 2012).

The elastic net is a model that combines both penalties (Zou & Hastie, 2005). The standard version tries to solve an optimization problem of the form defined by (32):

$$\min_{\boldsymbol{b}} \sum_{i=1}^{n}(y - \boldsymbol{x}^T\boldsymbol{b})^2 + \lambda_1 ||\boldsymbol{b}|| + \lambda_2 ||\boldsymbol{b}||^2 \tag{32}$$

The parameters $\lambda_1, \lambda_2 \geq 0$ control the size of each kind of penalty.

**Cook's distance**

When fitting a regression model it is possible that one or more observations might have a strong influence. A way to measure that is Cook's distance (Cook, 1979). Cook's distance is defined by (33):

$$D_i = \frac{\sum_{j=1}^{n}(\widehat{Y}_j - \widehat{Y}_{j(i)})^2}{k\, MSE} \tag{33}$$

where $D_i$ is Cook's distance for point $i$, k is the number of fitted parameters in the model, MSE is the mean squared error and $\widehat{Y}_j$ is the prediction rom the model for observation $i$ for the original model, and $\widehat{Y}_{j(i)}$ is the prediction for point $j$ where $i$ has been omitted.

### 4.1.9 Supervised principal component analysis

Principal components analysis (PCA) is an orthogonal linear transformation that is commonly used as a dimensionality reduction technique. The original formulation belongs to Pearson (1901). Let $\boldsymbol{X}$ be a matrix with each column representing a variable and each row an observation. Now, let $\boldsymbol{\Sigma} = \frac{X^T X}{n}$ be the covariance of the data, after the columns of the matrix $\boldsymbol{X}$ have been centered by subtracting the mean. The covariance matrix can be decomposed as shown in (34):

$$\boldsymbol{\Sigma} = \boldsymbol{W}\boldsymbol{C}\boldsymbol{W}^T \tag{34}$$

Where $C$ is a diagonal matrix, where each element is an eigenvalue of $\Sigma$, arranged in order of magnitude starting from the largest eigenvalue on the top left, and $W$ is a matrix where each row corresponds to an eigenvector. The eigenvectors are called "components" and the elements of the eigenvectors are called "loadings". The first component captures the greatest percentage of variance in the dataset, the second component captures the second greatest amount of variance and so on. Since the matrix $W$ is orthonormal, the components are uncorrelated with each other.

PCA is particularly useful as a dimensionality reduction technique. By keeping only the first $n$ eigenvectors, it is possible to get a reconstruction of the data, while retaining a large percentage of the variance. PCA is often used as an explanatory technique, since the components can be interpreted as latent factors.

While PCA finds orthogonal components in the direction of greatest variance, this says nothing about the relation of the components to the response variable. Supervised PCA (Bair, Hastie, Debashis, & Tibshirani, 2004) is a technique to produce principal components that are correlated with the response and can improve the performance of a supervised learning task.

The supervised PCA algorithm works as follows:

1. Standardize the features.
2. Fit a univariate regression model for each feature separately.
3. Remove all features with coefficients $|\beta| < \alpha$, where $\alpha$ is some threshold
4. Compute the principal components using the reduced dataset.
5. Keep $m$ components and use them to fit the model.

The algorithm has been shown to produce good results in settings with a high number of features, relative to the number of observations, and is also particularly effective at handling correlated features (Hastie, Tibshirani, & Friedman, 2009).

### 4.1.10 Correlation-based feature subset selection

Feature selection is a very common problem in machine learning and data mining. There is a variety of algorithms for feature selection. An algorithm that is particularly effective is the correlation-based feature subset selection algorithm (Hall, 1999).

The "merit" of a set of features for predicting a particular class is measured by equation (35):

$$r_{zc} = \frac{k\bar{r}_{zi}}{\sqrt{k + k(k-1)\bar{r}_{ii}}} \tag{35}$$

In this equation k is the number of features, $\bar{r}_{zi}$ is the mean of the correlations between the features and the outside variable, and $\bar{r}_{ii}$ is the mean inter-correlation between the features. In case of numerical features, any correlation measure can be used. In case the features are categorical, then the correlation can be substituted by the symmetrical uncertainty. Symmetrical uncertainty is normalized in [0, 1] and is defined by formula (36):

$$symmetrical\ uncertainty = 2\left[\frac{gain}{H(X) + X(Y)}\right] \tag{36}$$

where gain H(X) and H(Y) is the entropy of categorical variables X and Y respectively, and gain is the information gain defined by formula (37):

$$gain = H(X) + H(Y) - H(X,Y) \qquad (37)$$

where $H(X,Y)$ is the mutual information of X and Y.

Correlation-based feature subset selection is based on the idea that a good set of features should be highly correlated with the class, while the features are not correlated with each other. A dataset can be optimized over this measure through heuristics, such as greedy search, or metaheuristics such as genetic algorithms.

## 4.2 Evaluating predictive models

### 4.2.1 Evaluating regression models on skewed data

A very common measure of the predictive fit of a regression model is the RMSE.

However, the RMSE penalizes severely larger errors. This behavior might not be always desirable, especially in problems where the response variable follows a skewed distribution. Data sampled from skewed distributions will contain a few or more points with values relatively large when compared to the average point. An error on these point will have a large effect on the value of the RMSE.

There are other metrics that are less susceptible to this problem, such as the mean absolute error (MAE) and the mean relative error (MRE). The problem with these metrics is that they might still be difficult to interpret for skewed datasets, since they are still affected by outliers. Therefore, other measures might be more appropriate.

The Pearson correlation coefficient is not affected by outliers, so it can be a more appropriate measure in those cases. The correlation in this context is calculated between the predicted values and the true values. A scatterplot between these two sets of values should form a 45°straight line passing through the origin. The Pearson correlation coefficient measures the linear relationship between two variables so it can provide an estimate of the performance of the fit in this context. However, Pearson's correlation coefficient does not take into account systematic biases into the data. Its value will be 1 for any linear relationship, even if this is not a 45° line through the origin.

Therefore, when facing problems from a right skewed distribution a metric is needed in order to assess whether the predicted and the true values fall on a 45° line through the origin. The concordance correlation coefficient (Lin, 1989) is such a measure. The concordance correlation coefficient is defined by (38):

$$\rho_c = \frac{2\rho\sigma_x\sigma_y}{\sigma_x^2 + \sigma_y^2 + (\mu_x - \mu_y)^2} \qquad (38)$$

The concordance correlation coefficient can be used to assess the agreement between the predicted values of a statistical model and the actual values. The advantage over Pearson's correlation coefficient is that Pearson's correlation coefficient ignores any bias that there might be between the true and the predicted values. Pearson's correlation coefficient will assign a high value to any relationship $y = ax + b$, while the concordance correlation coefficient will penalize any relationship that deviates from $y = x$. This ensures a stricter evaluation of the agreement between predicted values and the response.

The concordance correlation coefficient and error metrics such as the RMSE or the MAE can deviate from each other. RMSE and MAE can be severely affected by a large error to a single case, ignoring small improvements over many other cases. The concordance correlation coefficient does not suffer so much from this problem. This is better illustrated in Figure 4.7 below:

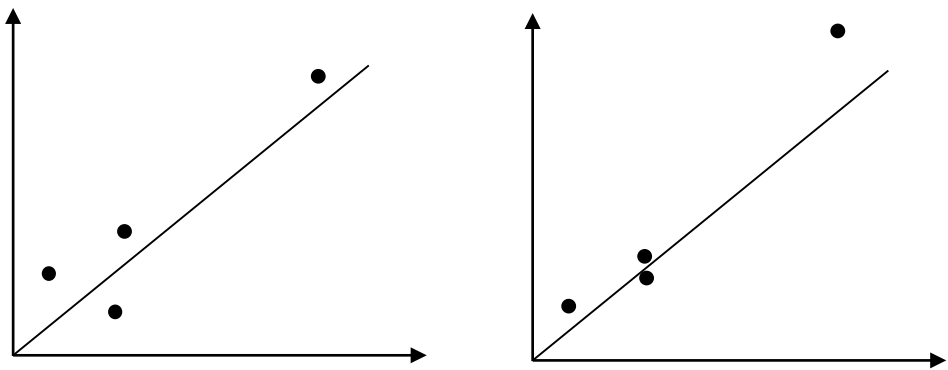

**Figure 4.7. Example of the difference between MAE and the concordance correlation coefficient.**

The two graphs represent plots of the true versus the predicted values. The black line is the 45° line through the origin. The point on the upper right part has the largest response out of all points. The right plot presents a case where the prediction on this point is worse, but improves for the three points with small response. The RMSE and the MAE can give the exact same error for both cases, because the error on the point with a large value counteracts the improvement on the cases with a smaller response. The concordance correlation coefficient, however, will improve in the second case.

**4.2.2   Evaluating classification models for data with unbalanced classes**

A common way to evaluate the success of a classifier is the accuracy, which is defined as the number of correct instances classified. However, accuracy might not give accurate results when dealing with skewed data where the classes are unbalanced.

The problem with unbalanced classes is that in many cases a very good accuracy score can be reached by simply guessing the majority class. So, if, for example, 60% of the data belong to class A, with the rest of the data split equally among classes B, C, D and E, then simply guessing A will yield an accuracy score of 60%. This makes it difficult to understand whether the classifier has learned a concept from the data, or is simply guessing the majority class.

A metric that can help with this problem is Cohen's kappa (Cohen, 1960). Cohen's kappa is defined by (39):

$$\kappa = \frac{\Pr(a) - \Pr(e)}{1 - \Pr(e)} \qquad (39)$$

Cohen's kappa was originally developed as a tool of inter-rater agreement (Carletta, 1996). Pr(a) is the percentage of agreements between the two raters, and Pr(e) is the percentage agreement that would have been achieved by chance alone. When measuring the accuracy of a classification algorithm the first rater is represented by the predictions of the algorithm and the second rater by the ground truth. Cohen's kappa ranges between 0 and 1, with a value of 1 indicating perfect agreement. A particular benefit of Cohen's kappa is that it can also work with multiclass classification problems.

## 4.3 Dynamic time warping

Comparing time series of different lengths is a non-trivial task. Measures of association such as Pearson's correlation coefficient assume an equal number of elements for every subject. Dynamic time warping (DTW) is an algorithm that allows the comparison of time series of different lengths by finding an optimal way to match two sequences of different length (Müller, 2007).

Consider two time series $R$ and $L$ of length $n$ and $m$, respectively and let $d(R_i, L_j)$ be some distance between the elements of the time series. Define $d(R_i, L_j)$ to be the distance between two elements. An optimal warping path is a set of matrix elements $w_1, w_2, w_s, ..., w_k$, where $w_s$ is any set of indices $(i, j)$ such that the total distance $\sum d(R_i, L_j)$ is minimized. For the purposes of the present work, the Euclidean distance was used.

The dynamic time warping distance between two elements $i$ and $j$ can be minimized through the recursive relation shown in (40):

$$DTW(i,j) = d(R_i, L_j) + \min\{DTW(i-1, j-1), DTW(i-1, j), DTW(i, j-1)\} \quad (40)$$

subject to the following boundary conditions show in (41) that ensure that the warping path starts at the first element of the series and ends at the last element of both series:

$$DTW(1,1) = 0 \quad (41)$$
$$DTW(n, m) = \infty$$

Additionally, Equation (42) implicitly enforces the following monotonicity condition, which ensures that the warping path will not go backwards in time:

$$if\ w_k = (i, j), then\ w_{k-1}(i', j'), where\ i' \leq i\ and\ j' \leq j \quad (42)$$

A proof of the optimality of DTW for finding the path with the minimum warping distance can be found in (Müller, 2007).

Figure 4.8 shows the alignment of two time series from the dataset through DTW.

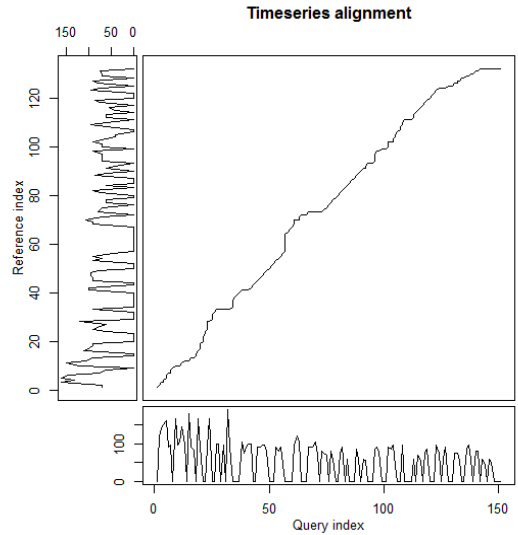

**Figure 4.8. Example of DTW for aligning two exposure time series. The main graph shows the path through the distance matrix.**

## 4.4 Software used

Various software packages and languages were used for conducting the research in this thesis. More specifically, data manipulation was conducted mainly in Python 2.7 with the "pandas" package and in R. Some additional manual tweaking had to be done in Excel.

Weka and Rapidminer were used mainly for the first investigation (Chapter 5) for testing out ideas quickly, but all the algorithms were run in R using the caret package.

Matlab was used mainly for its neural network toolbox when testing some ideas for the second investigation (Chapter 6). However, none of the results made it to the thesis. The code for the second investigation was eventually written in R.

The code for the third investigation (Chapter 7) was written in R.

# 5 Predicting the recovery time after a football injury using the UEFA injury recordings

*The goal of the first investigation is to predict the recovery time of football players after an injury based on information retrieved from the UEFA injury recordings. The response variable is the recovery time in days since the onset of injury. The problem is approached using three types of analysis. First a statistical investigation using generalized linear models is conducted, in order to get a better understanding of the dataset and assess the significance of the covariates. Then, a variety of machine learning algorithms (random forest, naïve Bayes, k-NN, neural networks, SVMs and Gaussian processes) are applied compared for the predictive model. Finally, feature selection was performed and then the machine learning algorithms are re-applied improving the final outcome. The results illustrate that the UEFA recordings contain useful information that can be used for predicting the recovery time after an injury.*

## 5.1 Introduction and motivation

### 5.1.1 Overview

A common issue regarding injuries is the time taken to return to play after an injury.

An accurate estimate of that time would benefit the manager who could plan team selection and strategy taking into account likely absences from the squad for the upcoming games. In the case of severe injury, the coach may be able to make suitable adjustments within the squad or sign new players to cover any absences if appropriate.

Secondly, an accurate estimate of when a player is expected to return to play would also aid the medical team at a football club. The medical team would have an evidence base to refer to when explaining the likelihood of absence to the club manager. Such a base may prevent any disagreements between medical staff and manager and potentially might also help inappropriate early returns to play.

Finally, the injured player would be helped as an accurate estimate would aid his expectations about when to return to play. There are psychological factors that might be involved in recovery and it is important to manage a player's expectations. A player could be demoralized if he expects much quicker recovery than should be expected. Furthermore, if the player is a key player of the team, his recovery could also influence the morale of other players.

The best current estimators of time to return to play following injury in a professional footballer has been the experience of the physician and epidemiological studies of injuries to this specific cohort. As a result, the prediction of time to return to play may be quite variable.

For example, "return to play" following anterior cruciate ligament reconstruction can range from 16 to 24 weeks (Bizzini, Hancock, & Impellizzeri, 2012). Similarly, different recommendations exist for concussions (Cantu, 1986; Dicker, 1991; Collins, Lovell, & McKeag, 1999; Lovell, Collins, & Bradley, 2004) and hamstring injuries (Mendiguchia & Brughelli, 2011). This problem has also been stated by Moen et al. (2014) regarding recovery from hamstring injuries which can range from 1 to 104 days. Therefore, it is clear that estimating the "return to play" after an injury is an area that needs to be researched further.

As part of the UEFA guidelines, teams in the premier league have to collect information on every injury that takes place. This information mainly concerns the description about an injury (e.g., whether the player was running, whether there was a collision, etc.) along with some extra pieces of information (e.g. whether the injury is recurrent).

The UEFA injury recordings are one of the few cases where professional football teams have to follow a pre-specified standard for data collection. Also, the UEFA injury recordings are quick and easy to fill out for an injury, so their adoption by new clubs is easy. These two points become very important if we take into account the problems of data collection in football as were discussed in section 3.1. A predictive model based on the UEFA recordings could be implemented very quickly by any team.

The purpose of this investigation is to build a predictive model for the recovery time of professional football athletes after an injury has taken place. The goal is to make the prediction based on information available at the time of injury.

Obviously, the model could benefit from the inclusion of more variables, such as a player's medical exams or training records.

However, this information was not available. This problem is directly related to the issues discussed in section 3.1. Even if this information was available for one club, it is likely that the information would not be available for another club or it would be available following a different recording standard.

Therefore, establishing a predictive baseline on a commonly accepted data standard becomes important for making progress in the prediction of the recovery time in football injuries.

The primary goal of this study is to test the degree to which this task is feasible. Once this was established, the next goal was to see whether some methods are more suited for this task compared to others. Finally, the datasets and the features themselves were processed using feature selection in order to find a subset of features that could help improve the performance of the models.

The results are presented, along with their significance for the football and sports analytics community.

#### 5.1.2 Research goals

The research goals of this investigation were the following:

- To identify whether the UEFA recordings, as a standard in recording injuries, can be useful for predicting the recovery time.
- To understand which variables from the UEFA recordings seem to be more relevant for the task.
- To compare different algorithms for this task.

### 5.2 Design and methods

#### 5.2.1 The datasets
The three datasets used for this study are:

- The Tottenham Hotspur Football Club (THFC) UEFA injury recordings dataset.

- The Wolverthampton Wanderers (WW) UEFA injury recordings dataset.
- The integrated dataset, which was a merge of the THFC and the WW datasets.

The common variables between the first two datasets are shown in Table 5.1 below.

**Table 5.1. List of common variables between all datasets.**

| Variable | Description |
|---|---|
| Activity | Describes whether the injury took place in the training field or in the game |
| Phase of play/Mechanism | Describes the exact way that the injury happened (e.g. running or shooting) |
| Injury | Description of the injury in broad terms (i.e. muscle, ligament, bone or tendon) without a specific diagnosis |
| Injured side | Describes whether the left or right side was injured |
| Body part injured | Where the player was injured |
| Recurrence | Describes whether the same injury has happened to the same player in the past |
| Days unavailable | The main variable of interest in our model. It specifies how many days a player stayed out of play after his injury |
| Severity | This is a categorical description of the "Days unavailable" variable |

The first dataset consists of a list of injuries at THFC which were recorded according to the UEFA guidelines. The total number of instances was 152 and the total number of variables was 10. The unique variables for this dataset are presented in Table 5.2. Note that the variable "injury" included in the dataset is not a final diagnosis, but a first general estimate such as "muscle strain" or "bone injury".

**Table 5.2. List of variables unique in the THFC dataset.**

| Variable | Description |
|---|---|
| Age | The age of a player |
| Stage of season | The stage of season (e.g. mid-season or off-season) when the injury occurred |
| Type | Describes whether the injury was due to overuse or it was an acute injury |
| Position (when injured) | The position of the player at the moment of injury (e.g. forward) |

The official diagnosis was left out of the model. Diagnoses consist of free text, and the medical staff takes some freedom in the degree of elaboration. For example, in this study's dataset there were some knee injuries that were described as "knee pain, unspecified", "patellofemoral pain" and "Left knee medial meniscus". These diagnoses could be elaborated even further, or they could be abstracted, by classifying them all as "knee injuries".

However, it is not entirely clear what degree of elaboration would actually help in the prediction of the response variable. Furthermore, it is not always clear when diagnoses typed in two different ways for two different entries actually refer to the same thing, or to an injury that for the purposes of a predictive model could be considered as equivalent.

Therefore, it was decided that the best way to advance research in that front would be to know what degree of accuracy can be achieved in the prediction of the response variable before including the diagnosis, so that future research could actually tackle the problem of trying to identify the correct level of abstraction needed for this task. This result could be used to establish a baseline which future, more elaborate models, will improve.

Something else to note is that all head injuries (5 in the THFC and 2 in the WW dataset) were removed. The reason is that head injuries are treated in a different way to other types of injuries, related to muscles and joints, due to the fact that the physiology of the brain is different to that of the rest of the body. In contrast to other types of injury, head injuries are accompanied by neuropsychological testing in order to assess brain functioning and suitability of return to play and can be accompanied by long-lasting cognitive deficits (McCrory, Makdissi, & Collie, 2005).

The second dataset consists of the UEFA recordings collected by WW. The unique variables for this dataset are shown in Table 5.3. The total number of instances was 78 and the total number of variables was 11.

**Table 5.3. List of variables unique in the WW dataset.**

| Variable | Description |
|---|---|
| Date of Injury | The date that the injury occurred |
| Footwear | Self-describing |
| Surface condition | Describes the condition of the ground at the moment of injury, e.g. wet or dry |
| Strapping | Whether the injury region was strapped or not right after the injury |
| Referee's decision | The referee's decision on whether there was a foul or not |

There were discrepancies between the two datasets, because the teams have some freedom on how they should note the information down. Table 5.4 shows the common variables and the dataset-specific variables.

**Table 5.4. Common and non-common variables between the two datasets.**

| Common variables | THFC only variables | WW only variables |
|---|---|---|
| Activity, Days unavailable, Severity, Recurrence, Body part injured, Phase of Play/Mechanism, Injury, Side, Stage of season/Date of injury | Age, Stage of season, Position, Type | Surface Condition, Referee's decision, Strapping, Footwear, Date of injury |

The third dataset consists of an integration of the two datasets and contained only the common variables. The total number of variables was therefore 7 and the total number of rows was 230. All variables were categorical with the exception of the response "Days unavailable", the variable "Age" for the THFC dataset and the variable date for the WW dataset.

The categories were harmonized by combining categories that referred to the same conceptual entity. For example, one dataset contained an entry regarding the "Phase of play" as "Collision – Tackled by other player", while the other dataset used "Tackled by other player". In the majority of cases it was clear to see how to harmonize the categories. In the cases where that was not clear, the original category from the corresponding dataset was kept intact.

A histogram of the recovery time (in days) is shown in Figure 5.1 for all datasets. It is evident that most of the injuries are less than 25 days and the histogram is skewed with a tail on the right.

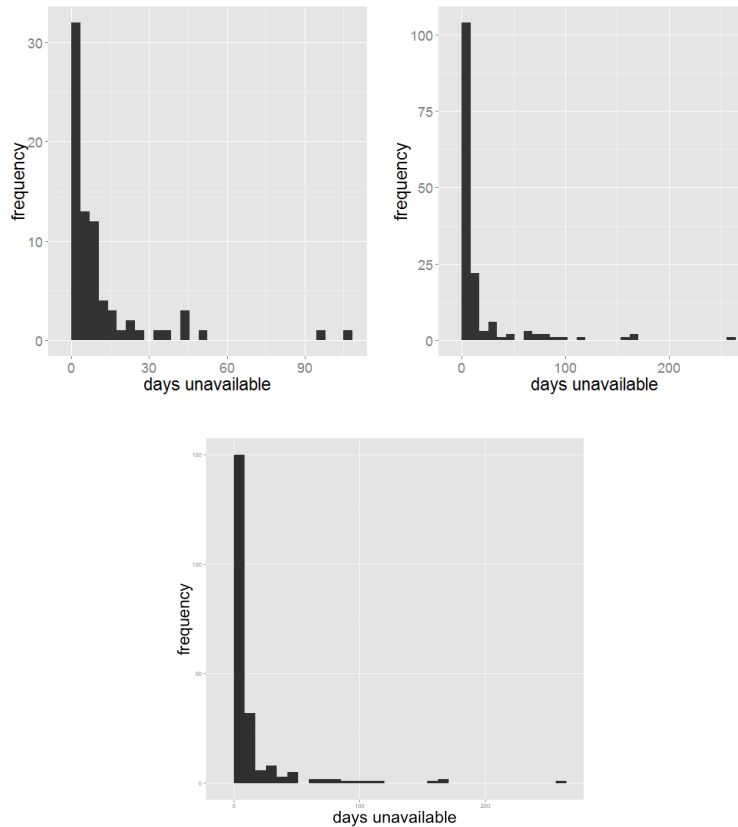

**Figure 5.1. Histogram of the response variable "Days unavailable" for the WW (left), THFC (right) and the integrated datasets.**

Table 5.5 below shows summary statistics for all three datasets.

**Table 5.5 Mean, median and standard deviation of the response variable for all datasets.**

|  | **Mean** | **Median** | **Sd** |
|---|---|---|---|
| **THFC** | 15.5 | 2 | 36.03 |
| **WW** | 10.78 | 5 | 18.47 |
| **Integrated** | 13.83 | 3 | 31.23 |

The variable "severity" is a classification for the "days unavailable" variable used by medical professionals and the UEFA injury standard. The categorization is as follows:

- Transient: 0-7 days lost
- Mild: 8-28 days lost
- Moderate: 29-84 days lost
- Severe: 84+ days lost

Table 5.6 below shows the frequency table for the "Severity category" for the THFC and the WW dataset.

**Table 5.6. Frequency table and proportions for the variable "Severity" for the THFC and the WW datasets.**

|  | WW | | THFC | | Integrated | |
|---|---|---|---|---|---|---|
|  | Count | Percent | Count | Percent | Count | Percent |
| **Transient** | 52 | 66.6% | 101 | 66.4% | 153 | 66.6% |
| **Mild** | 18 | 23% | 29 | 19.1% | 47 | 16.0% |
| **Moderate** | 6 | 7.8% | 14 | 9.2% | 20 | 8.6% |
| **Severe** | 2 | 2.5% | 8 | 5.3% | 10 | 4.3% |
| **Total** | 78 | 100,0 | 152 | 100.0% | 230 | 100% |

Figure 5.2 shows the bar chart for severity for each dataset respectively. It's clear that the majority of injuries were transient.

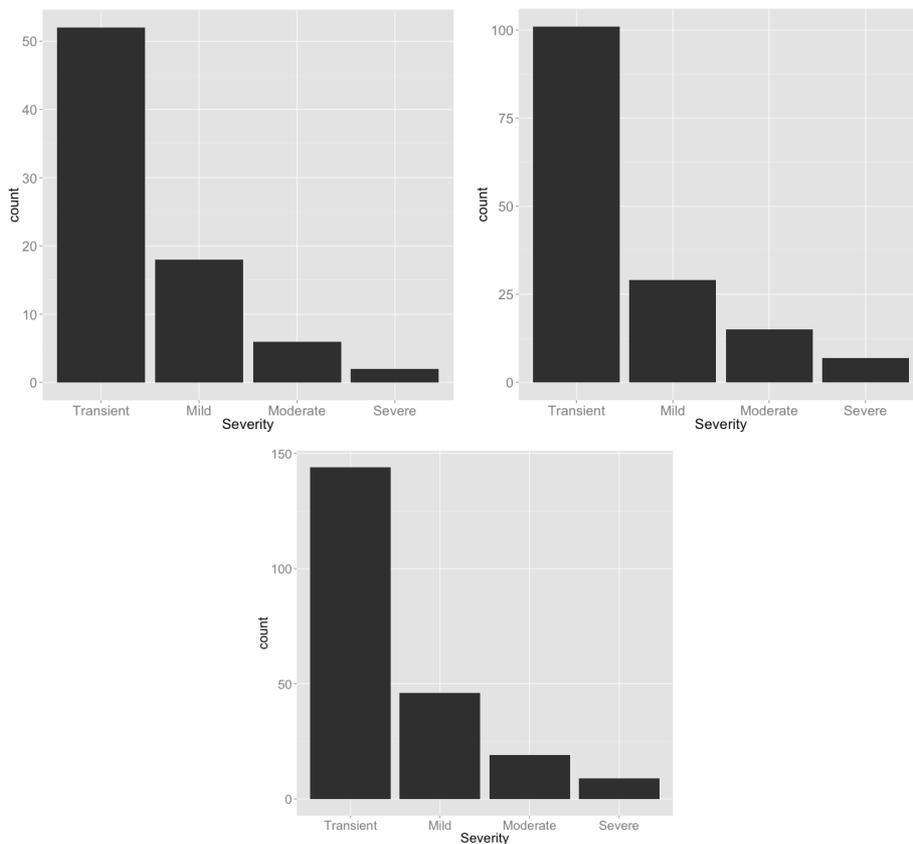

**Figure 5.2. Severity bar chart for the WW (left), THFC (right) and the integrated datasets.**

Something worth mentioning that this categorization has not been devised having a statistical classification task in mind, but rather it is based on medical practice. It is not improbable that other types of classification might be better for a predictive classification model, even if the current categorization is used within the football medical community.

It also worth noting that there is not even a consensus on how complete recovery is defined and no methodology exists to assess that (Hagglund, Walden, & Ekstrand, 2003). The players are usually considered rehabilitated once they receive clearance from the medical staff, but pressure to play on important games, or misdiagnoses might affect when they get reintroduced to play. Furthermore, the player's compliance towards the rehabilitation protocol can affect his recovery as well, but there was no record of compliance in these datasets.

### 5.2.2 Exploratory analysis and graphs for common features

This section will provide an exploratory analysis of the common features from both datasets.

**Recurrence**

Figure 5.3 shows bar charts of "recurrence" for the three datasets. The majority of cases are first episodes, but a substantial number of cases are recurrent injuries. For the integrated dataset, "early", "delayed" and "late" recurrence were simply converted to "recurrence".

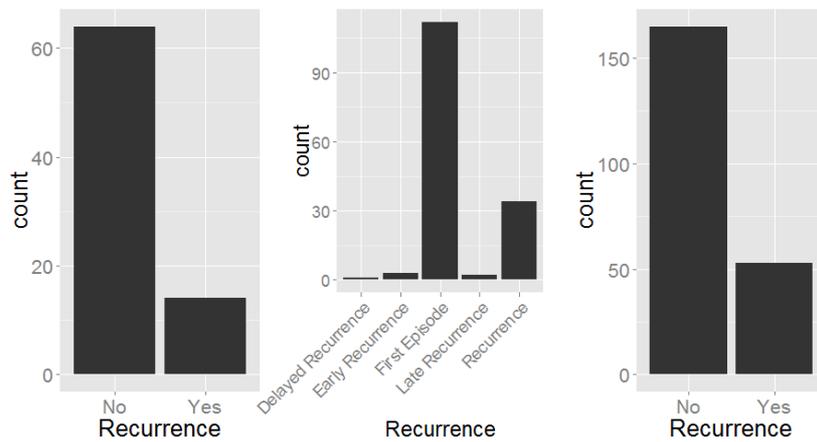

**Figure 5.3. Bar charts of "recurrence" for the WW, THFC and the integrated dataset (from left to right).**

There are some discrepancies as to how injuries were recorded in the two teams, with THFC opting for a more fine-grained description of recurrence. However, the overall picture is similar for both datasets as Table 5.7 shows.

**Table 5.7. Proportion of "recurrence" for all datasets.**

|  | WW | THFC | Integrated |
|---|---|---|---|
| **Recurrence=Yes** | 82.1% | 73.7% | 24.3% |
| **Recurrence=No** | 17.9% | 26.3% | 75.6% |

The existence of previous injuries is an important factor for subsequent injuries (Chomiak, Junge, Peterson, & Dvorak, 2000; Arnason, et al., 2004). It is reasonable to expect that subsequent injuries might affect the recovery time, if the recovery from previous injuries is not complete, or if the area has developed increased sensitivity.

Indeed, Table 5.8 below shows that recurrent injuries have larger recovery times across all datasets.

**Table 5.8. Recovery (mean+/-standard deviation) for recurrent and non-recurrent injuries across all datasets.**

|  | WW | THFC | Integrated |
|---|---|---|---|
| **Recurrence=Yes** | 22.6+/-34.1 | 37.2+/-55.6 | 27.8+/-47.9 |
| **Recurrence=No** | 8.1+/-11.7 | 9.1+/-25.6 | 9.3+/-21.8 |

**Injured side**

Figure 5.4 shows bar charts of the variable "injured side".

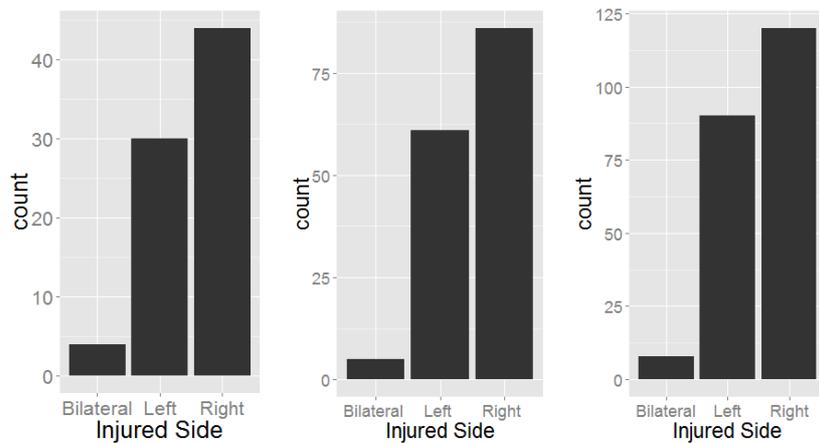

Figure 5.4. Bar chart of the "injured side" for the WW, THFC and integrated dataset (from left to right).

Table 5.9 shows the proportions of the injured sides for both datasets. The proportions of right and left footed footballers are similar in both.

Table 5.9. Proportion of the different types of "injured side" for all datasets.

|  | WW | THFC | Integrated |
|---|---|---|---|
| **Left** | 38.4% | 40.1% | 41.2% |
| **Right** | 56.4% | 56.5% | 55.0% |
| **Bilateral** | 5.1% | 3.3% | 3.6% |

The side of the injury might not seem as a very relevant variable, until we take into consideration that the players usually have a dominant leg. It has been documented that the dominant leg might be more prone to particular injuries (Ekstrand & Gillquist, 1983; Ekstrand, Hägglund, & Waldén, 2011). The dominance of one leg can lead to imbalance and imbalance between the two legs can be a risk factor for injuries (Yeung, Suen, & Yeung, 2009).

Unfortunately, information about imbalance was not available in the datasets. However, the side of the injury can be an indirect way of getting this information, if we take into account that the majority of the players had the right leg as the dominant one.

Table 5.10 shows that indeed there is not much differences between the recovery times for left and right sided injuries. Bilateral/other injuries have shorter recovery times, but this must be due to the fact that these are mainly upper body injuries, such as wrist or neck injuries, that usually are less serious than injuries in the lower extremities.

Table 5.10. Recovery (mean+/-standard deviation) for the variable "injured side" across all datasets.

|  | WW | THFC | Integrated |
|---|---|---|---|
| **Left** | 10.2+/-18.6 | 16.2+/-33.8 | 14.2+/-29.8 |
| **Right** | 12.0+/-19.1 | 15.6+/-38.5 | 14.1+/-33.2 |
| **Bilateral/other** | 3.8+/-4.9 | 1.5+/-1.2 | 3.1+/-3.8 |

**Activity**

Figure 5.5 shows a bar chart of the activity that was being conducted when the injury took place, while Table 5.11 shows the respective proportions. Most injuries take place during matches.

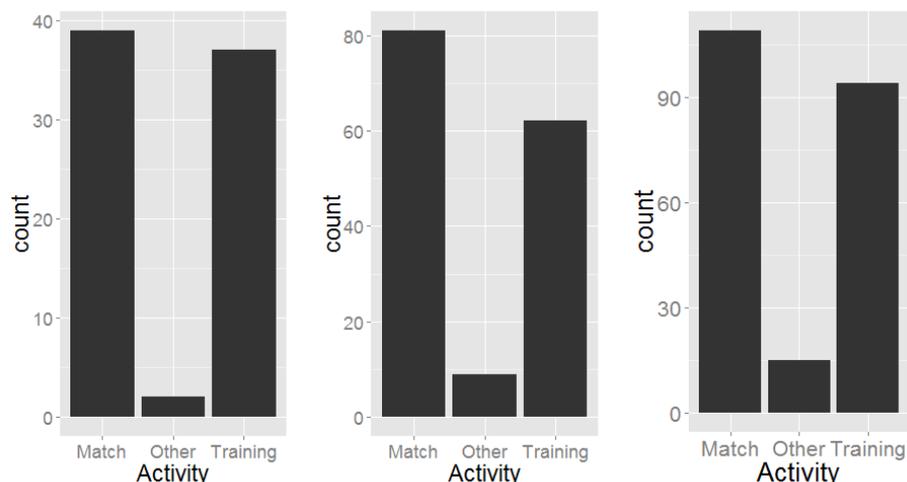

**Figure 5.5. Bar chart of "activity" for the WW, THFC and integrated datasets (from left to right).**

**Table 5.11. Proportion of "activity" across all datasets.**

|  | WW | THFC | Integrated |
|---|---|---|---|
| **Match** | 50% | 53.2% | 51.1% |
| **Training** | 47.4% | 40.7% | 41.9% |
| **Other** | 2.5% | 5.9% | 7.0% |

It seems that there are more injuries taking place during matches than training. This makes sense if we take into account the aggressive nature of the sport and the fact that many injuries take place as a result of collision. The UEFA elite clubs report for 2012/2013 (Ekstrand J., 2013) noted that the training to match ratio for the participating clubs had a mean of 3.5. However, there were 3.4 injuries per 1000 training hours averaged across all clubs, while the same mean for matches was 22 injuries per 1000 match hours, demonstrating the hazardous nature of football matches compared to training (Ekstrand J., 2013).

Table 5.10 below shows that for this data there were not huge differences for the recovery time between different activities.

**Table 5.12. Recovery (mean+/-standard deviation) for different types of activity across all datasets.**

|  | WW | THFC | Integrated |
|---|---|---|---|
| **Match** | 8.3+/-11.0 | 17.7+/-42.5 | 12.3+/-26.2 |
| **Training** | 13.6+/-24.1 | 14.2+/-30.8 | 15.7+/-36.5 |
| **Other** | 6.0+/-2.8 | 11.5+/-33.5 | 10.5+/-30.1 |

**Phase of play/Mechanism of injury**

Figure 5.6 shows a bar chart of the "phases of play" that are involved in an injury. The majority of injuries take place when running or sprinting.

There was not a completely clear correspondence between the two datasets. For example, the "N/A" category in the THFC dataset corresponded to injuries such as those that took place during weightlifting training. The "Other" category in the WW dataset had a wider meaning, and could include the categories "N/A", "Unknown Mechanism" from THFC.

Similarly, the "kicking" category in the WW dataset could correspond to either "shooting" or "other acute mechanism" in the THFC dataset.

So, the two datasets in that case were harmonized by simply adding up the categories.

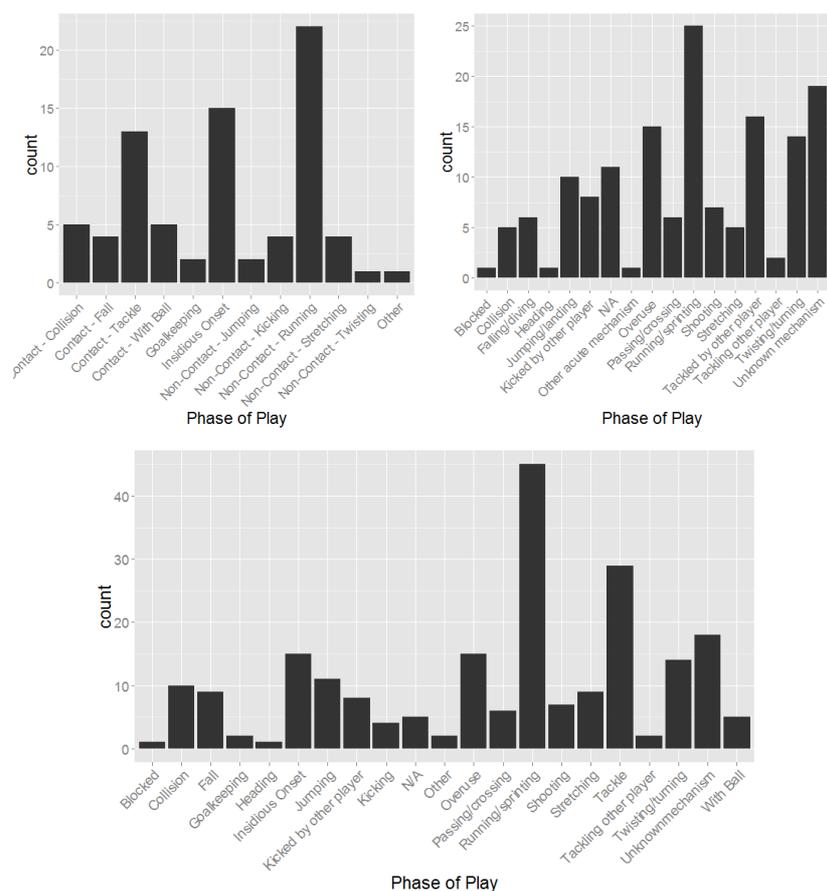

**Figure 5.6. Bar chart of "phase of play" for the WW, THFC and integrated datasets (from left to right).**

It makes sense to assume that different mechanisms will be correlated with different types of injuries. Headings for example should be related to concussions (Maher, Hutchison, Cusimano, Comper, & Schweizer, 2014), while getting tackled can be related to severe injuries, since foul play many times takes place as a result of a poor challenge. However, it is not entirely clear how the phase of play can affect the severity, since the same conditions could in theory produce injuries with different severity ratings.

Table 10.1 to Table 10.3 in the Appendix show the mean and standard deviation of the recovery per type of injury (the tables are not reproduced here due to their length). There are clear differences between the recovery times of different mechanisms of injury.

**Body part injured**

Figure 5.7 shows a bar chart of the body part that was injured for each dataset. It's clear that the majority of injuries takes places at the lower extremities. Most injuries are either at the ankle, the thigh, the lower leg or the knee.

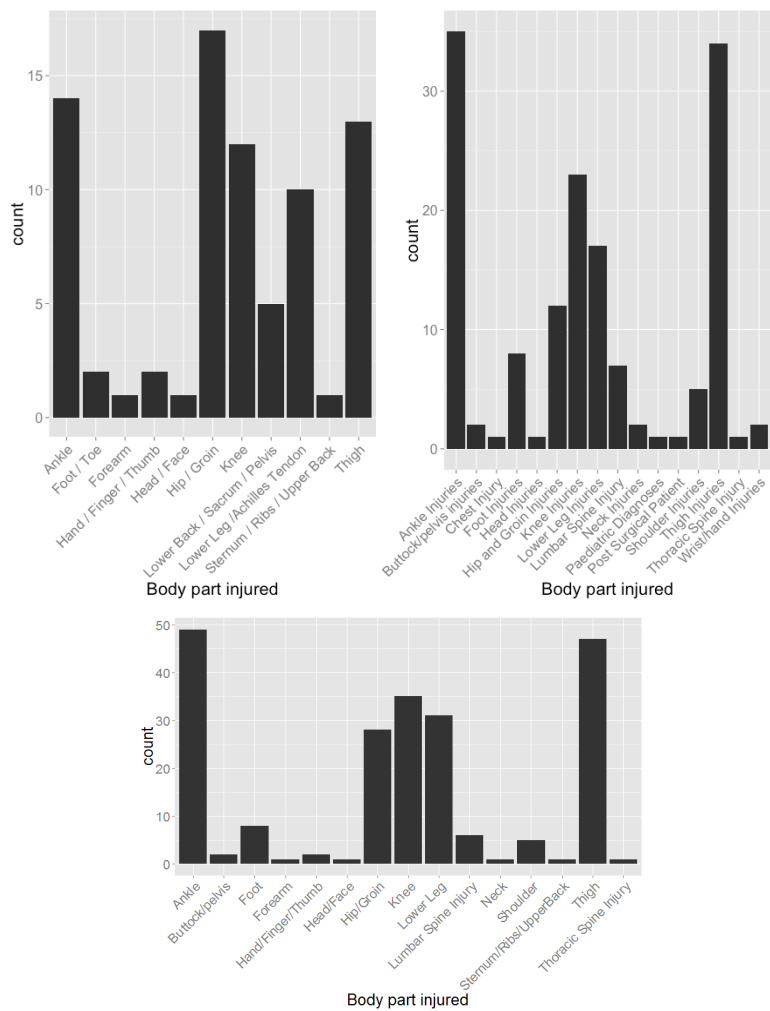

**Figure 5.7. Bar chart for "body part injured" for the WW, THFC and integrated datasets (from left to right).**

Regarding the harmonization of this category the two datasets had one-one correspondence. The differences were in the following categories:

- The WW dataset used the category "Sternum/Ribs/Upper back" to refer to any injury around that area. The THFC dataset used the categories "Shoulder", "thoracic spine injury" and "chest injury", being more detailed. The "chest injury" was merged with "Sternum/Ribs/Upper back" in the integrated dataset. The categories "shoulder" and "thoracic spine injury" were left as they are.

- The THFC dataset contained a single post-surgical patient and an injury with a pediatric diagnosis. In both cases, the injuries were related to the lower leg, so these injuries were included in the category "Lower Leg" in the integrated dataset.

The body part that is being injured is related to the severity of an injury (Chomiak, Junge, Peterson, & Dvorak, 2000). Some body parts might be more sensitive. Some others might be more prone to collisions in ways that can cause severe injuries. Also, some body parts, such as the knee joint, are constantly under more stress throughout a footballer's career, so they could be prone to overuse injuries.

Table 10.4 to Table 10.6 in the Appendix show the mean and standard deviation for recovery per body part. It is easy to see that there are differences in the recovery time for different body parts.

**Stage of season/Date of injury**

Figure 5.8 shows a bar chart of the "stage of season" during which the injury took place for the THFC dataset.

Most injuries took place during the early or the mid season. An important fact that stands out from the graph is that a significant number of injuries takes place during the pre-season. The pre-season is "special" in the sense that the players do not have any professional games, but they are under a heavy training schedule. So, the circumstances under which they can get injured during the pre-season can be quite different when compared to the main season.

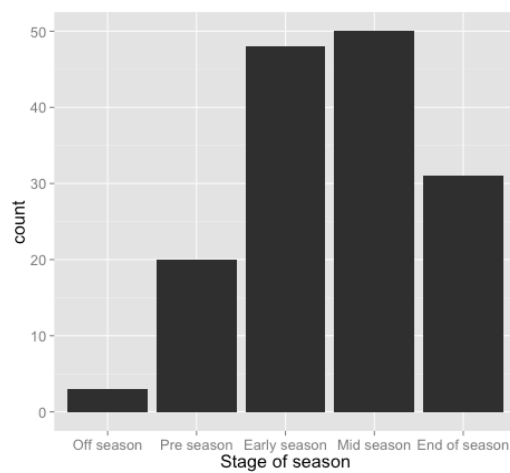

**Figure 5.8. Bar chart of the "stage of season" for the THFC dataset.**

The variable "date of injury" was converted to a day, with day=1 being the 1$^{st}$ of January and day=365 being the 31 of December. Figure 5.9 shows a histogram of this variable. The picture here is less clear than the picture for the THFC dataset. Injuries seem to follow an "on-and-off" pattern. This could come as a result of the fact that when players get injured the team might put more effort into preventing any more injuries from taking place. E.g. the coach might let a player rest more, the player might be less aggressive on the pitch, or the medical staff might put more focus on that player. As soon as the player returns to a level of fitness, this effort might be reduced, leading to further injuries.

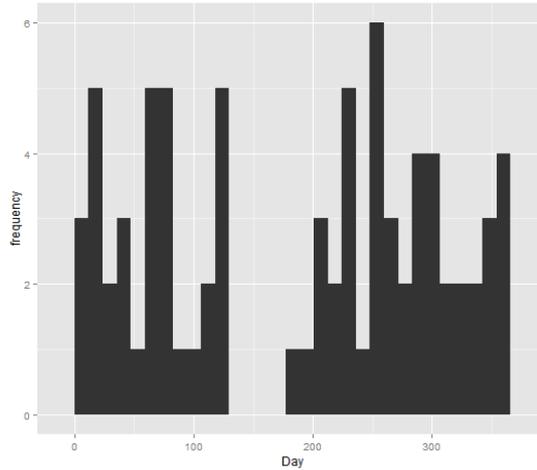

**Figure 5.9. Histogram of the day that the injury took place for the WW dataset.**

Harmonisation in this case was particularly straightforward since similar definitions are used accross all clubs when referring to the different stages of the football season. The off-season is between the last game of the previous season and the first week of July that signals the beginning of the preseason. The preseason lasts until the first game in August. Then, early, mid and end of season take up one third of the football season each.

Figure 5.10 for the integrated dataset shows a slightly different picture to the one from the THFC dataset. It seems that more injuries take place as the season progresses. There were many injuries in the WW that took place towards the end of the season, which in turn is reflected on this barchart.

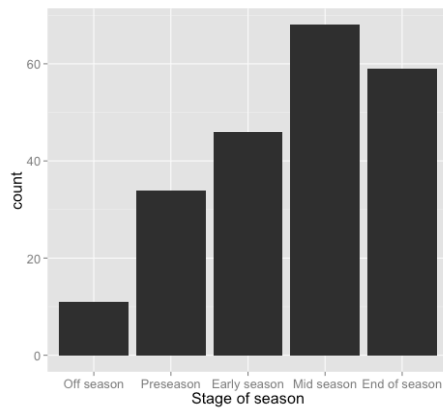

**Figure 5.10. Bar chart of the "stage of season" for the integrated dataset.**

Table 5.13 shows the mean and standard deviation of the recovery time for the THFC and the integrated datasets. What strikes out is the fact that the recovery time for the early season seems to take longer and to be much more variable.

**Table 5.13. Recovery (mean+/-standard deviation) for "stage of season" across the THFC and integrated datasets.**

|              | THFC          | Integrated     |
|--------------|---------------|----------------|
| **Off season**   | 8.0+/-13.8    | 15.3+/-19.2    |
| **Pre-season**   | 11.8+/-23.3   | 9.9+/-18.8     |
| **Early season** | 21.3+/-46.2   | 67.3+/-102.1   |
| **Mid-season**   | 12.5+/-28.8   | 13.4+/-27.3    |
| **End of season**| 14.2+/-37.1   | 9.6+/-26.7     |

Figure 5.11 shows a plot of the recovery time versus the day of injury in the year for the WW data. There does not seem to be a clear relationship, and the correlation between the two variables is -0.02.

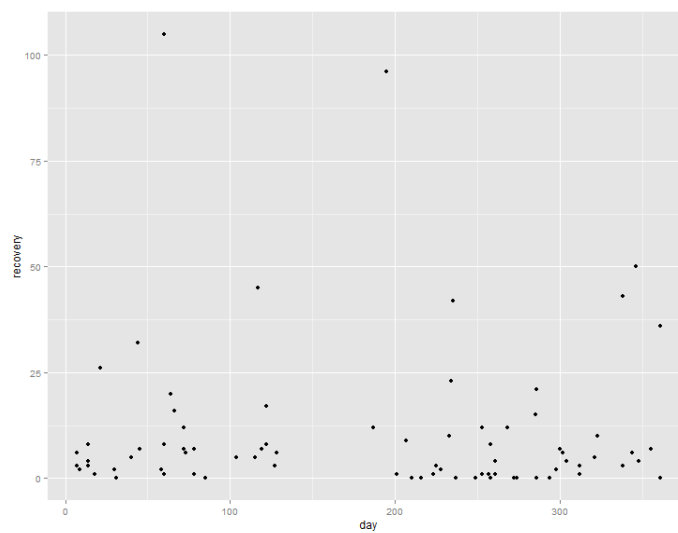

**Figure 5.11. Recovery versus "day of injury" for the WW dataset.**

## Injury

Figure 5.12 shows a bar chart of injuries for the THFC, the WW and the integrated dataset. The prominent category is muscle ruptures and strains, with ligaments sprains following. Common reasoning suggests that this should be the most important feature for prediction. Note, that this variable is not equivalent to a final diagnosis, since a final diagnosis is far more detailed. Harmonisation in this case was straightforward since there is correspondence between the categories in the two datasets.

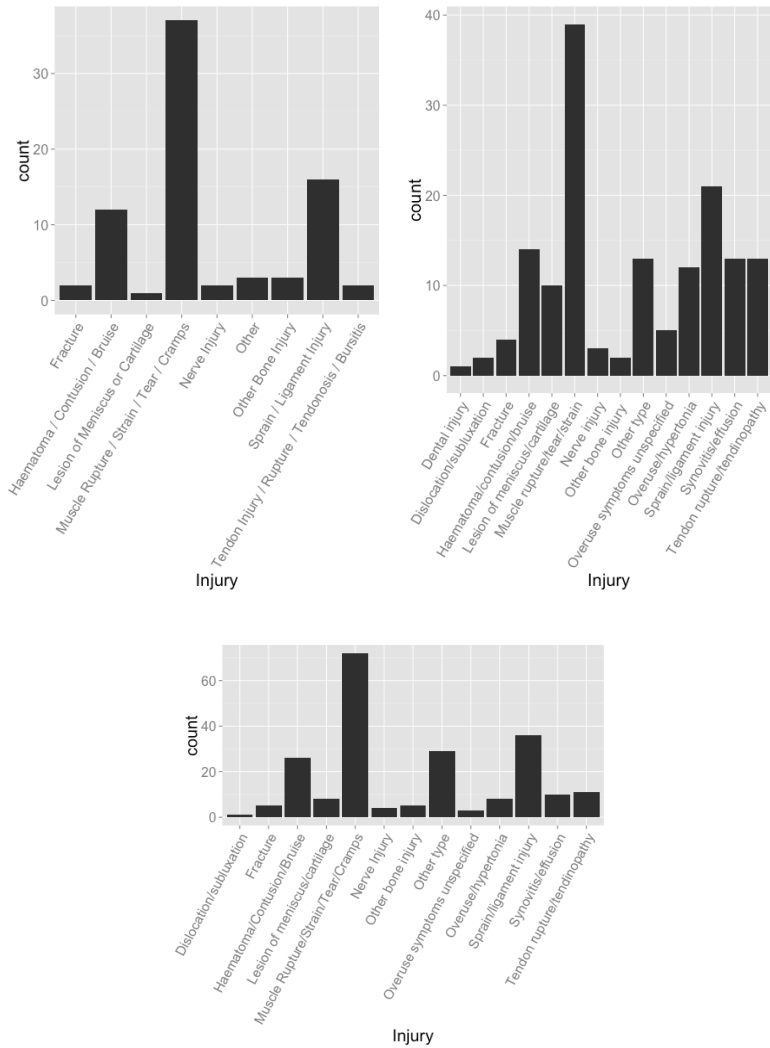

**Figure 5.12. Bar chart of injuries for the WW, THFC and integrated dataset.**

Obviously, the same picture holds for the integrated dataset as it can be seen in Figure 5.12.

### 5.2.3  THFC exploratory analysis and graphs

Figure 5.13 below shows a histogram of age. The majority of players are in their twenties, with the mean of the distribution being at 26.5. The distribution is slightly skewed towards the right. This comes as a result of the fact that it is impossible to include any players under 18 in the squad, while some players can play well in their late 30s if their physical condition allows it.

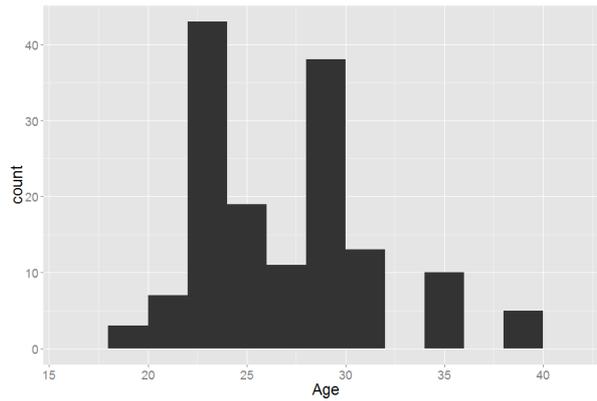

**Figure 5.13. Histogram of "age" for the THFC dataset.**

It is reasonable to expect age to be related to recovery. This is something that has been confirmed by previous studies (Cloke, et al., 2012). Figure 5.14, however, does not show a clear relationship between recovery and age. Perhaps the fact that all athletes are elite professionals and their bodies are in excellent physical condition reduces any effect that age might have on recovery.

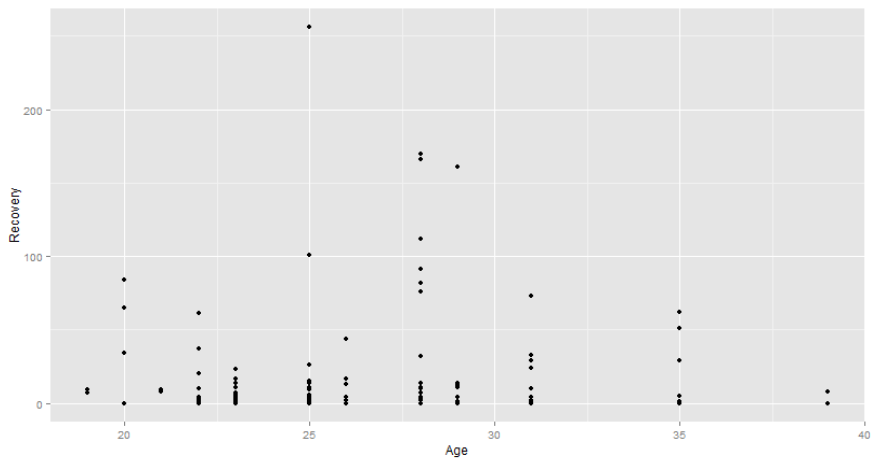

**Figure 5.14. Plot of the recovery time versus the age for the THFC dataset.**

Figure 5.15 shows a bar chart of the position when injured. The majority of injuries take place for midfielders and defenders. Position when injured can be different in training and in matches. In some cases, players can play different positions during training (e.g. a forward playing as a midfielder). This variable records the position at the time of injury and not the player's usual position.

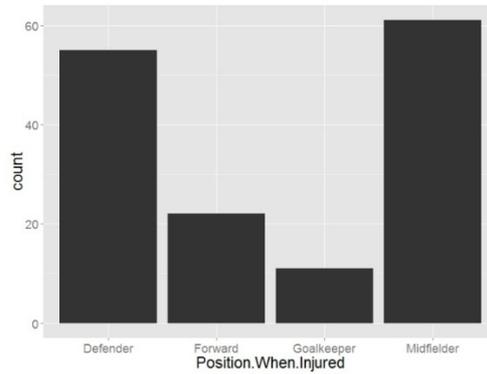

**Figure 5.15. Bar chart of "position when injured" for the THFC dataset.**

Figure 5.16 shows the bar chart of the type of injury. The majority of injuries are classified as traumatic.

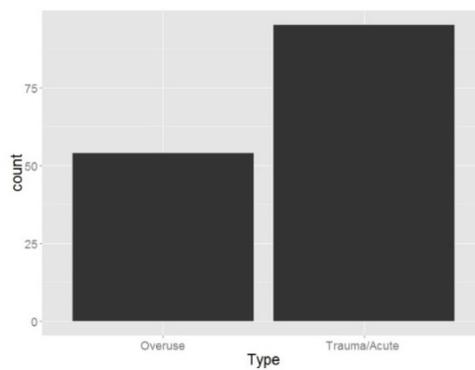

**Figure 5.16. Bar chart of classification of "injury type" for the THFC dataset.**

Figure 5.17 shows four bar charts of the type of injury with regards to position. From the plots we can see that defenders have a larger incidence of acute injuries, rather than overuse injuries. Midfielders' injuries are more balanced between the acute and overuse categories. This is likely to happen due to two factors. First, different positions pose different demands on the pitch. Forwards should be more likely to get tackled, rather than tackling themselves. Secondly, different positions require different physical characteristics from the players. The physical characteristics of a player might affect the types of injuries, as well, as the recovery time.

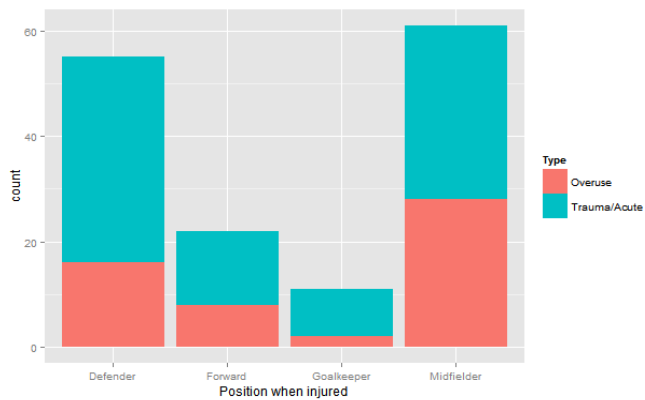

**Figure 5.17. Bar charts of position versus type of injury for the THFC dataset.**

Table 5.14 shows the relationship between recovery and the position when injured, where we can see that there are no striking differences between positions, with the exception of goalkeepers, that have shorter recovery periods.

**Table 5.14. Relationship between "position when injured" and the recovery time (mean and standard deviation).**

|  | Mean | Sd |
|---|---|---|
| **Defender** | 15.4 | 28.3 |
| **Forward** | 12.5 | 35.4 |
| **Goalkeeper** | 7.4 | 12.0 |
| **Midfielder** | 17.9 | 44.2 |

Similarly, there are no striking differences between acute and overuse injuries as we can see in Table 5.15.

**Table 5.15. Relationship between the "type of injury" and the recovery time (mean and standard deviation).**

|  | Mean | Sd |
|---|---|---|
| **Overuse** | 16.5 | 43.3 |
| **Trauma/Acute** | 14.3 | 30.6 |

### 5.2.4 WW dataset exploratory analysis and graphs

Figure 5.18 show the bar chart of "referee's decision" for the WW dataset. There is a large number of "N/A" values. Non-applicable in this case can mean either of two things. First, it can refer to injuries that took place during training. Secondly, it can refer to injuries within the game that were not the result of foul play. An example can be a traumatic injury that comes as a result of the player accelerating too fast. Foul play has been indicated as a risk factor for severe injuries (Chomiak, Junge, Peterson, & Dvorak, 2000).

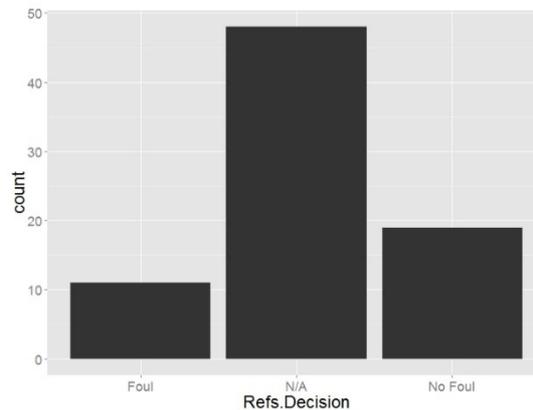

**Figure 5.18. Bar chart of "referee's decision" for the WW dataset.**

Table 5.16, however, shows that this is not the case for this dataset. There are no great differences between categories and N/A has the highest mean. The majority of N/A values correspond to injuries that took place in training, so maybe injuries in training are more severe for this particular dataset.

**Table 5.16. Recovery time (mean and standard deviation) versus "referee's decision" for the WW dataset.**

|         | Mean | Sd   |
|---------|------|------|
| Foul    | 6.0  | 7.3  |
| N/A     | 12.9 | 22.0 |
| No Foul | 8.1  | 11.3 |

Figure 5.19 shows a bar chart of "strapping" for the WW dataset. It seems that there is an approximately equal split between strapped and non-strapped injuries, with 48% of the injuries being strapped. Strapping is a common measure applied by physicians on injured joints, so some physicians might recommend strapping as soon as an athlete gets injured, without this necessarily being indicative of the injury's severity.

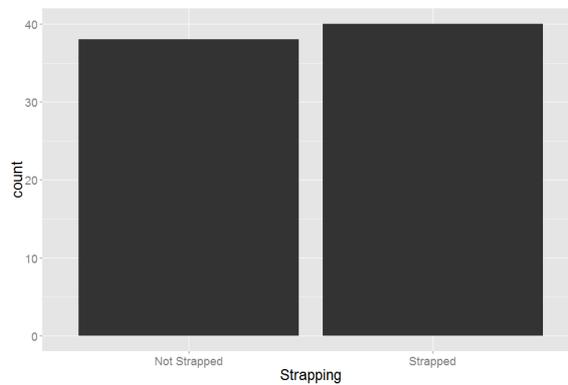

**Figure 5.19. Bar chart of "strapping" for the WW dataset.**

Table 5.17 shows that strapping might actually lead to shorter recovery times.

**Table 5.17. Recovery time (mean and standard deviation) versus "strapping" for the WW dataset.**

|             | Mean | Sd   |
|-------------|------|------|
| Not Strapped | 14.2 | 24.2 |
| Strapped    | 7.5  | 9.6  |

Figure 5.20 shows the bar chart of "surface condition" for the WW dataset. The wetness of the grass can change the traction which in turn can affect things such as the acceleration and the deceleration of the athlete, or the ease of maneuvering. It could the case that a type of grass predisposes athletes to certain type of injuries which in turn can influence severity.

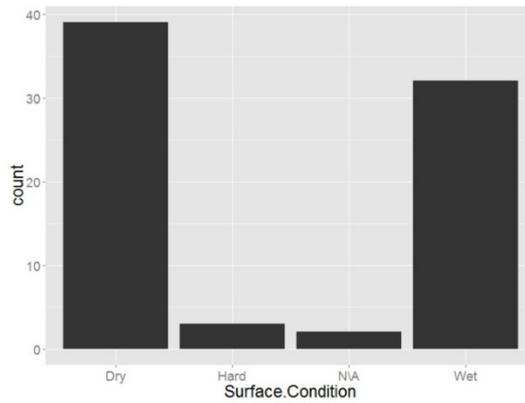

**Figure 5.20. Bar chart for "surface condition" for the WW dataset.**

Table 5.18 shows that there is not much differences between the different surface conditions.

**Table 5.18. Recovery time (mean and standard deviation) versus "surface condition" for the WW dataset.**

|  | Mean | Sd |
|---|---|---|
| **Dry** | 9.0 | 17.1 |
| **Hard** | 7.0 | 2.6 |
| **N/A** | 5.5 | 2.1 |
| **Wet** | 13.4 | 21.0 |

Figure 5.21 shows the type of footwear at the time of injury for the WW dataset. There are some cases that are unclassified. This is probably an omission of the people responsible for the data entry. For the purposes of this analysis, these cases were considered equivalent to non-applicable (N/A). Footwear can be important since it affects the traction (Wannop, Worobets, & Stefanyshyn, 2010) which, as it was discussed previously, might be related to specific injuries.

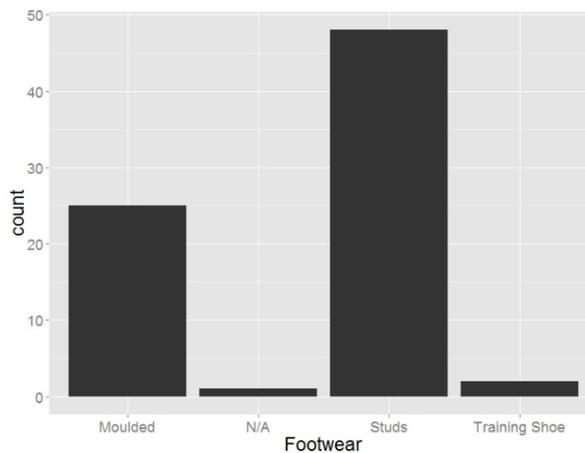

**Figure 5.21. Bar chart for the type of footwear for the WW dataset.**

Table 5.19 shows that there are not big differences between the different types of shoes. Also, training shoes have only 2 cases, while there is only one N/A case, so their descriptive statistics can be ignored.

**Table 5.19. Recovery time (mean and standard deviation) versus "footwear".**

|  | Mean | Sd |
|---|---|---|
| **Moulded** | 8.6 | 19.0 |
| **N/A** | 45.0 | 0 |
| **Studs** | 11.6 | 18.5 |
| **Training Shoe** | 5.5 | 2.1 |

### 5.2.5 Experimental protocol

The problem can be posed as either a regression or classification problem, depending on whether we are modeling the "Days unavailable" or the "Severity" variable. Both approaches were tried and different kinds of models were compared.

The following 4-step protocol was used which is depicted in Figure 5.22.

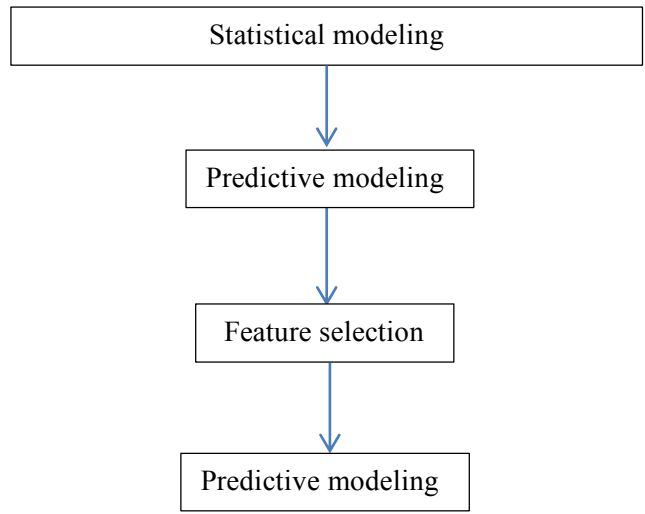

**Figure 5.22. Flowchart of the steps of analysis for the first investigation.**

First of all, the modeling problem is broken down in 2 parts: statistical modeling and predictive modeling. Statistical modeling was used in order to identify whether there is useful information in the data for the given task. This was conducted through significance testing either using the omnibus test at model level, or individual significance tests for each coefficient. This analysis helped to set the expectations for the predictive modelling, while also testing the importance of the covariates through significance testing.

For the analysis, two types of the generalized linear model were used: Poisson regression and negative binomial regression. A log link function was used for both of them. Also, ordinal regression was conducted having the "Severity" variable as the response. The motivation behind the use of ordinal regression was that the categories are ordinal by nature, so an ordinal method might be able to utilize this information.

Once the statistical analysis was done, predictive modeling was conducted. The predictive modeling was split down into 2 problems: regression and classification. For each problem, a set of methods was employed and compared with each other. All the tests were conducted on each dataset individually.

After predictive modelling, feature selection was conducted through random forests and genetic algorithms using correlation-subset feature subset selection. This procedure resulted in three new datasets with a reduced number of variables. The predictive models were re-applied and the results were compared to the previous results.

### 5.2.6 Parameter grids for the machine learning algorithms

#### *5.2.6.1 Regression*
Five different methods were used and evaluated: neural networks, support vector machines, random forests, k-NN and Gaussian processes. Random forests were chosen because they are good at dealing with datasets with lots of noisy and useless features, since they embed feature selection as was explained in section 4.1. If the random forests overperform compared to the rest of the classifiers which do not embed feature selection, then this could be an indication of features that cause problems in modeling. Also, the features were mainly categorical, and decision trees can naturally deal with them.

Support vector machines and Gaussian processes are two popular and successful kernel methods as was explained in section 4.1, commonly used for classification and regression. The kernels of choice were the radial basis function kernel and the polynomial kernel. These kernels were chosen for their proven effectiveness, and subsequent popularity, in the literature (Hastie, Tibshirani, & Friedman, 2009).

The degree of the polynomial kernel can also be useful in understanding if any interactions exist between the covariates as described in section 4.1.1. A kernel of degree $d$ takes into account all interactions between variables up to $d$. If $d > 1$ improve the performance, then it is likely that higher level interactions exist in the dataset. This can make for an interesting comparison with the statistical investigation.

Neural networks are particularly good for dealing with non-linearities. It is not particularly clear whether non-linear interactions between variables exist in the dataset. The performance of neural networks against other models in this case can be a kind of test for the existence of complex relationships.

K-NN is a simple non-parametric method and was used as a benchmark against the performance of the rest of the algorithms.

Each method was executed with many different parameter sets. Grid search was used in order to find the best parameters. Table 5.20 below shows the parameters that each method used and their value ranges. The optimization objective for all algorithms was the minimization of the RMSE.

The neural network was trained using backpropagation with weight decay.

**Table 5.20. Parameter grid for the models.**

| Model | Parameters |
|---|---|
| Random forests | Number of trees $\in \{50,100,\ldots,5000\}$ |
| Neural Networks | Epochs $\in \{500,1000,\ldots,3000\}$, Hidden Neurons $\in \{5,10,\ldots,50\}$, Decay $\in \{0,10^{-10},10^{-9},\ldots,1\}$ |
| SVM, Gaussian Process | RBF kernel with degree $\in \{0.0001,0.001,\ldots,5\}$ |
| | Polynomial kernel with degree $\in \{1,2,3,4\}$ |
| | $C \in \{0,0.5, 1,\ldots,100\}$ |
| | Scale $\in \{0.1,0.5,1,\ldots,10\}$ |
| k-NN | $k \in \{1,2,\ldots,20\}$ |

### *5.2.6.2 Classification*

The same models with the same parameter grids that were used for regression were used for classification as well, along with the addition of naïve Bayes. The naïve Bayes assumption is particularly strong, but it, as it was discussed in section 4.1.6 if it holds, then naïve Bayes is the optimal classifier under both quadratic and zero-one loss. The simplicity of naïve Bayes can make it a good benchmark for more complex algorithms, such as random forests.

## 5.3 Evaluating performance

### 5.3.1 Regression

All the tasks were evaluated using the RMSE calculated over 10 iterations of 10-fold cross validation.

Along with the RMSE the correlation and the concordance correlation coefficient were calculated as well. Careful inspection of individual predictions was showing that the RMSE was severely affected by only a few errors. This comes as a result of the RMSE penalizing severely larger errors and the data being skewed. Therefore, it was decided to use the two correlation measures in addition to RMSE, in order to get a better understanding of the performance of the models. This is the same problem as the one outlined in section 4.2.1.

### 5.3.2 Classification

The primary measure for assessing the performance of the classifier was the classification accuracy. However, the classes are unbalanced and so accuracy might not give the true picture of the results. Therefore, Cohen's kappa was used as an additional measure of the classifiers' performance. An additional advantage of Cohen's kappa is that it can naturally deal with multiclass problems.

## 5.4 Statistical modeling

### 5.4.1 Poisson regression

Before the Poisson model could be used, it is important to check its assumptions. The Poisson model assumes that the mean is equal to the variance. This assumption was violated for all datasets. For the THFC data the mean was 15.5 and the variance 1298.821. For the WW dataset the mean was 10.8 and the variance 341.3. For the integrated dataset the mean was 13.8 and the variance 975.8. Therefore, the robust estimator of the standard error proposed by White (1980) was used in order to compute the p-values for the coefficients.

Table 5.21 show the results of the omnibus test for the Poisson regression model against the intercept only model. Table 5.22 shows the results of the p-values. The p-values for the factors have been derived based on a likelihood ratio test between a model that includes the factor and one that omits it. The covariates that are statistically significant at the 5% level are colored red. It is clear that not all variables are significant at the 5% level.

**Table 5.21. Omnibus test for the Poisson regression model for all datasets.**

| Dataset | Likelihood Ratio Chi-Square | Degrees of freedom | Significance |
|---|---|---|---|
| THFC | 5238.965 | 62 | <0.001 |
| WW | 1114.589 | 50 | <0.001 |
| Integrated | 5443.331 | 73 | <0.001 |

**Table 5.22. P-values for the coefficients of the Poisson regression models for all datasets.**

| Source | p-value THFC | p-value WW | p-value integrated |
|---|---|---|---|
| (Intercept) | 0.948 | <0.001 | <0.001 |
| Recurrence | <0.001 | <0.001 | <0.001 |
| Activity | 0.258 | 0.190 | 0.564 |
| Position (When Injured) | 0.006 | - | - |
| Phase Of Play/Mechanism | <0.001 | <0.001 | <0.001 |
| Body Part Injured | 0.876 | <0.001 | <0.001 |
| Type | 0.127 | - | - |
| Injury | <0.001 | <0.001 | <0.001 |
| Side | 0.221 | <0.001 | 0.011 |
| Stage Of Season | 0.039 | - | 0.316 |
| Age | 0.354 | - | - |
| Footwear | - | 0.066 | - |
| Strapping | - | 0.005 | - |
| Surface Condition | - | <0.001 | - |
| Referee's decision | - | 0.043 | - |
| Date | - | 0.074 | - |

Figure 5.23 shows the fitted versus the predicted values while Table 5.23 shows the performance metrics. The fit seems quite good for the WW dataset and reasonably good for the THFC and the integrated dataset. Pearson's correlation and the concordance correlation coefficient are close to 1 for both datasets. However, the performance on the integrated dataset is worse, something which can be evidenced by all three metrics.

**Table 5.23. Performance metrics for Poisson regression across all datasets.**

| Dataset | RMSE | Correlation | Ccc |
|---|---|---|---|
| THFC | 11.4 | 0.94 | 0.94 |
| WW | 4.9 | 0.96 | 0.96 |
| Integrated | 19.0 | 0.79 | 0.75 |

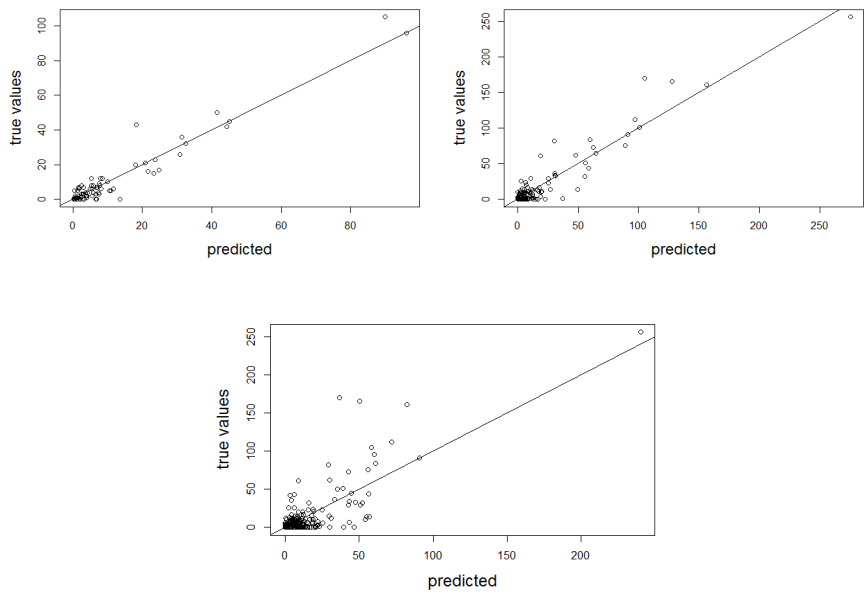

**Figure 5.23. Plot of the predicted values for the recovery time Poisson regression and the actual values for the WW, THFC and integrated datasets respectively (from left to right).**

### 5.4.1.1 Diagnostic plots

Figure 5.24 shows the deviance residuals against the predicted values for the Poisson regression models. The assumptions of the model for the WW dataset seem to be met, since no significant over- or under-dispersion is seen, but this is not the case for the THFC and the integrated datasets. The variance is increasing moving from left to right.

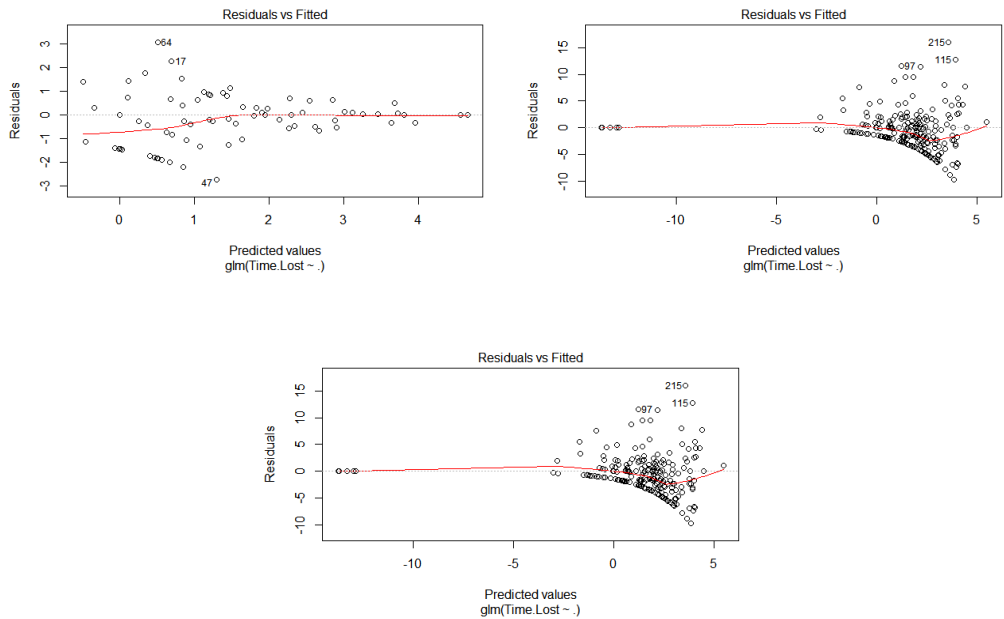

**Figure 5.24. Residuals vs fitted plot for the Poisson regression model fitted to the WW, THFC and integrated dataset respectively (from left to right).**

This type of deviation in Figure 5.24 suggests that there must be some kind of interaction between variables that is taking place. Some additional investigation was done in order to discover if an interaction term between variables could improve the plot. Unfortunately, due to the limited amount of data available, not all interaction terms can be tried out, since the model becomes unidentifiable. For example, the coefficients of the reasonable interaction between "injury" and "phase of play" cannot be estimated.

An interaction term that seems to improve the residual plot for the THFC data and can be estimated was the interaction between body area and injury. The residual plot for the updated model is depicted at Figure 5.25. It is still far from perfect, but still better than the residual plot at Figure 5.24, since the variance has been reduced on the right side of the plot.

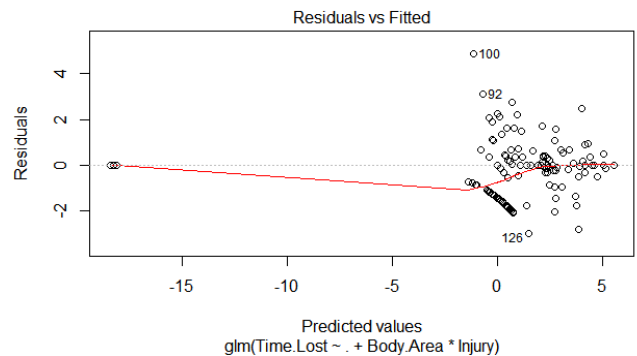

**Figure 5.25. Updated residual vs fitted plot for the THFC data, including the interaction of "body area" with "injury".**

### 5.4.2 Negative binomial regression

The results of the likelihood ratio test of the model against an intercept only model are shown in Table 5.24. The model performs statistically significant better than an intercept only model.

**Table 5.24. Results of the omnibus test for negative binomial regression across all datasets.**

| Dataset | Likelihood Ratio Chi-Square | Degrees of freedom | p-value |
|---|---|---|---|
| THFC | 351.002 | 62 | <0.001 |
| WW | 101.15 | 50 | <0.001 |
| Integrated | 383.978 | 73 | <0.001 |

Table 5.25 shows the results of the significance testing for each individual variable. The picture is similar to the one for Poisson regression. Regarding the THFC dataset, the only difference is the variable "Body Part Injured" which is now significant. Regarding the WW dataset, only four variables are now significant, but these are common with Poisson regression. Regarding the integrated dataset, the only difference is that now the "Stage of Season" is significant, but "Side" is not.

**Table 5.25. P-values the coefficients of the negative binomial model across all datasets.**

| Source | p-value THFC | p-value WW | p-value integrated |
|---|---|---|---|
| (Intercept) | 0.660 | 0.298 | <0.001 |
| Recurrence | <0.001 | 0.003 | <0.001 |
| Activity | 0.352 | 0.096 | 0.203 |
| Position (When Injured) | 0.037 | - | - |
| Phase Of Play/Mechanism | 0.004 | 0.024 | <0.001 |
| Body Part Injured | <0.001 | 0.002 | <0.001 |
| Type | 0.309 | - | - |
| Injury | <0.001 | 0.344 | <0.001 |
| Side | 0.845 | 0.055 | 0.270 |
| Stage Of Season | 0.035 | - | <0.001 |
| Age | 0.888 | - | - |
| Footwear | - | 0.057 | - |
| Strapping | - | 0.145 | - |
| Surface Condition | - | 0.357 | - |
| Referee's decision | - | 0.466 | - |
| Date | - | 0.124 | - |

Figure 5.26 shows scatterplots of the predicted values versus the actual values and Table 5.26 shows the performance metrics. A few things stand out. First of all, the overall fit seems reasonable. Like with Poisson regression, the best fit seems to be achieved for the WW data. The fit for the THFC dataset contains one very large outlier, while the fit for the integrated dataset is still reasonably good at a correlation of 0.55 between predicted and true values. All three metrics are worse than the ones by Poisson regression. Also, like in Poisson regression, the model has the worst performance on the integrated dataset.

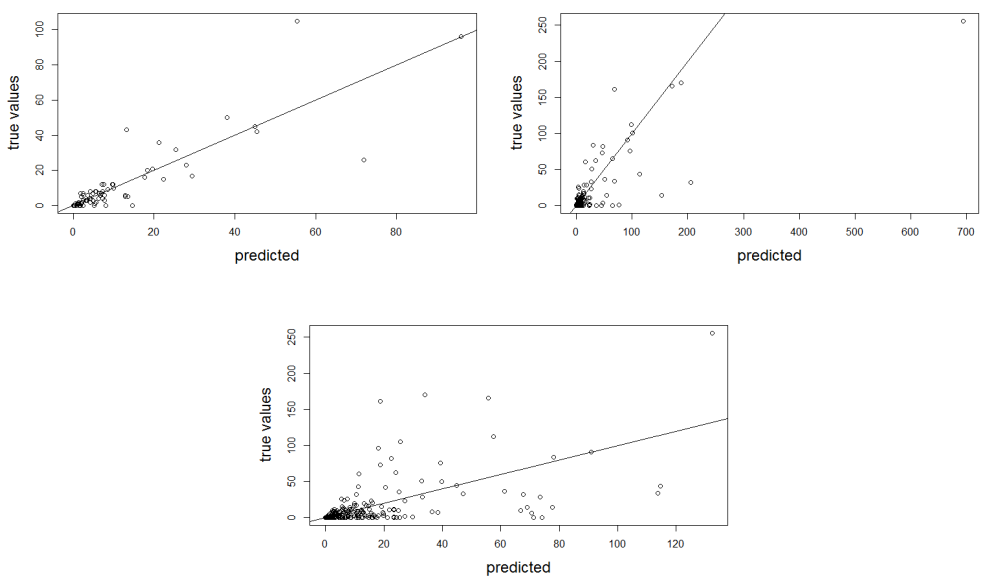

**Figure 5.26. Scatterplot of the log of the mean predictor versus the response variable for the WW, THFC and integrated dataset respectively (negative binomial model).**

**Table 5.26. Performance metrics for negative binomial regression across all datasets.**

| Dataset | RMSE | Correlation | Ccc |
|---|---|---|---|
| THFC | 43.45 | 0.78 | 0.65 |
| WW | 9.38 | 0.85 | 0.85 |
| Integrated | 26.22 | 0.55 | 0.51 |

Something noteworthy is the existence of an outlier in the THFC dataset. This was an instance that corresponded to a lesion of meniscus and 256 days lost for the player. However, Cook's distance for this point is 0.006 which places it in the 96$^{th}$ place out of the 152 data points in terms of the influence it has. Therefore, this data point does not seem to influence the fit considerably.

### 5.4.2.1 Diagnostic plots

Figure 5.27 below shows the diagnostic plots for the negative binomial models. The picture here is pretty similar to the picture for Poisson regression. The WW dataset provides a reasonably good fit. The models for the THFC and the integrated datasets seem to be affected by an interaction term.

Unlike in Poisson regression, the interaction between body part and injury did not seem to improve the residual plots.

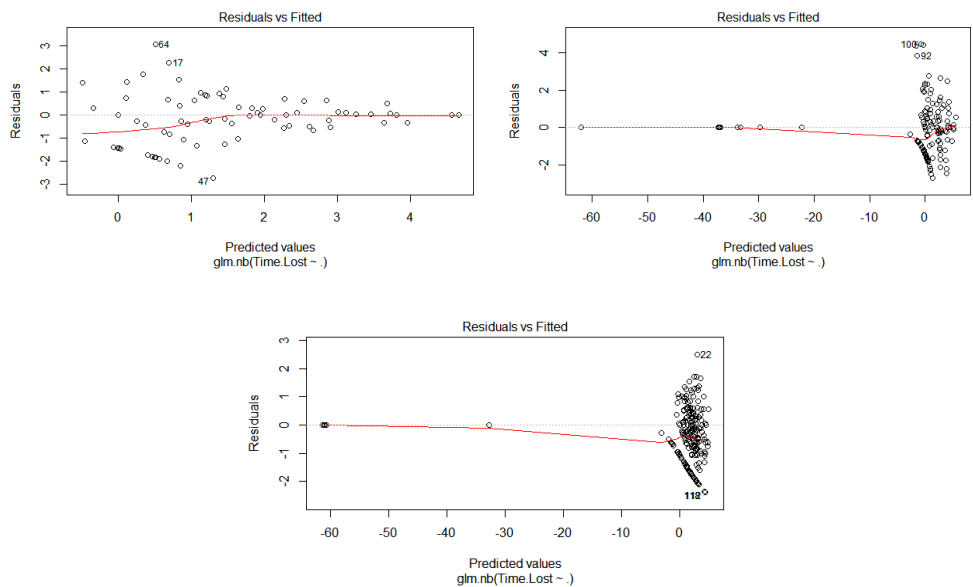

**Figure 5.27. Residual vs fitted values plot for the negative binomial model fitted to the WW, THFC and integrated dataset respectively.**

### 5.4.3 Ordinal regression

Table 5.27 and Table 5.28 show the results of the ordinal regression models.

**Table 5.27. Omnibus test for the ordinal regression models across all datasets.**

| Dataset | Likelihood Ratio Chi-Square | Degrees of freedom | Significance |
|---|---|---|---|
| THFC | 83.274 | 62 | 0.037 |
| WW | 127.319 | 41 | <0.001 |
| Integrated | 243.73 | 75 | <0.001 |

**Table 5.28. P-values for the ordinal regression coefficients across all datasets.**

| Source | p-value THFC | p-value WW | p-value integrated |
|---|---|---|---|
| Recurrence | 0.637 | 0.035 | 0.026 |
| Activity | 0.312 | 0.684 | 0.814 |
| Position (When Injured) | 0.936 | - | - |
| Phase Of Play/Mechanism | 0.817 | 0.576 | 0.605 |
| Body Part Injured | 0.994 | 0.449 | 0.999 |
| Type | 0.960 | 0.371 | - |
| Injury | 0.974 | - | 0.834 |
| Side | 0.756 | 0.040 | 0.899 |
| Stage Of Season | 0.315 | - | 0.019 |
| Age | 0.977 | - | - |
| Footwear | - | 0.313 | - |
| Strapping | - | 0.526 | - |
| Surface Condition | - | 0.411 | - |
| Referee's decision | - | 0.958 | - |

The omnibus tests suggest that there is a reasonably good fit for the model. However, no single variable seems to be statistically significant.

Table 5.29 shows the accuracy and the kappa statistic for the ordinal regression models. Accuracy seems to be relatively good, but the metric is inflated by the fact that most injuries are transient. The kappa statistic gives a more accurate picture. The fit seems to be less than good for the THFC and the WW datasets.

**Table 5.29. Performance metrics for ordinal regression across all datasets.**

| Dataset | Accuracy | Kappa |
|---|---|---|
| THFC | 71.0% | 0.31 |
| WW | 75.0% | 0.40 |
| Integrated | 73.8% | 0.37 |

### 5.4.4 Discussion of the statistical analysis

The models illustrate that the datasets contain information regarding the response variable.

Three things stand out from the analysis. First, it seems that there is some interaction term that for the THFC data, for which we don't have enough data to estimate. The fit is reasonably good for all datasets, but the assumptions are not fully satisfied for the THFC dataset and this carries on to the integrated dataset.

Secondly, it seems that not all variables are equally important. Some variables have been marked as statistically significant for all datasets, while others seem to be less useful. However, the significant variables in all datasets seem to be more or less the same.

Thirdly, the ordinal regression model does not perform very well. This could be due to the way the "Severity" variable is encoded, since there is no guarantee that this discretization scheme is the best one for statistical modelling. Therefore, approaching the problem as a regression problem might be a more sensible strategy than approaching it as a classification problem.

Table 5.30 shows the number of times that significance test was significant at the 5% level for the Poisson regression and the negative binomial models out of all the times a covariate appears in the datasets. The ordinal regression models were omitted from this table, due to their mediocre performance. The variables that were significant more than 50% of the tests are colored red.

**Table 5.30. Number of coefficient tests were significant at the 5% level.**

| Source | Importance |
|---|---|
| Recurrence | 6/6 |
| Activity | 1/6 |
| Position When Injured | 2/2 |
| Phase Of Play | 6/6 |
| Body Part Injured | 5/6 |
| Type | 0/2 |
| Injury | 5/6 |
| Side | 2/6 |
| Stage Of Season | 3/4 |
| Age When Injured | 0/2 |
| Footwear | 0/2 |
| Strapping | 1/2 |
| Surface Condition | 1/2 |
| Referee's decision | 1/2 |
| Date | 0/2 |

There are 6 variables that stand out. The choice of these variables makes sense based on the exploratory analysis and the literature outlined in section 5.2.

## 5.5 Predictive modeling results

### 5.5.1 Predictive regression results

The best results parameter settings achieved for each method are shown in Table 5.31. These parameter settings where evaluated again for each classifier by running 10 rounds of 10-fold cross validation. The mean RMSE, Pearson correlation and concordance correlation across all 100 runs are reported, along with the standard deviation, at Table 5.31. Regarding SVM and Gaussian processes, only the best kernel is reported, as measured by all three metrics.

**Table 5.31. Regression results for the THFC dataset.**

| Method | Parameters | RMSE(test) | Correlation(test) | Ccc(test) |
|---|---|---|---|---|
| k-NN | k=5 | 33.54+/-16.79 | 0.225+/-0.387 | 0.196+/-0.324 |
| SVM | Polynomial kernel, degree=3, C=0.5, scale=0.01 | 28.8 +/- 0.365 | 0.511+/-0.408 | 0.269+/-0.244 |
| Gaussian Process | Polynomial kernel, degree=3, scale=0.01 | 29.5+/- 0.846 | 0.428+/-0.393 | 0.226+/-0.218 |
| Neural Network | Neurons=45, epochs=2500, decay=0.001 | 32.585 +/- 2.02 | 0.261+/-0.211 | 0.229+/-0.182 |
| Random Forest | Trees=70 | 30.1+/-0.155 | 0.471+/-0.341 | 0.406+/-0.337 |

**Table 5.32. Regression results for the WW dataset.**

| Method | Parameters | RMSE(test) | Correlation(test) | Ccc(test) |
|---|---|---|---|---|
| k-NN | k=3 | 22.21+/-9.46 | 0.03+/-0.26 | 0.023+/-0.253 |
| SVM | Polynomial kernel, degree=3, scale=0.01, C=1 | 36.4+/-34 | 0.189+/-0.409 | 0.120+/-0.243 |
| Gaussian Process | polynomial kernel, degree=2, scale=0. 1 | 24.5+/-17.3 | 0.183+/-0.424 | 0.142+/-0.308 |
| Neural Network | Neurons=45, epochs=2500, decay=0.01 | 18.1+/-10.2 | 0.21+/-0.338 | 0168+/-0.220 |
| Random forest | Trees=100 | 17.3+/-11.4 | 0.201+/-0.467 | 0.182+/-0.382 |

**Table 5.33. Regression results for the integrated dataset.**

| Method | Parameters | RMSE(test) | Correlation(test) | Ccc(test) |
|---|---|---|---|---|
| k-NN | k=1 | 34.37+/-11.57 | 0.248+/-0.325 | 0.199+/-0.263 |
| SVM | Polynomial kernel, degree=2, scale=0.1, C=0.25 | 28.8+/-11 | 0.25+/-0.31 | 0.175+/-0.279 |
| Gaussian Process | Polynomial kernel, degree=3, scale=1 | 27.2+/-13.1 | 0.285+/-0.257 | 0.151+/-0.158 |
| Neural Network | Neurons=45, epochs=2000, decay=0.01 | 31.6+/-11.4 | 0.270+/-0.31 | 0.181+/-0.245 |
| Random forest | Trees=30 | 29.8+/-11.2 | 0.251+/-0.317 | 0.167+/-0.231 |

### 5.5.2 Classification predictive models results

Table 5.34 shows the results from the classification. The baseline prediction (predicting "Transient" which was the majority category) provides an accuracy of 66.4%.

**Table 5.34. Classification results for the THFC dataset.**

| Algorithm | Optimal parameter | Test accuracy | Cohen's kappa |
|---|---|---|---|
| Random forest | Trees=70 | 66.6%+/-3.05% | 0.256+/-0.173 |
| Naïve Bayes | None | 66.7%+/-4.73% | 0 |
| k-NN | k=5 | 64%+/-6.56% | 0.053+/-0.114 |
| SVM | Polynomial kernel, degree=3, scale=0.01,C=0.25 | 64.5%+/-4.52% | 0.179+/-0.131 |
| Neural Network | Neurons=35, decay=0.1, epochs=2500 | 65.69%+/-5.72% | 0.211+/-0.157 |
| Gaussian Process | Polynomial kernel, degree=3, scale=0.01 | 67.9%+/-2.08% | 0.209+/-0.197 |

**Table 5.35. Classification results for the WW dataset.**

| Algorithm | Optimal parameter | Test accuracy | Cohen's kappa |
|---|---|---|---|
| Random forest | Trees=25 | 67.4%+/-7.57% | 0.134+/-0.139 |
| Naïve Bayes | None | 70.08%+/-9.29% | 0 |
| k-NN | k=7 | 66.43%+/-9.29% | 0.013+/-0.171 |
| SVM | Polynomial kernel, degree=3, scale=0.01,C=0.25 | 67.06%+/-6.87% | 0.225+/-0.234 |
| Neural Network | Neurons=35, decay=0.001, epochs=2500 | 66.74%+/-9.51% | 0.170+-0.464 |
| Gaussian Process | Polynomial kernel, degree=3, scale=0.01 | 62.22%+/-6.4% | 0.167+/-0.170 |

**Table 5.36. Classification results for the integrated dataset.**

| Algorithm | Optimal parameter | Test accuracy | Cohen's kappa |
|---|---|---|---|
| Random forest | Trees=30 | 63.8%+/-6.68% | 0.169+/-0.121 |
| Naïve Bayes | None | 66.1%+/-1.38% | 0 |
| k-NN | k=5 | 60.09%+/-7.08% | 0.0566+/-0.128 |
| SVM | Polynomial kernel, degree=3, scale=0.1,C=0.1 | 65.2%+/-5.05% | 0.109+/-0.135 |
| Neural Network | Neurons=45, decay=0.01, epochs=2500 | 57.8%+/-6.61% | 0.211+/-0.157 |
| Gaussian Process | Polynomial kernel, degree=3, scale=0.1 | 61.53%+/-6.29% | 0.119+/-0.131 |

### 5.5.3 Discussion of the predictive modeling results

The regression task produces some interesting results. The mean correlation and the concordance correlation coefficient are higher than 0. However, the metrics suffer from very large variance in all datasets. Also, none of the metrics gets above 0.5. These results provide evidence that the task is feasible, but there is room for improvement. The best model overall is random forest. This can be an indication that there are noisy features in the dataset.

The results for the classification model are much worse. The main reason has to be the way that the "Severity" variable has discretized the recovery time. The discretization scheme might be medically relevant, but might not necessarily be helpful for a predictive model. The accuracy of all classifiers is about the same, but Cohen's kappa outlines a different picture. Random forests, SVM and MLP perform better than Naïve Bayes and k-NN on the THFC and the integrated dataset, even if they have about the same accuracy with these algorithms.

All these, indicate, like the results for the regression do, that there is information in the dataset for our goal, even if the performance is far from excellent.

Something that stands out is that random forest seems to perform way better than the other classifiers on the THFC dataset. Perhaps, this is an indication that there are many useless features in the dataset. For that purpose, the next step in the analysis was feature selection.

## 5.6 Feature selection

Based on the statistical analysis, it seems that there are some features that might be more important than others. This was also reinforced by the superiority of random forests on the task compared to the other methods, which can be an indication of noisy features in the dataset. Therefore, feature selection was conducted in order to find a better subset of features to improve performance.

### 5.6.1 Random forests

A first attempt at feature selection was conducted by using random forests. Feature selection through random forests was discussed in section 4.1.5. The importance was calculated by the node impurity for each feature summing over all trees. Node impurity in this particular case was measured by the residual sum of squares. In that particular case 500 trees were used. Finally, the variables are ranked from the variable that leads to the biggest decrease in node impurity, to the variables that leads to the least.

The number of trees used was the optimal number of trees for each dataset as described in the previous section. The node impurity was measured by the residual sum of squares. The response variable was the recovery time as number of days. The results are shown at Figure 5.28 below.

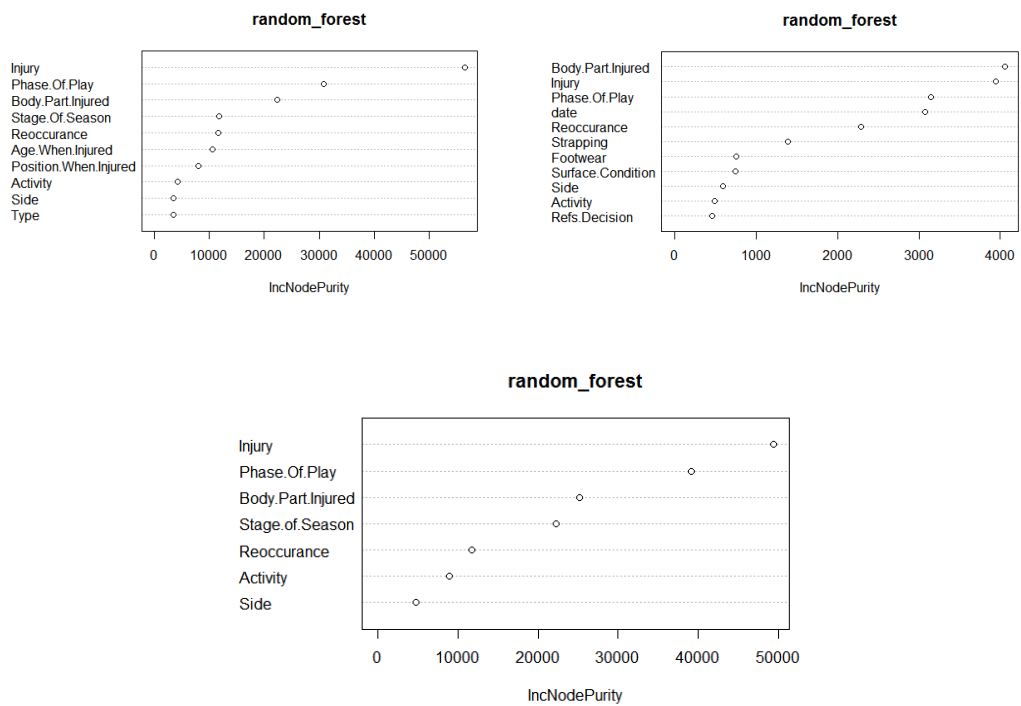

**Figure 5.28. Random forest feature importance results for the THFC, WW and integrated datasets respectively (from left to right).**

The two datasets produce very similar results, which also agree with the results from the statistical analysis regarding the significance of the variables. The most important features are "Injury", "Phase of play" and "Body part injured" for all datasets. "Recurrence" also seems to be quite important. The results are close to the ones derived by the statistical analysis in section 5.4.

### 5.6.2 Genetic algorithms with correlation-based feature selection

Genetic algorithms are commonly used for feature selection, especially when the number of features is large, because they allow exploration of the whole space (Devillers, 1996). The parameters of the genetic algorithm were the following:

- Population: 50
- Generations: 1000
- Crossover probability: 0.7
- Mutation probability: 0.05
- Selection: Roulette wheel

Each member of the population was described by a binary vector of the form [0, 1, 0, 1, 1…] where the length was equal to the number of features. Each vector was randomly initialized.

The fitness of the members was estimated using the Correlation-based Feature Subset Selection algorithm. The procedure was performed through 10-fold cross validation. The results are shown in Table 5.37. The columns show the percentage of times (across all folds) that a feature survived in the best individual of the last generation. The results are close to the results obtained by the random forest feature selection and the significance testing. The most important features for the THFC dataset are the "injury", the "body area" and "recurrence". The most important features for the WW dataset are the "injury" and "recurrence". For the integrated dataset, the most important features are the "phase of play", the "injury", the "body area" and the "recurrence".

Table 5.37. Results of the feature selection using correlation-based feature subset selection.

| Attribute | number of folds (%) THFC | number of folds (%) WW | number of folds (%) integrated |
|---|---|---|---|
| Activity | 0% | 0% | 0% |
| Phase of play | 10% | 10% | 100% |
| Age | 0% | - | - |
| Injury | 100% | 100% | 100% |
| Position when injured | 20% | - | - |
| Type | 0% | - | - |
| Injured Side | 0% | 20% | 0% |
| Body Area | 100% | 60% | 100% |
| Recurrence | 100% | 100% | 100% |
| Stage of Season | 0% | - | 60% |
| Footwear | - | 10% | - |
| Surface Condition | - | 50% | - |
| Strapping | - | 10% | - |
| Refs decision | - | 60% | - |
| Date (day of injury) | - | 0% | - |

### 5.6.3 Important features

Taking into account the results of feature selection, the regression algorithms were trained again on the datasets removing the following variables:

- THFC: type, side, activity, position when injured, stage of season
- WW: phase of play, activity, date
- Integrated: side, activity

Therefore, the THFC dataset now consisted of 5 variables (phase of play, injury, body part injured, recurrence, age) the WW dataset consisted of of 8 variables (injury, injured side, body part injured, recurrence, footwear, surface condition, strapping, referee's decision) while the integrated dataset had 5 variables (recurrence, body part injured, phase of play/mechanism, injury, stage of season/date). The results are shown in the tables below (Table 5.38 - Table 5.43). The three last columns of each table show the results of the previous run where all of the features had been included.

**Regression**

Table 5.38. Results for the THFC dataset. Removed: type, side, activity, position when injured, stage of season.

| Method | Parameters | RMSE (new) | RMSE (old) | Correlation (new) | Correlation (old) | Ccc (new) | Ccc (old) |
|---|---|---|---|---|---|---|---|
| k-NN | k=8 | 30.17+/-18.23 | 33.54+/-16.79 | 0.243+/-0.325 | 0.225+/-0.387 | 0.167+/-0.238 | 0.196+/-0.324 |
| SVM | Polynomial kernel, degree=3, C=0.5, scale=0.01 | 27.0+/-14.5 | 28.8+/-14.7 | 0.584+/-0.262 | 0.511+/-0.408 | 0.455+/-0.23 | 0.269+/-0.244 |
| Gaussian Process | Polynomial kernel, degree=3, scale=0.01 | 34.0+/-11.6 | 29.5+/-15.4 | 0.472+/-0.284 | 0.428+/-0.393 | 0.409+/-0.284 | 0.226+/-0.218 |
| Neural Network | Neurons=20, epochs=2500, decay=0.1 | 34.7+/-8.6 | 32.585+/-10.1 | 0.418+/-0.259 | 0.261+/-0.211 | 0.364+/-0.236 | 0.229+/-0.182 |
| Random Forest | Trees=70 | 30.3+/-18.9 | 30.1+/-21.4 | 0.463+/-0.323 | 0.471+/-0.341 | 0.362+/-0.288 | 0.406+/-0.337 |

Table 5.39. Results for the WW dataset. Removed: phase of play, activity, date.

| Method | Parameters | RMSE (new) | RMSE (old) | Correlation (new) | Correlation (old) | Ccc (new) | Ccc (old) |
|---|---|---|---|---|---|---|---|
| k-NN | k=1 | 15.58+/-6.84 | 22.21+/-9.46 | 0.479+/-0.448 | 0.03+/-0.26 | 0.322+/- | 0.023+/-0.253 |
| SVM | polynomial kernel, degree=4, scale=0.5, C=2.5 | 15.3+/-8.95 | 36.4+/-34 | 0.44+/-0.367 | 0.189+/-0.409 | 0.293+/-0.276 | 0.120+/-0.243 |
| Gaussian Process | polynomial kernel, degree=4, scale=0. 5 | 15.4+/-8.69 | 24.5+/-17.3 | 0.414+/-0.348 | 0.183+/-0.424 | 0.277+/-0.254 | 0.142+/-0.308 |
| Neural Network | Neurons=35, epochs=2500, decay=0.01 | 18.08+/-7.03 | 18.1+/-10.2 | 0.368+/-0.346 | 0.21+/-0.338 | 0.237+/-0.271 | 0168+/-0.220 |
| Random forest | Trees=10 | 16.6+/-9.26 | 17.3+/-11.4 | 0.0955+/-0.405 | 0.201+/-0.467 | 0.05+/-0.249 | 0.182+/-0.382 |

**Table 5.40. Results for the integrated dataset. Removed: Side and activity.**

| Method | Parameters | RMSE (new) | RMSE (old) | Correlation (new) | Correlation (old) | Ccc (new) | Ccc (old) |
|---|---|---|---|---|---|---|---|
| k-NN | k=1 | 32.80+/-8.83 | 34.37+/-11.57 | 0.228+/-0.345 | 0.248+/-0.325 | 0.216+/-0.305 | 0.199+/-0.263 |
| SVM | Polynomial kernel, degree=2, scale=0.1, C=0.25 | 32.6+/-12.1 | 28.8+/-11 | 0.313+/-0.262 | 0.25+/-0.31 | 0.234+/-0.212 | 0.175+/-0.279 |
| Gaussian Process | Polynomial kernel, degree=4, scale=1 | 31+/-12.6 | 27.2+/-13.1 | 0.293+/-0.24 | 0.285+/-0.257 | 0.230+/-0.24 | 0.151+/-0.158 |
| Neural Network | Neurons=10, epochs=2000, decay=0.1 | 33.9+/-11.1 | 31.6+/-11.4 | 0.371+/-0.351 | 0.270+/-0.31 | 0.333+/-0.315 | 0.181+/-0.245 |
| Random forest | Trees=30 | 29.4+/-12.2 | 29.8+/-11.2 | 0.27+/-0.281 | 0.251+/-0.317 | 0.190+/-0.223 | 0.167+/-0.231 |

**Classification**

**Table 5.41. Results for the THFC dataset. Removed: type, side, activity, position when injured, stage of season.**

| Method | Parameters | Accuracy (new) | Accuracy (old) | Kappa (new) | Kappa (old) |
|---|---|---|---|---|---|
| Random Forest | Trees=70 | 64.5%+/-10.8% | 66.6%+/-3.05% | 0.21+/-0.18 | 0.256+/-0.173 |
| Naïve Bayes | None | 66.5%+/-3.1% | 66.7%+/-4.73% | 0 | 0 |
| k-NN | k=7 | 66%+/-5.2% | 64%+/-6.56% | 0.03+/-0.11 | 0.053+/-0.114 |
| SVM | Polynomial kernel, degree=2, scale=0.1, C=0.5 | 69.1%+/-6.8% | 64.5%+/-4.52% | 0.23+/-0.09 | 0.179+/-0.131 |
| Neural Network | Neurons=15, epochs=2500, decay=0.01 | 61.3%+/-9.1% | 65.69%+/-5.72% | 0.26+/-0.08 | 0.211+/-0.157 |
| Gaussian Process | Polynomial kernel, degree=3, scale=0.1 | 63.9%+/-3.2% | 67.9%+/-2.08% | 0.21+/-0.08 | 0.209+/-0.197 |

**Table 5.42. Results for the WW dataset. Removed: phase of play, activity, date.**

| Method | Parameters | Accuracy (new) | Accuracy (old) | Kappa (new) | Kappa (old) |
|---|---|---|---|---|---|
| Random Forest | Trees=50 | 60.6%+/-14.2% | 67.4%+/-7.57% | 0.02+/-0.2 | 0.134+/-0.139 |
| Naïve Bayes | None | 68.9%+/-7.1% | 70.08%+/-9.29% | 0 | 0 |
| k-NN | k=7 | 67%+/-7.5% | 66.43%+/-9.29% | 0 | 0.013+/-0.171 |
| SVM | Polynomial kernel, degree=3, scale=0.1, C=1 | 63.8%+/-20.4% | 67.06%+/-6.87% | 0.27+/-0.42 | 0.225+/-0.234 |
| Neural Network | Neurons=10, decay=0.0001, epochs=2500 | 57.6%+/-13.1% | 66.74%+/-9.51% | 0.16+/-0.13 | 0.17+-0.464 |
| Gaussian Process | Polynomial kernel, degree=3, scale=0.01 | 55.12%+/-19.3% | 62.22%+/-6.4% | 0.12+/-0.15 | 0.167+/-0.17 |

**Table 5.43. Results for the integrated dataset. Removed: Side and activity.**

| Method | Parameters | Accuracy (new) | Accuracy (old) | Kappa (new) | Kappa (old) |
|---|---|---|---|---|---|
| Random Forest | Trees=200 | 63.8%+/-7.3% | 63.8%+/-6.68% | 0.21+/-0.18 | 0.169+/-0.121 |
| Naïve Bayes | None | 66%+/-2.4% | 66.1%+/-1.38% | 0 | 0 |
| k-NN | k=7 | 65.2%+/-5.9% | 60.09%+/-7.08% | 0.09+/-0.1 | 0.0566+/-0.128 |
| SVM | Polynomial kernel, degree=3, scale=0.1, C=1 | 59.1%+/-10.8% | 65.2%+/-5.05% | 0.11+/-0.15 | 0.109+/-0.135 |
| Neural Network | Neurons=15, decay=0.001, epochs=2500 | 61.3%+/-7.5% | 57.8%+/-6.61% | 0.24+/-0.16 | 0.211+/-0.157 |
| Gaussian Process | Polynomial kernel, degree=3, scale=1 | 64.2%+/-7.6% | 61.53%+/-6.29% | 0.20+/-0.15 | 0.119+/-0.131 |

### 5.6.4 Discussion of feature selection results

Regarding regression, it is clear that feature selection has clearly aided the algorithms in the task. Three things stand out from the modeling of the new datasets. First, after feature selection, the mean concordance correlation coefficient for SVMs, Gaussian Processes and neural networks has increased in mean and reduced in variance. The best model for the THFC and the WW datasets is the SVM, while for the integrated dataset is the neural network.

Secondly, random forest seems to perform worse on the THFC and WW datasets after feature selection. The performance on the integrated dataset remains about the same. Before feature selection, random forest had the best performance. This is an indication that the embedded feature selection of random forest does not help in this case, since a good subset of features has already been chosen. This provides further validation for the choice of features.

Thirdly, even though the concordance correlation coefficient has improved in most cases, the RMSE might have not, and in some cases it might even get worse. However, since the response variable is skewed, the RMSE can be influenced very easily by mistakes in instances where the recovery time was long.

The picture is less clear for classification. Feature selection seems to have helped in the THFC dataset, with neural networks achieving relatively good performance. Some small improvements can also be seen for the integrated dataset. However, the performance on the WW dataset is not good. The SVM shows an improved kappa statistic, but at the cost of larger variance.

## 5.7 Discussion

### 5.7.1 Evaluation of the results

It is possible to achieve a reasonably good accuracy at the regression task. Without feature selection the models do not perform very well, but the performance improves greatly after feature selection is conducted.

Similarly, the classification results are not very good, but they improve for the THFC and the integrated datasets after feature selection. The poor performance in the classification task might come as a result of the way that the recovery time is being discretized. This conversation comes back to one of the points discussed in section 3.1, regarding the use of non-statistically relevant jargon in sports data. The discretization scheme applied in this case might help the practitioner assess an athlete better, but does not aid the prediction accuracy of classification algorithms.

It seems that the SVM outperforms slightly the other methods for regression and is also good for classification. Also, it seems that polynomial kernel seems to be more successful in this task when compared to the RBF kernel both for SVMs and Gaussian processes. The kernels seem to work best when the degree is between 2 and 4. This clearly indicates the existence of second (or higher) order interactions, something that had been evidenced in the statistical analysis as well.

### 5.7.2 Feature elimination

The performance improved when some features were simply removed. These features were:

- THFC: type, side, activity, position when injured, stage of season.
- WW: phase of play, activity, date
- Integrated: side, activity

Some plausible explanations as to why some features might be redundant follow.

Regarding the THFC dataset, the variable "type" probably contains information that is already contained in the "injury" variable. Most injuries will either fall in the "Acute" or "Overuse" category. For example, the category "haematoma/contusion/bruise" corresponds always to an acute injury. Similarly, a large proportion of injuries of type "overuse" (28% of the total overuse injuries), correspond to the "overuse symptoms unspecified" or "overuse/hypertonia" categories of the injury variable.

The "stage of season" for the THFC dataset and the "date" for the WW dataset seem to be useless. This probably indicates that fatigue and physical stress might not get accumulated in the bodies throughout the season, at least in a way that can affect recovery.

The variable "activity" is useless for all datasets. It had already been discussed that Table 5.12 does not show any big differences on the recovery time between different activities, so this result is not surprising.

The variable "position when injured" is an interesting case. Some types of injuries can be more common for some positions than others (Hunt & Fulford, 1990). However, the position itself might not be relevant with recovery. Secondly, the THFC recorded the position at that moment, which in training can sometimes be different to the player's true position. This could have reduced the amount of information in this variable that was relevant for predicting recovery.

Something interesting is that the "injured side" seems to be useless for the THFC dataset, but not for the WW dataset. Similarly, the "phase of play/mechanism" seems to be useless for the WW dataset but not for the THFC dataset. Perhaps the fact that these variables can deteriorate the performance of the predictive models in one dataset but not the other might have more to do with the way the variables were being recorded within each club.

## 5.8 Conclusion

### 5.8.1 Overview

This research dealt with the question of whether it is possible to predict the recovery time after an injury in professional football based on the UEFA injury recordings, while it also testing different methods against each other for this task. The results illustrate that it is possible to reach some degree of accuracy in this task, but the size of the dataset, and maybe the variables themselves, limit the performance.

However, this work paves the way for future research that can include bigger and more complicated datasets and can also be extended by protocols that can combine experts' opinions. Future research will build on top of the current results in order to provide a functional system for assessing injuries in professional football.

There are many types of injuries in football that can occur under different circumstances. Future research should use datasets from other football clubs in order to verify and expand the current results. Ideally, datasets from football clubs from different countries should be obtained, since the style of play in each country, along with other factors (e.g. a country's climate), could influence the response variable. Additional information that could be used to improve the model includes anthropometric and medical information such as the height, weight or medical blood tests of players.

An interesting feature of this task is that the models could be included in a diagnostic protocol. After each injury, the medical staff will conduct detailed medical tests in order to diagnose the injury. The models presented in the previous sections (such as the polynomial kernel SVM) could accompany a diagnosis, providing some additional support for the experts' estimates.

### 5.8.2 Limitations

The study suffers from some limitations.

The main limitation in the study itself is that the model is still using only simple descriptive predictors. Intrinsic information is definitely required for an accurate model. For example, studies have indicated that MRI (Ekstrand, Healy, Walden, Lee, & English, 2012; Hallén & Ekstrand, 2014) as well as the passive straight leg raise (Moen, et al., 2014), can be useful when predicting the time to return to play for hamstring injuries.

A methodological limitation is the way that the data was being recorded. As it was mentioned, there were discrepancies between the two clubs, since it seems that there was some freedom in recording down data for some particular variables. The methods and algorithms used worked for all datasets. However, tests with data from other clubs would be required in order to understand whether the results generalize well.

A final limitation is that there are championship specific factors involved, which would require data from other championships in order to be assessed. For example, playing or coaching style might influence recovery. Nutrition is another factor. Data from different leagues and countries would help assess better the generalization of the models and any conclusions regarding the importance of the various variables in the UEFA injury recordings.

# 6 Predicting injuries in professional football using exposure records

*A common problem in football training, and other sports is understanding the relationship between the training and match schedule and injury. The purpose of this investigation was to predict the time (in days) to the first injury of the season based on training and match exposure records. For that purpose, a Gaussian process model was used equipped with a dynamic time warping covariance kernel. The results illustrate the feasibility of the proposed task.*

## 6.1 Introduction and motivation

### 6.1.1 Overview

A factor that is commonly accepted to affect injury incidence is exposure, which is defined as the time that a player spends either in training or in game. Studies in the past in the English (Odetoyinbo, Wooster, & Lane, 2008) and Spanish leagues (Rey, Lago-Pe, Lago-Ballesteros, & Casais, 2010) have studied how an increased number of football fixtures can lead to increased fatigue. This makes sense, since increased match exposure is more likely to lead to events such as collisions with another players or accidents.

The collection procedures for exposure in football have been standardized (Fuller, et al., 2006) and the time is recorded in minutes. When studying the connection between injuries and exposure a common metric is the number of injuries per 1000 hours of training and match. For example, this metric was used by the UEFA injury study for 2009 (Ekstrand, Hägglund, & Waldén, 2009). Ekstrand et al. (2006) also used the same metric for comparing injuries on natural and artificial turf. Dupont et al. (2010) used the number of injuries per 1000 hours of exposure when studying the effect of 2 football matches in a week. Arnason et al (2004) and Mallo and Dellal (2012) used the same way to assess various risk factors (such as height, weight, or leg extension power) for injuries. Hagglund et al. (2003) used the same metric for comparing the risk of injury over two decades.

The number of injuries per 1000 hours of exposure has also been used when studying the relationship between injuries and exposure on a nationwide level. The injury incidence in Icelandic football has been estimated to be around 35 per 1000 hours of match and 6 per 1000 training hours (Arnason, Gudmundsson, Dahl, & Jóhannsson, 1996) with similar statistics being reported for Major League Soccer (Morgan & Oberlander, 2001), English clubs (Hawkins & Fuller, 1999), the Saudi Professional League (Almutawa, Scott, George, & Drust, 2013), and the Swedish football league (Waldén, Hägglund, & Ekstrand, 2005).

What these studies miss, however, is a more fine-grained understanding of how exposure within club level can affect the incidence of injury. The training and match schedules can differ in many ways such as the maximum number of minutes trained or the rest periods. However, using a metric such as the number of injuries per 1000 hours of training/match conceals this information and does not aid the coach when designing a new training schedule that should avoid causing fatigue.

The purpose of this investigation is to build a predictive model for injuries that is based on the training and match exposure records of the athletes. This can give the opportunity to a practitioner to experiment with different types of training and match schedules and understand how they affect the likelihood of injury.

It was decided that the best course of action was to build a model of the time that it takes for an athlete to be injured for the first time in the season, ignoring the rest of the season. This choice is based on two reasons.

First, training injuries in soccer peak during July and match injuries peak in August (Hawkins, Hulse, Wilkinson, Hodson, & Gibson, 2001), most probably because the soccer players return from a long resting period.

Secondly, this approach was chosen because the data from that period (from the beginning of the off-season until the first injury) can provide us with the most cohesive dataset. When the off-season starts, the players are relatively fresh. Their bodies have recovered from the previous season and the medical staff will take extra care in the first weeks of training to make sure that the players get introduced into the season in an appropriate physical condition.

As soon as the season starts, the physical demands increase in different ways for each player. Injuries throughout the season can change the underlying phenomenon that is being modeled. The player might get additional sensitivity at the body part where he was injured previously. Also, the kind of exposure might change, as compared to the early season, since training might also include specific exercises for rehabilitation.

A Gaussian process model was used in order to predict the number of days to injury. The model used a covariance kernel based on Dynamic Time Warping, which can compare the similarity between subjects with exposure records of different lengths. The results illustrate the clear link between exposure and injury incidence.

### 6.1.2 Research goals
The research goals for this investigation were the following:

- Create a predictive model for the first injury that will take place from the onset of the season.
- Find a way to compare the similarity between the exposure records of different subjects.

## 6.2 Data

### 6.2.1 Initial dataset
The dataset consisted of the exposure records of 35 professional football players from the first team of Tottenham Hotspur Football Club. The dataset included the following variables:

- Player name
- Training exposure: The exposure time in minutes that the player trained for that day. No further classification of training (e.g. weight training or football drills) was done.
- Match exposure: The minutes the player spent playing on the pitch.
- Injured: Whether the player was injured or not.
- Day since $8^{th}$ of July: The number of days that passed since the $8^{th}$ of July 2012. That day is used as a reference point, because it was the beginning of the pre-season period.

Something important to note is that the training exposure consisted only of ball training and no weightlifting training. Training in football is usually split between sessions where the ball is used and weightlifting. If both training methods had been used, then it would be important to use different training exposure records for each, since the physical demands on the body can be different for each.

Table 6.1 below shows a few example rows from the dataset.

**Table 6.1. Example of the original dataset.**

| Player name | Day (since 8th July 2012 | Training exposure (mins) | Match exposure (mins) | Injured |
|---|---|---|---|---|
| Player 1 | 1 | 90 | 0 | No |
| Player 1 | 2 | 120 | 0 | No |
| Player 1 | 3 | 0 | 0 | Yes |
| Player 2 | 1 | 0 | 90 | No |
| Player 2 | 2 | 120 | 0 | No |
| Player 2 | 3 | 0 | 20 | No |
| Player K | 1 | 120 | 0 | No |
| Player K | 2 | 0 | 90 | No |
| Player K | 3 | 0 | 0 | Yes |

An issue with the dataset was that the days that were recorded were days that were noted either as training sessions or match sessions. Days off were not being recorded. However, these days are relevant for modelling since they constitute recovery days. Therefore, for that purpose, if a particular day was missing from the dataset, it was inserted and the training and match exposure times were set to 0.

Figure 6.1 below shows an example of an exposure series of a player after the extra days were added.

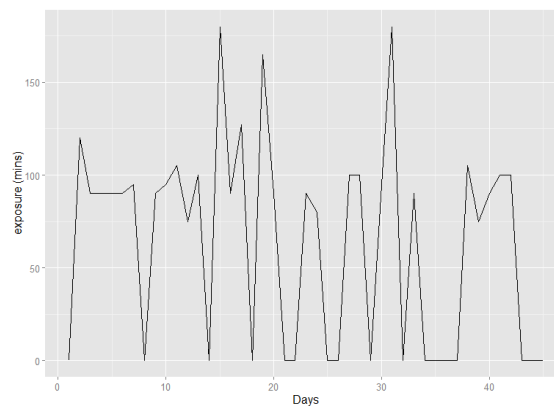

**Figure 6.1. Example of a training exposure series.**

There was lots of variation in the exposure times. This could come as a result of players having to leave the club to train with the national team or some players training for longer or shorter periods on the same day compared to other players.

The training exposure across all players had a mean of 35.83 minutes and standard deviation of 42.75 minutes. The mean as a measure of match exposure is less informative, because players in the main squad can play up to 90 minutes, but other players can play for 20-30 minutes, or even not at all. This explains the skewness in the histograms. Figure 6.2 shows histograms for training and match exposure for all players.

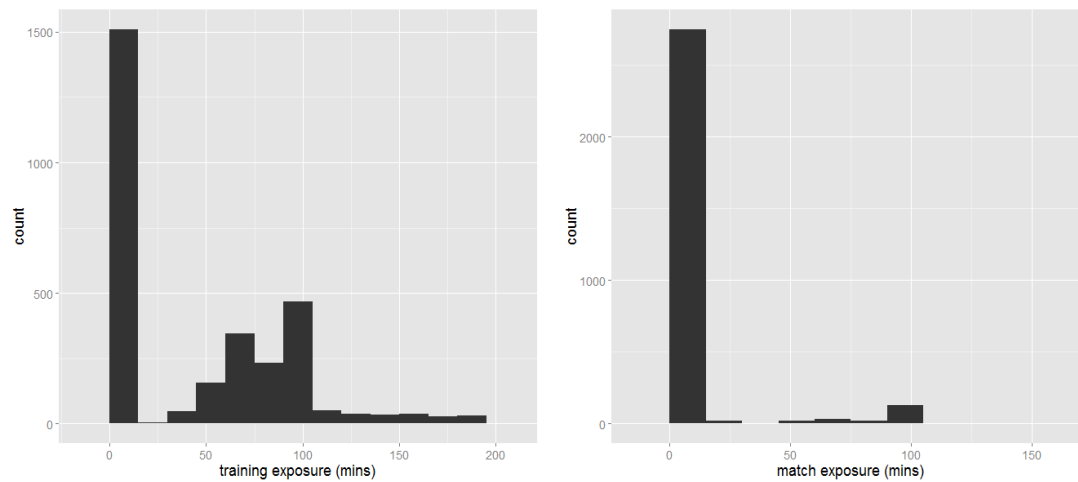

**Figure 6.2. Histograms of training and match exposure.**

The majority of values for both training and match exposure are 0. The match exposure has greater variability with very low training times, to training times that can range to 3 hours. Longer training times usually come as a result of multiple training sessions within the same day.

The match exposure is more structured. The majority of exposure times (when a match takes place and the exposure is non-zero) is 90 minutes. There are cases with exposure times less than zero, which correspond to substitutions within the game. Exposure times longer than 90 minutes usually correspond to games in the pre-season that are done for training purposes.

### 6.2.2 Cleaning up and preprocessing the data

In order for the model to make sense, a number of days should pass before a player gets injured. If a player gets injured too early (e.g. on the second day of the season), then it is unlikely that the reason of injury can be attributed specifically to the training schedule. Rather it could be due to factors such as the player not being fit or an accident.

Another issue was using injuries in the model that can be attributed to the training schedule. Collision injuries or transient injuries (e.g. a hand bruising) cannot be treated as such. An intrinsic injury was defined as an injury not-related to collision or contact of any sort. This definition was discussed and its validity confirmed by the medical team of THFC.

From the initial dataset of 35 players, 6 players were removed, leaving 29 players in the dataset. Out of the players that were removed 3 were goalkeepers. The goalkeeper is a special position, since the physical demands are different to that of other players on the field. Therefore, we decided to remove any goalkeepers, since a model of injuries on goalkeepers might be different to that of the rest of the team.

From the rest of the players that were removed, two of them were players that were injured within the first 3 days of training. It is likely that these injuries were caused due to other factors (such as already existing conditions) and cannot be directly attributed to the training itself.

The final player that was removed had an injury as a result of collision during a match. This is a special case that cannot be attributed to fatigue, so the case was removed in order to ensure that the dataset remains consistent with the focus of this study.

Finally, there were a few cases of players that had an injury but the injury was transient, and they player did not lose any days from training. It is very difficult to attribute this incidence to fatigue. Therefore, this injury was removed from the dataset, and we used the next injury that took place for this player. There were 5 cases like that in the dataset.

Out of the 29 players that remained, 21 players had been injured. The rest of the players left the club before the end of the season, and until that moment no injury had been recorded for them. Therefore, 9 players constitute right-censored cases. The response variable ranged from 14 days to 312, with the mean being 91.73 and the median 49.5.

The normality of the response is one of the assumptions behind Gaussian processes. Transforming the response through the log function takes the histogram closer to the normal distribution as can be seen on in the boxplots at Figure 6.3.

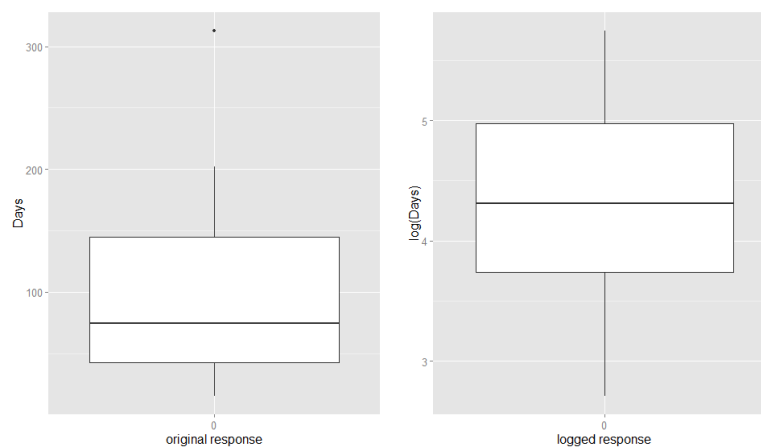

**Figure 6.3. Boxplots for comparison between the original response and the logged version.**

## 6.3 Methods of analysis

### 6.3.1 Assessment

The model was assessed with the following strategy. Let $T_i$ be the day of injury for subject $i$. For each subject $i$ in the training set the whole exposure record is used until time $T_i$. However, for each test subject $j$ the model is evaluated from $T_j$ to $T_j - 12$. So, the complete exposure record is used down to the exposure record up to 12 days before the day of injury. The value 12 was chosen because the smallest response variable was 14. For each parameter setting and for each $T_j - a, a \in \{0, \ldots, 12\}$, the concordance correlation coefficient and the MAE were calculated by using leave-one-out cross-validation.

This scheme is better illustrated in the figure below (Figure 6.4). The two athletes have been injured on different days (athlete A on day 5 and athlete B on day 3), but on each occasion we remove the last $a$ cases from their records. So, if $a = 0$ the full record from each athlete is used. If $a = 1$, then we will take records from 4 days before injury for athlete A and 2 days before injury from athlete B. Therefore, the testing scheme evaluates the ability of the model to predict injuries for different points into the future, based on the current data. From a practitioner's perspective the goal is to understand the time span over which the predictions of the model can be trusted.

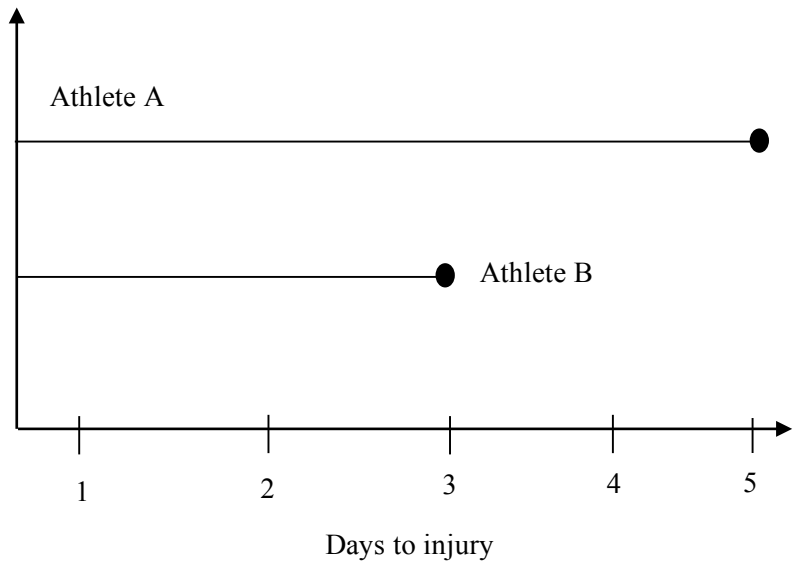

**Figure 6.4. Example of the testing scheme. The circle signifies the day of injury.**

Assume a real world scenario, where a practitioner wants to know whether an athlete is currently safe from injury. Define the current day to be T. A prediction $f^*$ for which $f^* < T$ can be interpreted as saying that the athlete should have already been injured, or, otherwise, is in an imminent danger of getting injured. A prediction $f^* > T$ can be interpreted as saying that the model assumes that the player will be injured in the future, and is safe from injury for now.

### 6.3.2 Models

The model of choice in this investigation was a Gaussian process model. A Gaussian process (explained in more detail in section 4.1.1) requires the specification of a covariance kernel. A challenge with this data is that the exposure records from two different subjects have different lengths. This makes the use of common kernels, such as the polynomial kernel, impossible. For that purpose, a covariance kernel was used that is based on dynamic time warping (DTW).

DTW (which was explained in detail in section 4.3) allows the comparison of time series of different lengths, a task which is not feasible with other measures of distance or association, such as the Euclidean distance or Pearson's correlation coefficient.

This property of DTW has made it useful in a number of studies. DTW has been used in financial time series (Gudmundsson, Runarsson, & Sigurdsson, 2008; Mager, Paasche, & Sick, 2008), decision trees (Rodriguez & Alonso, 2004) and brain activity classification (Chaovalitwongse & Pardalos, 2008). It has also been used in the past with support vector machines for audio speech recognition (Hiroshi, Ken-ichi, Mitsuru, & Shigeki, 2001). It has also been used with Gaussian processes for the recognition of bat species and flight calls (Damoulas, Henry, Farnsworth, Lanzone, & Gomes, 2010).

DTW does not necessarily produce a positive a semi-definite covariance kernel (Hansheng & Bingyu, 2007). In practice, non-positive definite kernels have been used, as well, as long as they provide good results for the task at hand.

For example, it has been demonstrated in practice that conditionally positive semidefinite kernels, such as the power kernel, can be particularly useful for some applications, such as image recognition (Boughorbel, Tarel, & Boujemaa, 2005). The sigmoid kernel is another kernel which is a popular choice for applications, even though it is not positive semi-definite (Hsuan-Tien & Chih-Jen, 2003).

In SVMs a positive semidefinite kernel ensures that the optimization problem is convex and the solution unique. In any other case, it is unclear whether a global optimum has been reached. In Gaussian process regression, this problem does not exist. There is a closed form solution for the predictive mean and variance of the model. So, a solution can always be found, unless the covariance kernel of the training instances $K(X, X + \varepsilon I)^{-1}$ is non-invertible.

For this research, it is still possible to use DTW as a kernel as long as the variance is not negative, since the kernel suggested can still be positive semi-definite for some datasets. A common trick (Hansheng & Bingyu, 2007) that has been employed in the past is to simply use the DTW distance inside an RBF kernel, instead of the Euclidean distance. So, $\exp\left(-\frac{\|x-y\|^2}{2\sigma^2}\right)$ become $\exp(-DTW(\boldsymbol{x},\boldsymbol{y})/2\sigma^2)$. The term $2\sigma^2$ is constant. By setting $\gamma = 1/2\sigma^2$ so we can define the following kernel:

$$K'(\boldsymbol{x},\boldsymbol{y}) = \exp(-\gamma DTW(\boldsymbol{x},\boldsymbol{y})), \gamma > 0 \tag{43}$$

The constant $\gamma$ should only take values that lead to positive variance, and this is something that is placed as an additional constraint.

At the present study each subject is characterized by two time series: one for the match and one for the training exposure records. In order to take both time series into account, we define the covariance kernel for the Gaussian process as a weighted average of the kernels in (44) for the two types of exposure:

$$K_{ij} = \frac{K'\left(\boldsymbol{x}_i^{training}, \boldsymbol{x}_j^{training}\right) + K'\left(\boldsymbol{x}_i^{match}, \boldsymbol{x}_j^{match}\right)}{2} \tag{44}$$

The choice of using an average between the two DTW distances is arbitrary. However, in practice this produced good results, while it also seemed to produce positive variances in the majority of cases.

The unknown parameters are the parameter $\gamma$ and the noise $\varepsilon$. Runs were conducted with 1000 different values of $\gamma$ in the range [0.00002, 0.2] and with different values for $\varepsilon$ in the range [0.0001, 0.01]. Parameter sets that produced negative variances were discarded.

## 6.4  Results and discussion

Figure 6.5 shows the performance of the best model for each $T - a, a \in \{0, \dots, 12\}$.

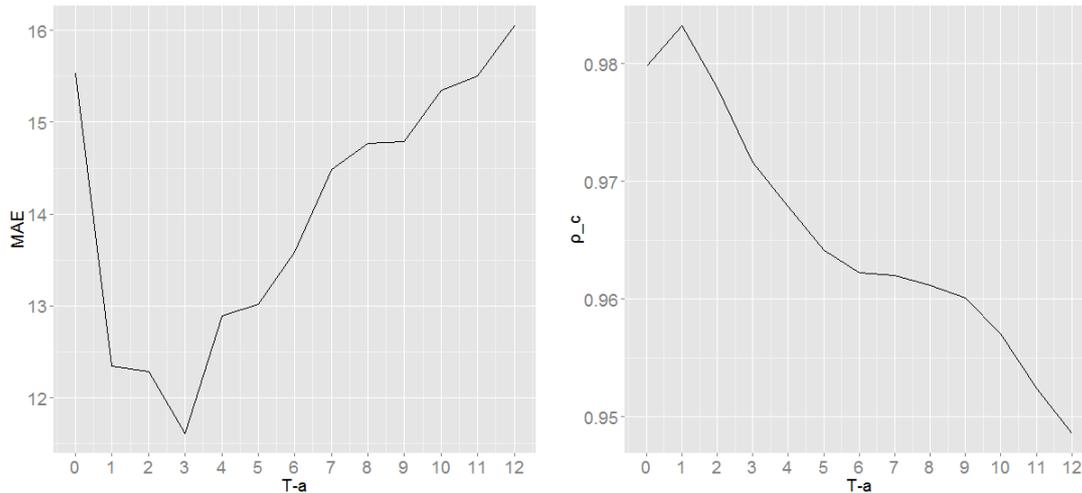

**Figure 6.5. MAE and concordance correlation coefficient for the best model from T to T-12.**

Firstly, the performance of the model is very good overall. The concordance correlation coefficient fluctuates from about 0.95 to 0.98, while the best MAE is about 11.

Secondly, something else that can be observed is that when as the records shorten more and more the performance of the model deteriorates on average. This is to be expected, since information is reduced in the dataset.

An interesting phenomenon, however, takes place from $T$ to $T-3$ as the concordance correlation coefficient and the MAE provides somewhat different results. According to the concordance correlation coefficient, the performance of the model from $T$ to $T-2$ is approximately the same, with $T-1$ having the best results. The MAE, however, seems to be minimized for $T-3$. The MAE starts high, then suddenly drops off, and then slowly rises again.

It seems that the models' predictions are fairly reliable, even when looking into 12 days into the future, since the concordance correlation coefficient does not fall below 0.95. However, the clear deteriorating trend indicates that there must be a threshold beyond which predictions cannot be trusted. Unfortunately, this could not be checked with the current dataset.

A detailed table of all the results for this analysis can be found in the Appendix (Table 10.10, Figure 10.1 and Figure 10.2).

## 6.5 Conclusions

This investigation provided evidence that predicting injuries from exposure records is a feasible task. This research demonstrated a way to use a Gaussian process model with a DTW kernel in order to handle exposure records of different lengths and predict the time until the first injury of the season.

Clearly the study suffers from some limitations as well. First, a larger dataset and a longer truncation range for the exposure records could provide a better assessment of the models. Also, the dataset is originating from only a single football club. More tests across different clubs should be done in order to validate the model.

Secondly, the professional athletes in this dataset did not participate in weightlifting training. Weightlifting training is a common method of training in many football clubs. The model would need to be updated in order to accommodate this kind of training if this research is carried out with datasets from other clubs.

Finally, the model predicts only the first injury incidence from the beginning of the season. It is clear that the next step would be to predict injuries throughout the season as well, since the applicability of this model is restricted only to the first few months of the season.

# 7 Predicting intrinsic injury incidence using in-training GPS measurements

*Wearable GPS technology can provide a wealth of information regarding the performance of an athlete. This information could be used in order to discover early signs of fatigue or overtraining which can lead to injury. This research compares various methods (random forest, SVMs, Gaussian Processes, neural networks, supervised PCA, ridge logistic regression, k-NN) for predicting intrinsic injuries from GPS data. Two different binary classification approaches were followed. Supervised PCA performs quite well on the task, achieving a mean kappa statistic of 0.21 and 0.14 in the two approaches respectively. Supervised PCA is also used for extracting components from the data which correlate with injury, and help summarize the large number of variables (69 in total).*

## 7.1 Introduction and motivation

Global positioning systems (GPS) are nowadays commonly used in many sports such as Australian rules football, hockey, rugby and cricket (Aughey, 2011; Cummins, Orr, O'Connor, & West, 2013). The benefits of such systems is the easy collection of variables that are very difficult (or impossible) to acquire otherwise. This includes, for example, the total distance covered in a training session, the number of sprints, the accelerations and the decelerations. These are variables that can correlate well with injury. Gabbett and Ullah (2012) reported that higher distances (in any speed) and high-speed running are related to injury in elite sports.

From the perspective of injury prediction, a GPS system, used in training, can be particularly useful for getting a more accurate picture of the load of each individual athlete. Studies for that have appeared in the last few years. For example, Aughey (2010) used GPS data to track the fatigue of players in Australian football. Wisbey et al. (2010) used GPS data to track the physical demands during games in the Australian Football League. Young et al. (2012) used GPS data to detect muscle damage in Australian football players.

GPS data has also been used for tracking the physiological demands of rugby players (Cunniffe, Proctor, Baker , & Davies, 2009). Similarly, Coughlan et al. (2011) used GPS data to track the activities of rugby player during a match, suggesting that such data, used in conjunction with video analysis, can help clarify the mechanism of many injuries.

In soccer, Casamichana et al. (2013) used GPS data as indicators of training load. They discovered that there is a close relationship between the GPS-collected variables and the rate of perceived exertion.

Currently, there has been no research in Premier League regarding the potential of GPS data to predict injuries. This comes partly due to the banning of the use of GPS units inside the pitch. However, GPS data can be used in training. The aforementioned research indicates that GPS data can be used to track fatigue or other physiological indicators which could possibly correlate with intrinsic injury incidence. The UEFA injury study for 2013/2014 (Ekstrand J. , 2014) reports that out of all injuries, 51.1% can be attributed to either overuse, tendon injury or muscle strains and cramps all of which are the result of intrinsic factors, such as inflammation.

The purpose of this investigation was to build a predictive model for predicting intrinsic injuries based on GPS data. Two different approaches were used in order to convert the GPS data collected from training to a binary classification problem of predicting whether an injury will take place or not. Various classification models were used (SVM, Gaussian process, random forest, naïve Bayes, ridge logistic regression, supervised PCA) with supervised PCA having the best performance as measured by the kappa statistic (explained in section 4.2.2). The principal components extracted from the supervised PCA are also used in order to reduce the dimensionality of the dataset and create interpretable components that correlate with injury.

## 7.2 Data

### 7.2.1 Initial dataset

The data consisted of GPS records of 29 professional football player from the first team of THFC from the period July - December 2014 and a record of the days during which they were injured. The GPS system was used only during training and not during matches, because of the premier league regulations. The GPS system used was the STATSports Viper[10].

There were 68 GPS variables recorded along with the duration of the training session, giving 69 variables in total. The large number of variables comes as a result of using zones for measuring a single variable. Each zone describes a speed range relative to the maximum speed for each athlete. There are 6 zones in total. An example is shown at Table 7.1 below:

Table 7.1. Example of different zones defined as regions of the maximum speed.

| Zone     | Z1    | Z2     | Z3     | Z4     | Z5     | Z6   |
|----------|-------|--------|--------|--------|--------|------|
| Boundary | <35%  | 35-45% | 45-55% | 55-65% | 65-75% | >75% |

The initial number of rows for the dataset was 1310.

Figure 7.1 shows a correlogram for the variables, where variables with high correlation have been clustered together. There are various clusters that can be identified throughout the graph. First, there are clusters of variables that are measuring the same entity in different zones. For example, impacts on zones 1 to 6 are correlated with each other, as well as decelerations from zones 1 to 6.

There are also some other correlations that can be observed. Accelerations and decelerations are correlated. Similarly, the total number of steps and the total loading are correlated with distance. Also, the impacts are correlated with dynamic stress load.

---

[10] http://statsports.ie/

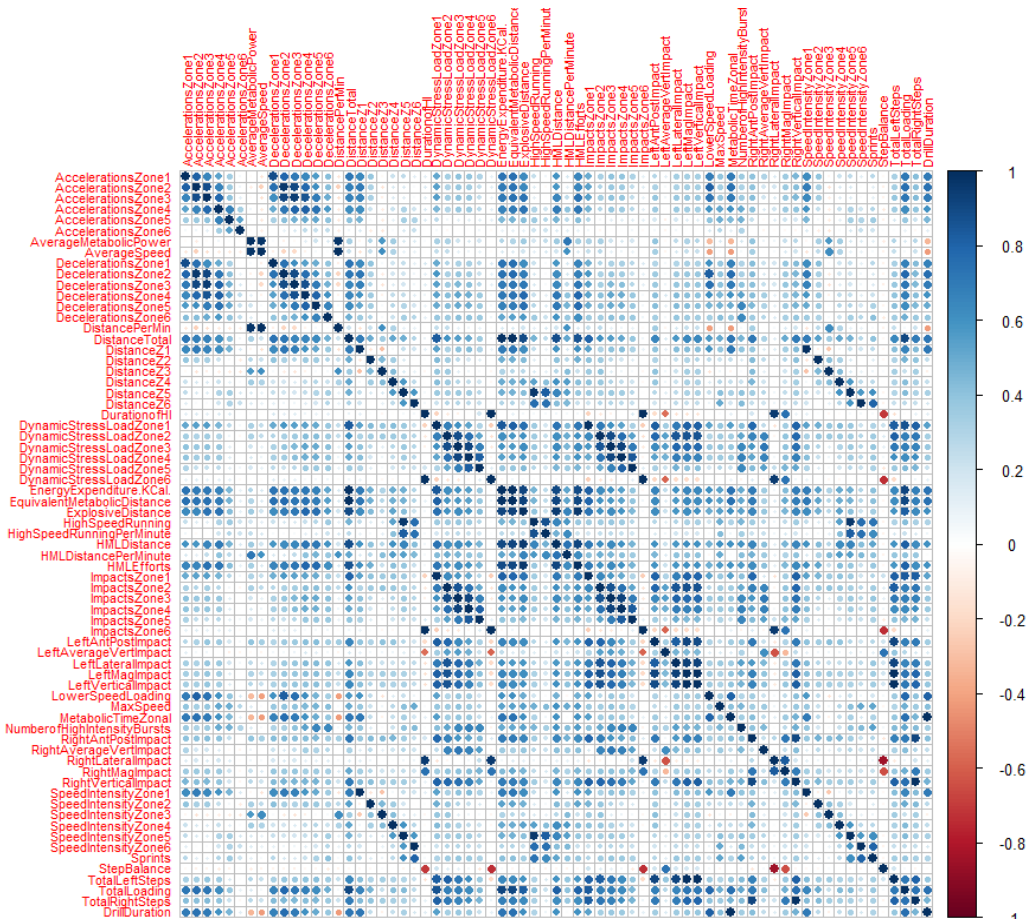

**Figure 7.1. Correlogram for the GPS variables.**

The main variables in the dataset are described below. A full detailed list is in the Appendix. A common theme in the histograms of the variables is that they are usually have some skewness on the right side, or some outliers. These correspond to training sessions of high intensity or high duration.

All the figures shown below are based on the dataset for approach B (explained in section 7.2.2).

**Distance covered**

A normal value for the distance covered for a 90 minute match ranges between 9 and 13km. Figure 7.2 shows the histogram of the distance covered. The distribution is approximately symmetric with a slightly heavier right tail.

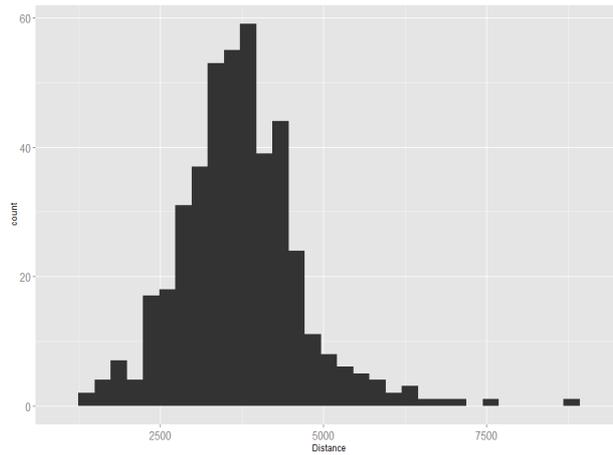

**Figure 7.2. Histogram for the "distance covered".**

Figure 7.3 below shows a boxplot of "distance covered" versus injury where it can be seen that there are indeed differences between the two groups, with injured players covering greater distance on average.

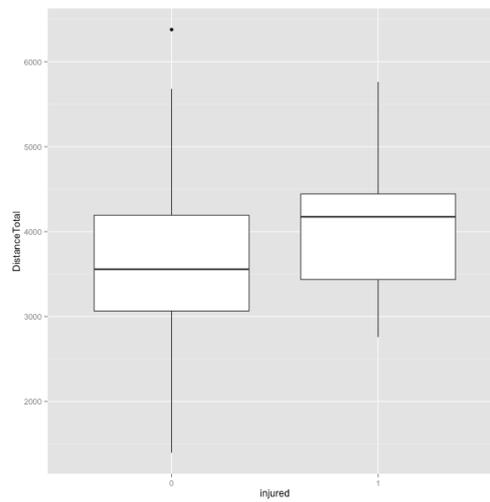

**Figure 7.3. Boxplot of "total distance" vs injury.**

**High speed running**

"High speed running" measures the distance covered when a player is moving in Zones 5 or 6. Figure 7.4 shows the histogram for the variable. The distribution is skewed on the right, with the bulk of the training sessions containing only a moderate amount of high speed running sessions.

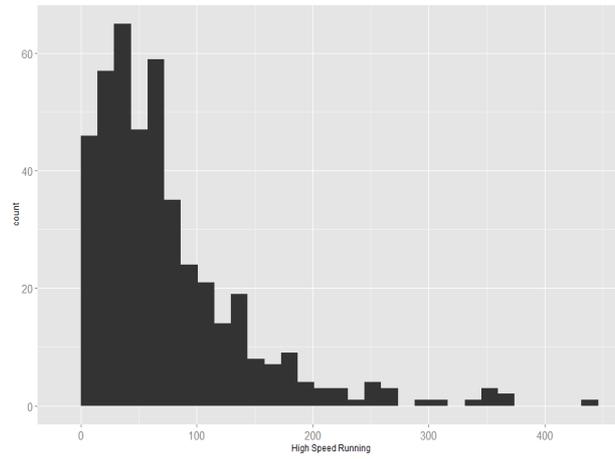

**Figure 7.4. Histogram for "high speed running".**

Figure 7.5 shows a boxplot of "high speed running" versus injury. There are not great differences between the two classes.

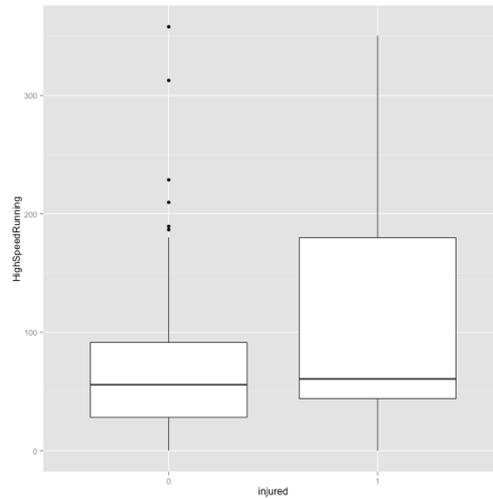

**Figure 7.5. Boxplot of "high-speed running" versus injury.**

**Speed intensity**

"Speed intensity" measures the amount of total physical exertion for the session based on speed and is a variable produced by STATSports Viper. This was calculated by applying an exponential weighting function to the speed data. The data is split in 0.1 intervals and the average speed for each interval is recorded. The variable "speed intensity" is produced as a weighted sum over all datapoints.

Figure 7.6 shows a histogram for "speed intensity". The distribution is roughly normal with a tail on the right. Speed intensity is broken down in 6 different zones, and the histogram has been created by aggregating over all zones.

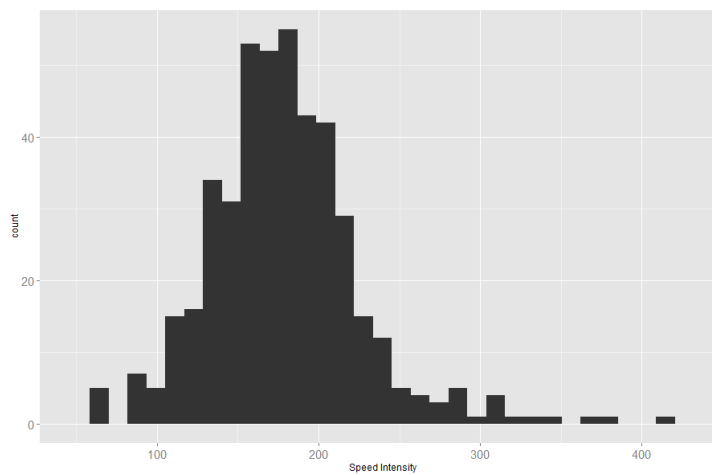

**Figure 7.6. Histogram of "speed intensity".**

Figure 7.7 shows a boxplot of "speed intensity" vs injury where it can be easily seen that the mean of the two categories is different. Higher speed intensities seem to be related to a higher number of injuries.

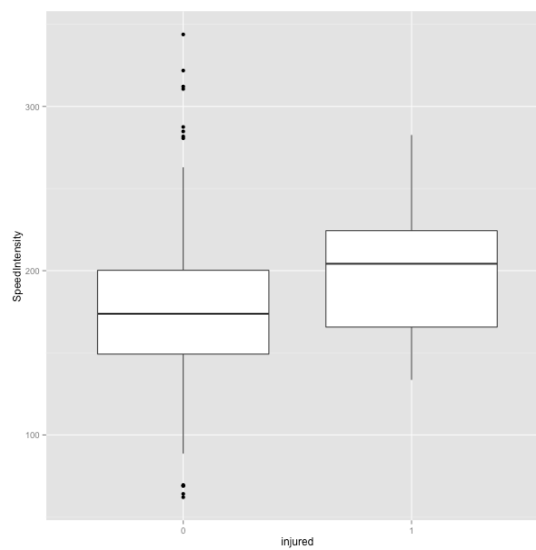

**Figure 7.7. Boxplot of "speed intensity" vs injury.**

**Accelerations and decelerations**

A movement is recorded as an acceleration or deceleration only when the measurement exceeds 0.5m/s$^2$ and when its duration is longer than ½ second. Figure 7.8 shows histograms for accelerations and decelerations. The two plots are very similar, but not identical, since there is not a one-to-one correspondence between them. Some accelerations can be accompanied by very slow decelerations, which do not count towards the "decelerations" variable, and some fast decelerations might take place after low speed running or walking. Accelerations and decelerations are broken down in zones. The histograms have been created by aggregating the points across all zones.

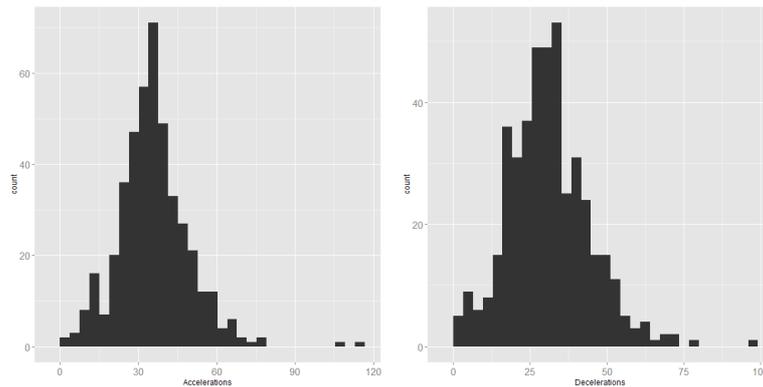

**Figure 7.8. Histograms of "accelerations" (left) and "decelerations" (right).**

Figure 7.9 shows the boxplots of "accelerations" and "decelerations" against injury. There do not seem to be great differences between the injured and non-injured populations.

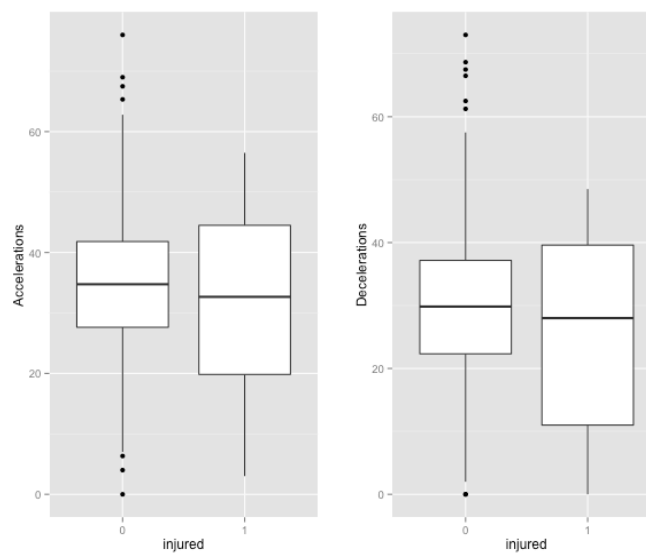

**Figure 7.9. Boxplot of "accelerations" (left) and "decelerations" (right) vs injury.**

**Metabolic power variables**

Metabolic power measures the rate at which energy is expended. The unit of measurement is Watts per kilogram W/Kg. Viper produces three main metabolic power variables: "equivalent metabolic distance" (EMD), "high metabolic power distance" (HML distance) and "average metabolic power" (AMP) and two secondary "metabolic distance zonal" and "metabolic time zonal".

*Equivalent Metabolic Distance (EMD)*

"EMD" is a combination of speed and acceleration/deceleration data. Viper first calculates W/Kg produced from accelerations and decelerations throughout the session. Then it converts this to the number of meters that the player would have to cover at constant speed to expand the same amount of energy. This number is added to the actual distance covered to produce the EMD value. The variable "metabolic distance zonal" is a record of the total distance used in the calculation of EMD, while "metabolic time zonal" is a record of the total time in which this distance was covered.

Figure 7.10 shows a histogram for "EMD". There is some skewness on the left, and some outliers on the right, and the tails drop off quite rapidly.

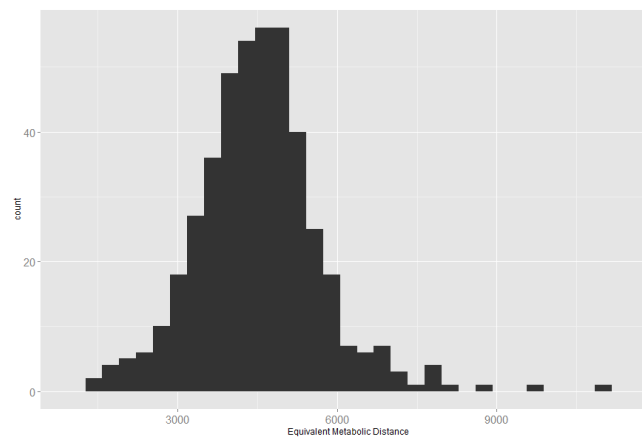

**Figure 7.10. Histogram of "EMD".**

Figure 7.11 shows a boxplot of "EMD" vs injury. There are not any great differences between the two groups.

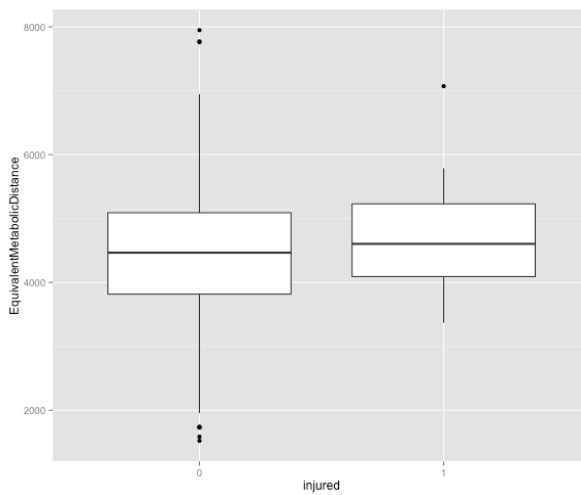

**Figure 7.11. Boxplot of "EMD" vs injury.**

*High Metabolic Power Distance (HML distance)*

"HML distance" uses the same methodology as "EMD" but uses the "distance covered" when metabolic power is above 25W/Kg which equates to "high speed running". The equivalent distance is calculated and then added to the original "high speed running" value to produce "HML distance". Figure 7.12 shows a histogram of "HML distance". The values seem to be relatively centered around 600, with some symmetry around the center.

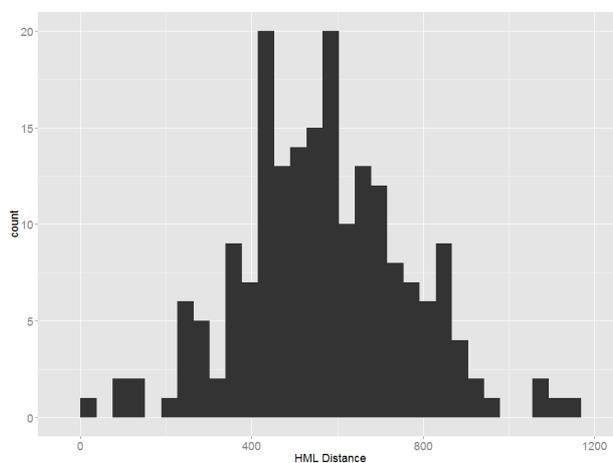

**Figure 7.12. Histogram of "HML" distance.**

Figure 7.13 shows a boxplot of "HML distance" vs injury. No big differences are observed between the two groups.

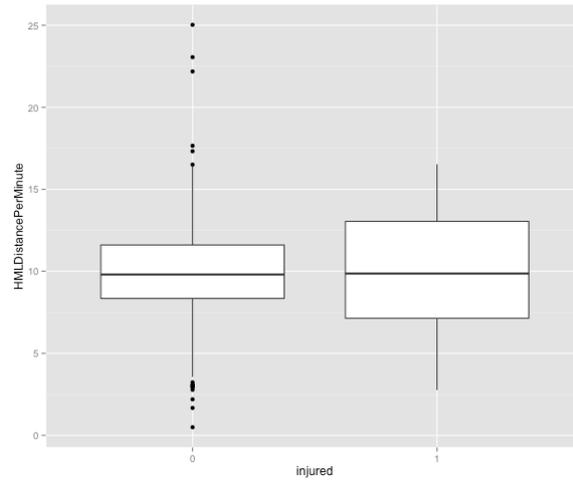

**Figure 7.13. Boxplot of "HML distance" vs injury.**

*Average Metabolic Power (AMP)*

"Average metabolic power" is the average W/Kg for the whole session. Figure 7.14 shows the distribution for the average metabolic power. It can be seen that the distribution is approximately symmetrical with a few outliers on both sides.

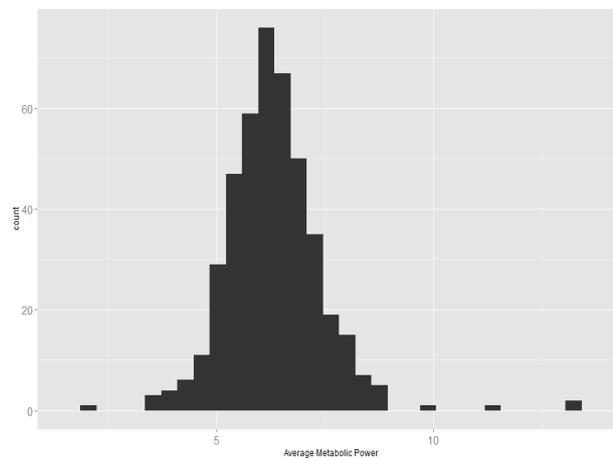

**Figure 7.14. Histogram of "AMP".**

Figure 7.15 shows a boxplot of "AMP" vs injury. There are no big differences between the two groups.

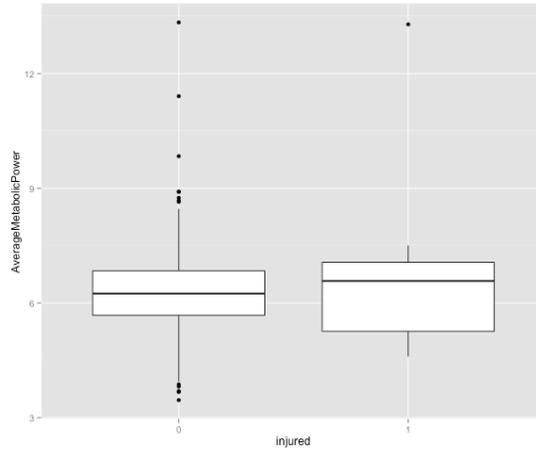

**Figure 7.15. Boxplot of "AMP" vs injury.**

**Dynamic Stress Load**

"Dynamic stress load" is a variable created by the Viper software and functions as a measure of fatigue. It uses a combination of accelerometer and GPS data along with impacts and a weighting function for the impacts based on the idea that an impact of 4g is more than twice as hard on the body as an impact of 2g.

Figure 7.16 shows the distribution for dynamic stress load. The has a tail to the right. "Dynamic stress load" is measured in zones so the histogram has been created by aggregating through all zones.

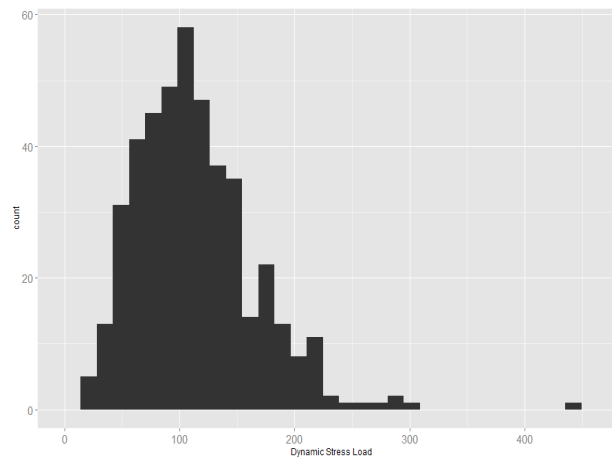

**Figure 7.16. Histogram of "dynamic stress load".**

Figure 7.17 shows a boxplot of "dynamic stress load" vs injury. The boxplots for the two groups show that the median for the injured group is higher.

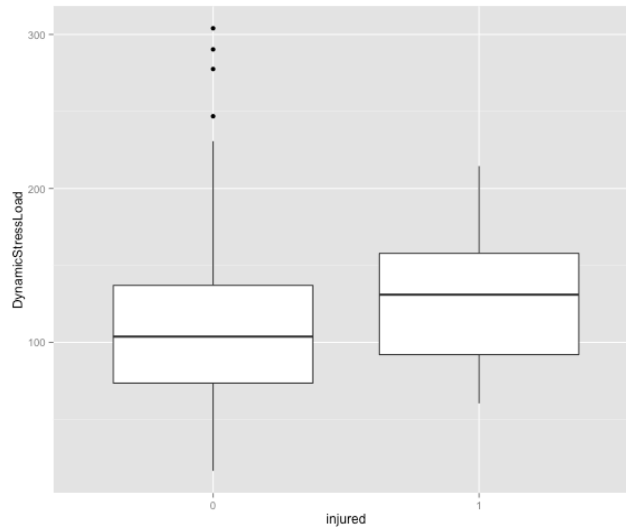

Figure 7.17. Boxplot of "dynamic stress load" vs injury.

**Impacts**

The purpose of the variable "impacts" is to detect every single movement, whether it is jumping or taking a single step. An impact is measured whenever the accelerometer records a value of magnitude above 2g in a 0.1 second period.

The are three different types of impact variables. First, impacts are broken down in zones. Secondly, impacts are broken down in left and right impacts and these are subcategorized in lateral, vertical and posterior/anterior, depending on which side and which axis of the body they take place. Finally, the magnitude of the impact in g is being measured for both sides.

Figure 7.18 shows a histogram of "impacts" created by aggregating across all zones. The distribution is roughly symmetrical with a few outliers on the right.

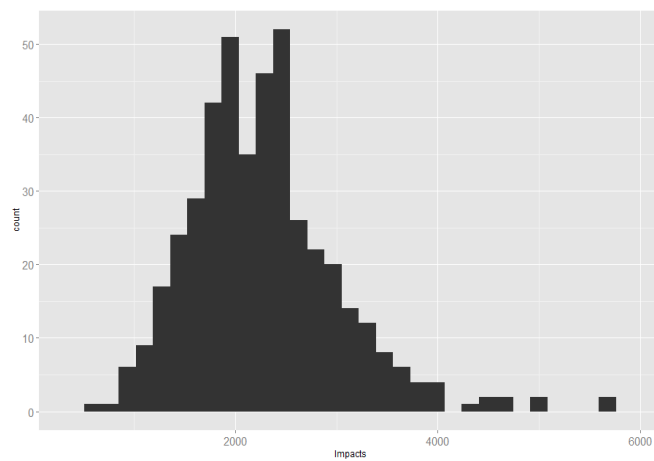

Figure 7.18. Histogram of "impacts".

Figure 7.19 shows a boxplot of impacts vs injury. The mean number of impacts is greater for the injured athletes, but there are data points with a very large number of "impacts" where the athlete was not injured, so it is difficult to identify whether this variable is important or not.

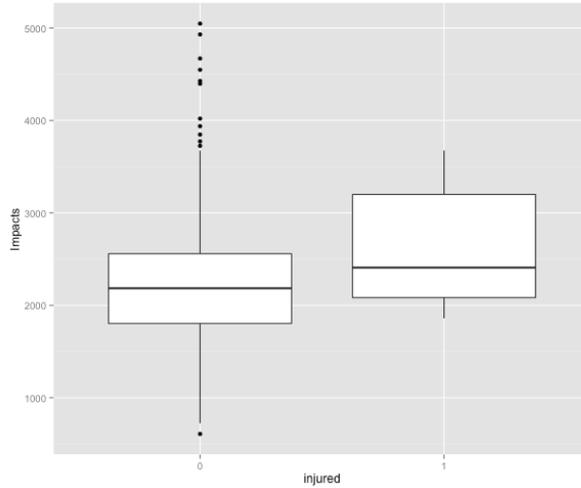

**Figure 7.19. Boxplot of "impacts" vs injury.**

Figure 7.20 below shows two histograms for the magnitude of left impacts and right side impacts respectively. We see that the two histograms follow a similar pattern. There is a short tail on the right and a single peak. The main difference is that the histogram for the right magnitudes has a few outliers. These outliers could correspond to events such as collissions which could possibly lead to injury.

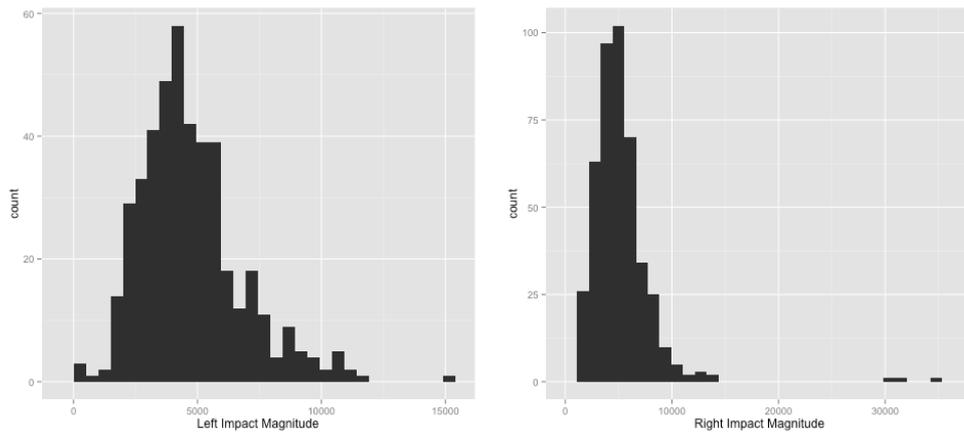

**Figure 7.20. Histograms for "left impact magnitude" and "right impact magnitude".**

In Figure 7.21 we see two boxplots of right and left impacts versus injury. It is clear that injuries are accompanied by larger magnitudes on average, even though there are some points where the magnitude was significantly high but no injury took place.

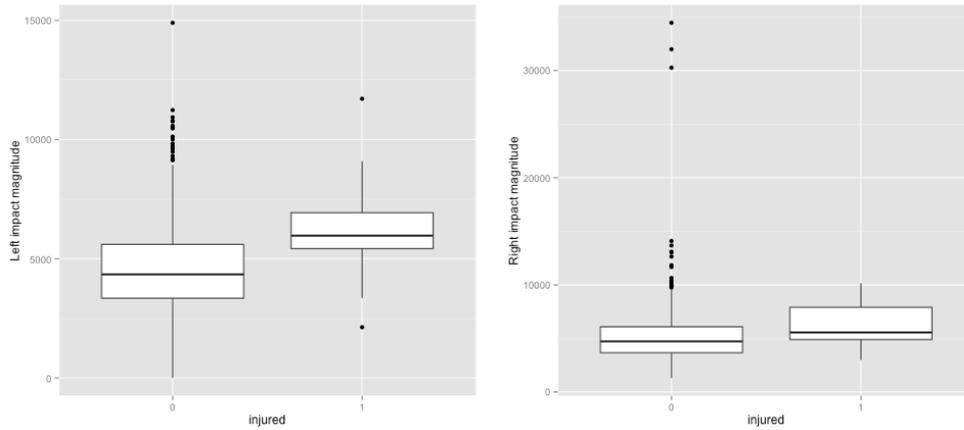

**Figure 7.21. Boxplots of right and left impact magnitude versus injury**

**Total Loading**

This variable uses accelerometer data in order to measure the forces applied to the player over the whole session. It is defined as the sum of the accelerometer values (measured in g) taken in three directions, sampled one hundred times per second. The total is divided by 1000 to scale down the values. Figure 7.22 shows the histogram for "total loading". The distribution is roughly symmetrical with a few outliers towards the right.

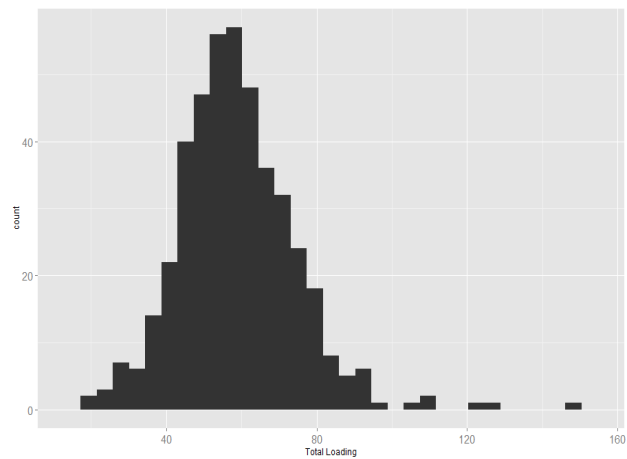

**Figure 7.22. Histogram of "total loading".**

Figure 7.23 shows a boxplot of "total loading" vs injury. The boxplots do not look much different to each other.

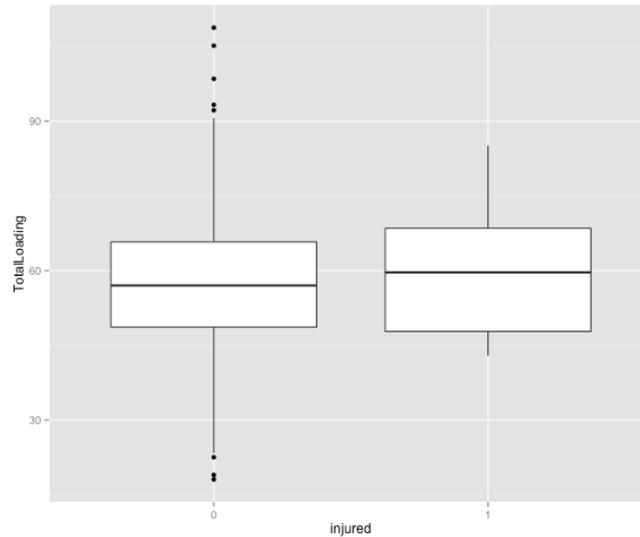

**Figure 7.23. Boxplot of "total loading" vs injury.**

**Lower Speed Loading**

This variable is equivalent to "total loading" but captures only activity that is below a certain threshold and static activity. For this data, the threshold had been set as to any activity in zones 1 or 2. Figure 7.24 shows a histogram for the distribution of "low speed loading". There is a left tail and outliers on the right, but the distribution is roughly symmetric.

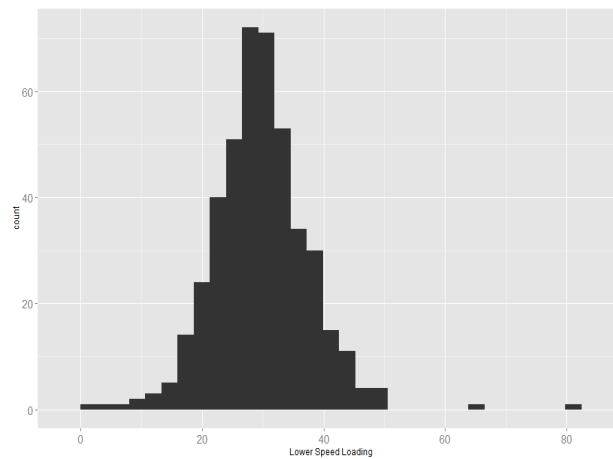

**Figure 7.24. Histogram of "lower speed loading".**

Figure 7.25 shows a boxplot of "lower speed loading" vs injury where no big differences are observed between groups.

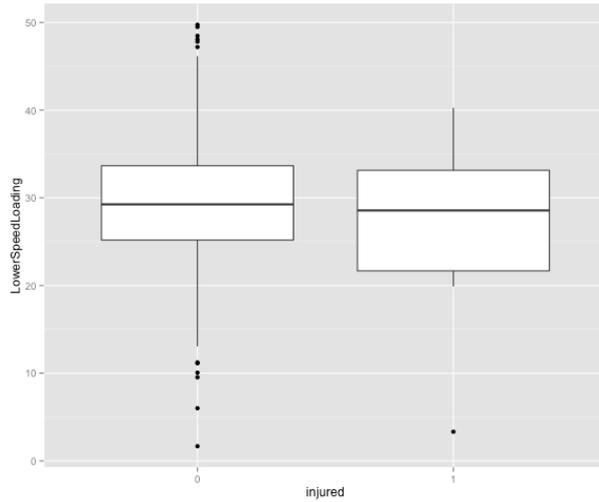

**Figure 7.25. Boxplot of "lower speed loading" vs injury.**

**Sprints**

Sprints are based on actions above a certain speed threshold that last for at least 1 second. The sprint stops when the speed falls below 80% of the sprint threshold. Because of that, not all accelerations or decelerations count as sprints. Figure 7.26 shows the histogram of "sprints". There is a long right tail, indicating that a large number of sprints has taken place only in few sessions.

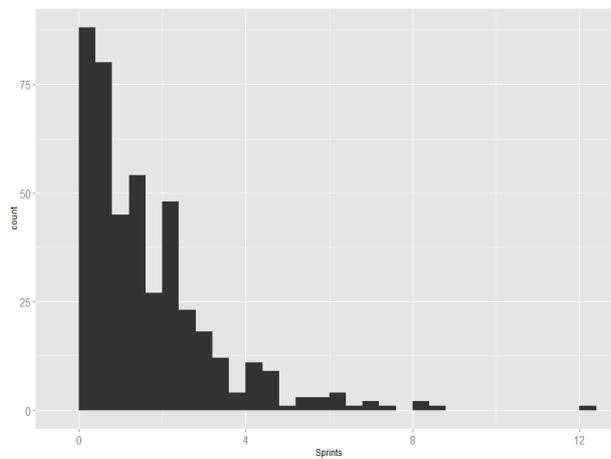

**Figure 7.26. Histogram of "sprints".**

Figure 7.27 shows a boxplot of "sprints" vs injury. Injuries seem to be associated with a higher number of sprints during training sessions.

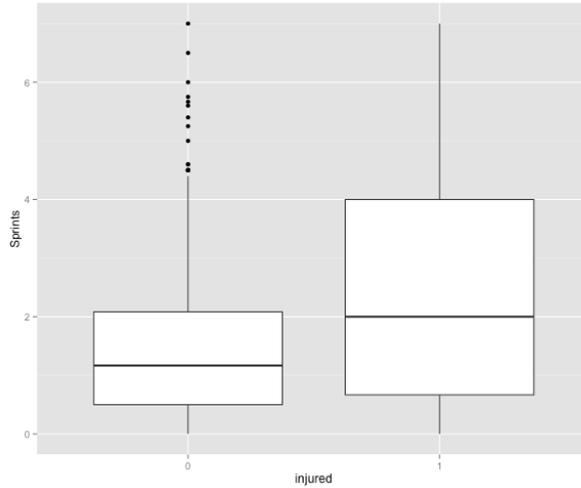

**Figure 7.27. Boxplot of "sprints" vs injury.**

**Energy Expenditure**

This metric gives the total energy associated with running only including accelerating and decelerating activity measured in kcal. It is scaled by the weight of the player measured in kilograms. Figure 7.28 shows the histogram of "energy expenditure". The values are very concentrated around the mean. Once again, there are outliers on the right hand side.

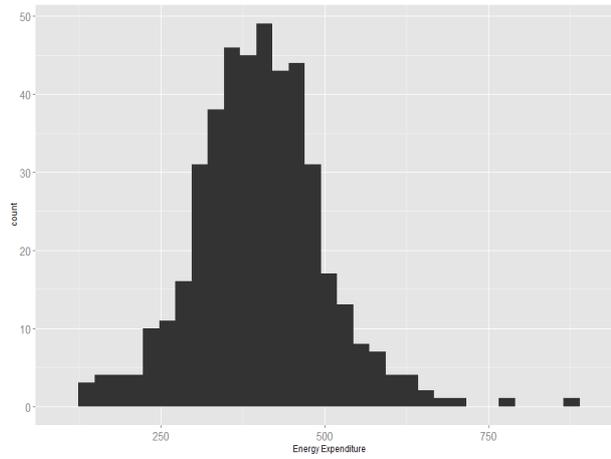

**Figure 7.28. Histogram of "energy expenditure".**

Figure 7.29 shows a boxplot of "energy expenditure" vs injury. The boxplots look similar for both groups.

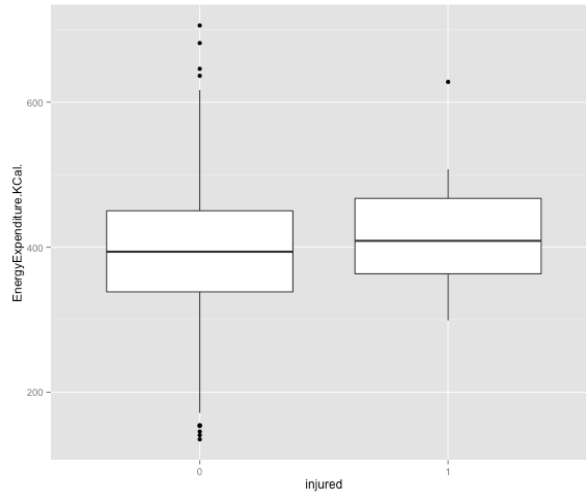

**Figure 7.29. Boxplot of "energy expenditure" vs injury.**

**Step Balance**

"Step balance" measures the impact that take place on each side of the body. A value of 0 indicates a perfectly balanced session. Positive values indicate more pressure on the dominant foot, while negative values indicate pressure on the non-dominant foot.

Figure 7.30 shows the histogram of the variable. The values are very concentrated around 0, with long tails that fall of rapidly. It is to be expected that in the majority of the training sessions the actions are relatively balanced between the two sides, with the dominant foot maybe getting a bit more stress.

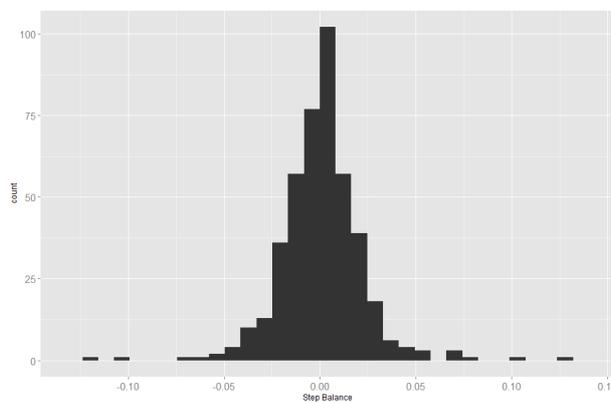

**Figure 7.30. Histogram of "step balance".**

Figure 7.31 shows a boxplot of "step balance" vs injury. "Step balance" for the non-injured cases is characterized by a larger number of outliers, but other than that there does not seem to be anything to suggest a significant difference between the two boxplots.

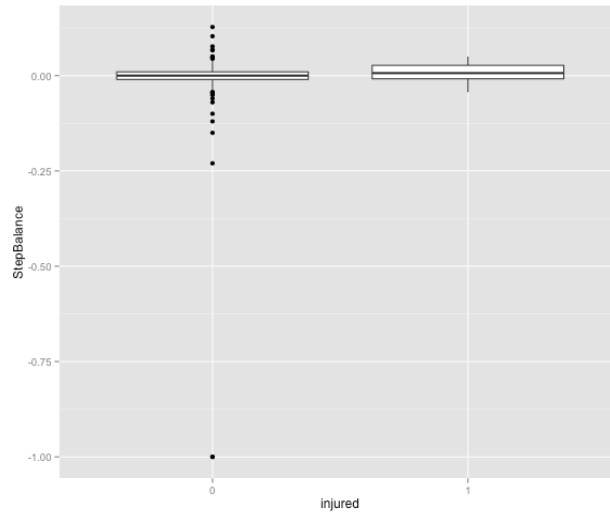

**Figure 7.31. Boxplot of "step balace" vs injury.**

### 7.2.2 Defining the problem

There are different ways that the dataset can be used in order to study the problem. Two different approaches were chosen.

**Approach A: Weekly prediction for injured players**

First, all the players that had not been injured were removed. Secondly, out of the players that had been injured, only those that had intrinsic injuries were kept. An intrinsic injury was defined as an injury not-related to collision or contact of any sort, the same definition used in Chapter 6. These injuries might be possible to be predicted by GPS variables, whereas collisions or contact-related injuries cannot. This left a total of 11 players in the dataset.

The dataset was then grouped per player and week, so that each variable corresponded to a player $i$ and a week $j$. A binary "injury" variable indicated whether the player had been injured in a week or not.

The aggregation over weeks was chosen for two reasons. First, a week is a natural split for scheduling in football. Every week is characterized by at least one match, and the training schedule is designed on a weekly basis. Secondly, other natural splits would be daily, or summing up weeks (e.g. the last two weeks). The problem with daily splits (that is, each row in the dataset represents a day of training) are twofold. First, the number of rows would become too large, but the number of injuries would still stay small. Secondly, it is unlikely that the GPS data from a single day are indicative of fatigue or overtraining. It is more likely that an average over at least a few days is more informative.

The mean was used as an aggregation summary to aggregate the instances over the week. For each player that had been injured, there were at least 2 training sessions, before the injury took place.

The final dataset consisted of 206 rows and 68 GPS variables, plus the response variable and the duration of the training session. Out of the 206 rows, there were 11 cases of injury (corresponding to the weeks that the player had been injured).

While it can be disputed that the choice to remove all athletes that were not injured can lead to a loss of information, the problem as it stands contains highly imbalanced classes. It was decided to first study an "easier" version of the problem, before using all the athletes in approach B.

Therefore, the problem, as it is posed, is whether it is possible to predict a week during which an injury will occur based on GPS data gathered from training, focused only on injuries that could have potentially come as a result of overtraining, fatigue or any other factor that might be identifiable during training or caused by it.

**Approach B: Weekly prediction for all athletes**

Approach A compares the weekly averages of the injured players only. Obviously, while there is some motivation for doing that, as it was explained previously, the case still remains that there might be loss of information.

In approach B, the process is identical to approach A, but all the players are included, even the ones that had not been injured. This led to a dataset of 426 rows and 70 variables in total (the same as in approach A), including the class variable.

## 7.3 Models

The methods used were: supervised PCA, random forest, ridge logistic regression, SVMs, Gaussian processes, naïve Bayes, neural networks and k-NN.

Supervised PCA is a particularly good technique for reducing the dimensionality of the problem with respect to a given response variable. Random forest and ridge logistic regression were chosen because of their ability to handle a high number of noisy features. Ridge logistic regression was chosen over L1 regularization, because the latter can be problematic when many of the variables are highly correlated as stated in section 4.1.8. Naïve Bayes and k-NN are simple algorithms and are being used as benchmarks for the more complicated methods.

The evaluation of the models was conducted through 10 rounds of 10-fold cross-validation. A parameter grid was used in order to find the optimal setting. Table 7.2 shows the parameter grid used for each model.

**Table 7.2. Parameter grid for each model.**

| Model | Parameters |
| --- | --- |
| Supervised PCA | Alpha $\in \{0, 0.01, \ldots, 0.5\}$, num_components $\in \{2:50\}$ |
| Ridge logistic regression | Ridge $\in \{10^{-15}, 10^{-14}, \ldots, 0.1\}$ |
| Random forest | Number of trees $\in \{50, 100, 150, \ldots, 5000\}$ |
| Naïve Bayes | - |
| SVM, Gaussian process | RBF kernel with degree $\in \{0.0001, 0.001, \ldots, 5\}$<br>Polynomial kernel with degree $\in \{1, 2, 3, 4\}$<br>$C \in \{0, 0.5, 1, \ldots, 100\}$<br>Scale $\in \{0.1, 0.5, 1, \ldots, 10\}$ |
| Neural networks | Epochs $\in \{500, 1000, \ldots, 3000\}$, Hidden Neurons $\in \{5, 10, \ldots, 50\}$, Decay $\in \{10^{-10}, 10^{-9}, \ldots, 1\}$ |
| k-NN | $k \in \{1, 2, \ldots, 20\}$ |

## 7.4 Results

### 7.4.1 Approach A

Table 7.3 shows the parameters for the best models. The best performance is achieved by supervised PCA followed by ridge logistic regression. K-NN fails completely in this task, since both the kappa and the precision are zero.

Table 7.3. Best model parameters for approach A. Accuracy and kappa are shown as mean+/-standard deviation.

| Model | Parameters | Accuracy | Kappa | Precision | Recall |
|---|---|---|---|---|---|
| Supervised PCA | Alpha=0.01, components=41 | 88.8%+/-1.8% | 0.21+/-0.04 | 0.55+/-0.09 | 0.33+/-0.07 |
| Ridge Logistic Regression | Lambda=$10^{-4}$ | 89.83%+/-5.73% | 0.16+/-0.33 | 0.105+/-0.02 | 0.182+/-0.06 |
| Random Forest | Trees=5000 | 94.17%+/-0% | 0.06+/0.04 | 0.06+/-0.01 | 0.03+/-0.01 |
| Naïve Bayes | - | 81.3%+/-1.7% | 0.04+/-0.08 | 0.107+/-0.03 | 0.273+/-0.08 |
| SVM | Polynomial kernel degree=2, C=1, scale=0.5 | 93.1%+/-3.37 | 0.12+/-0.9 | 0.15+/-0.02 | 0.18+/-0.04 |
| Gaussian Process | Polynomial kernel degree=2, noise=1.5, scale=1 | 94.67%+/-1.41% | 0.12+/-0.1 | 0.33+/-0.02 | 0.09+/-0.02 |
| Neural Networks | Neurons=20, epochs=1500, decay=0.0001 | 93.33%+/-3.6% | 0.03+/-0.02 | 0.07+/-0.05 | 0.20+/-0.13 |
| k-NN | k=1 | 94.67%+/-0% | 0.0 | 0.0 | 1.0 |

Naïve Bayes is a simple technique, making a very strong conditional independence assumption, so the fact that it fails in this case is not noteworthy. K-NN in this case performs very poorly. It is possible that this is due to the curse of dimensionality, since k-NN is particularly sensitive to it as it was discussed in section 4.1.7.

It is also interesting that random forest fails completely. This is probably due to the fact that all variables are numerical. Random forest has to find a threshold to split each numerical attribute, since it cannot deal with numerical features directly.

Neural networks do not perform well. SVM and ridge logistic regression seem to be working to some extent, however their performance is still worse than that of supervised PCA. Supervised PCA has a mean kappa statistic of 0.21 with a standard deviation of 0.04, whereas ridge regression (the second best) has a mean kappa statistic of 0.16. This makes sense, since PCA is particularly suited in cases where many of the features are correlated with each other.

Figure 7.32 shows a levelplot of the parameter alpha and the number of components for supervised PCA, with the color representing the kappa statistic. What can be seen in the figure is that the method works better when smaller values of alpha and more components are used. It seems that the best values are achieved when alpha is less than 0.05 and more than 20 components are used.

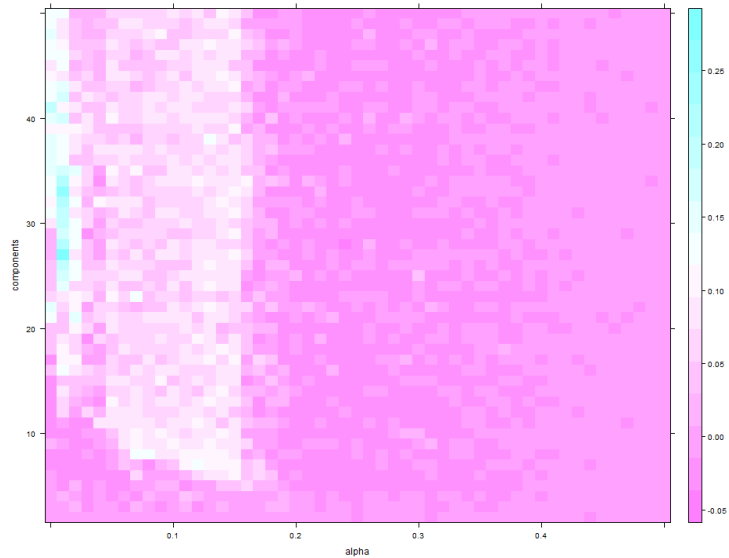

**Figure 7.32. Levelplot of alpha vs components for approach A. The color represents the kappa statistic.**

### 7.4.2 Approach B

Table 7.4 shows the results for approach B. The performance of the classifiers is worse than the performance in approach A. This is probably due to the fact that the two classes have become more imbalanced. What can be seen is that once again supervised PCA is the best method, as measured by the kappa statistic, followed by Gaussian processes and ridge logistic regression.

**Table 7.4. Best model parameters for approach B. Accuracy and kappa are shown as mean+/-standard deviation.**

| Model | Parameters | Accuracy | Kappa | Precision | Recall |
|---|---|---|---|---|---|
| Supervised PCA | Alpha=0.09, components=40 | 97.07%+/-3.31% | 0.14+/-0.09 | 0.19+/-0.06 | 0.20+/-0.8 |
| Ridge Logistic Regression | Lambda=$10^{-6}$ | 94.27%+/-2.72% | 0.09+/-0.04 | 0.03+/-0.02 | 0.1+/-0.01 |
| Random Forest | Trees=5000 | 95.42%+/1.73% | 0.02+/-0.05 | 0.03+/-0.05 | 0.375+/-0.12 |
| Naïve Bayes | - | 74.91%+/-7.40% | 0.06+/-0.14 | 0.057+/-0.03 | 0.545+/-0.05 |
| SVM | Polynomial kernel degree=2, C=10, scale=1 | 95.42%+/-2.21% | 0.04+/-0.01 | 0.1+/-0.02 | 0.09+/-0.04 |
| Gaussian Process | Polynomial kernel degree=3, noise=25, scale=1 | 94.51%+/-3.48% | 0.12+/-0.04 | 0.14+/-0.06 | 0.18+/-0.06 |
| Neural Network | Neurons=30, epochs=1500, decay=0.001 | 95.42%+/-5.67% | 0.04+/-0.03 | 0.05+/-0.04 | 0.1+/-0.7 |
| k-NN | k=1 | 97.42%+/-0% | 0.0 | 0.0 | 1.0 |

Figure 7.33 shows the levelplot for the parameter search for supervised PCA for approach B. We can see that the best solutions exist in the upper left corner for values of alpha less than 0.2 and 30 or more components.

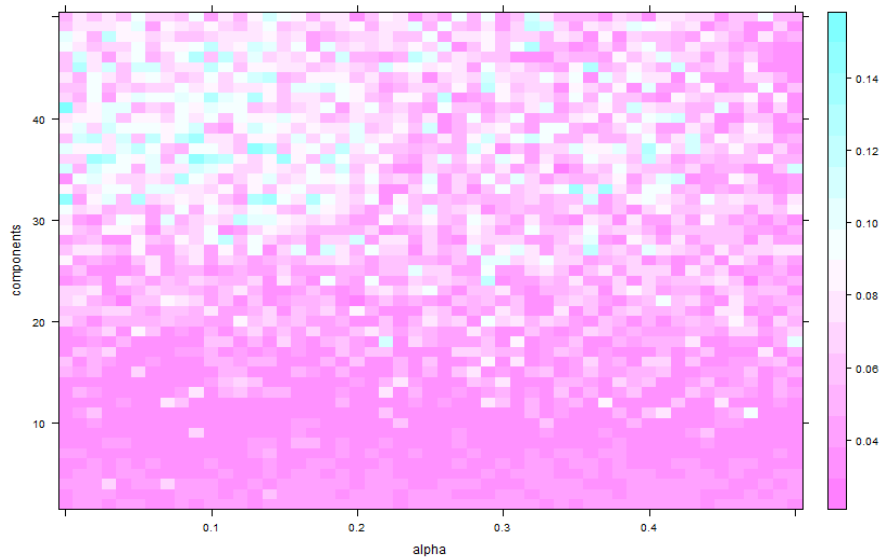

**Figure 7.33. Levelplot of alpha vs components for approach B. The color represents the kappa statistic.**

## 7.5 Further comments on PCA

The principal components can be used in order to get a better understanding of the dataset and how the variables relate to injury. This section will analyze the components of the best supervised PCA models for both approaches.

**Approach A**

Figure 7.34 shows the scree plot for the components. It seems that the first 4 components explain the greatest percentage of the variance, with the scree plot following rapidly at component 5 onwards. Therefore, it was decided that 4 components should be adequate for the analysis.

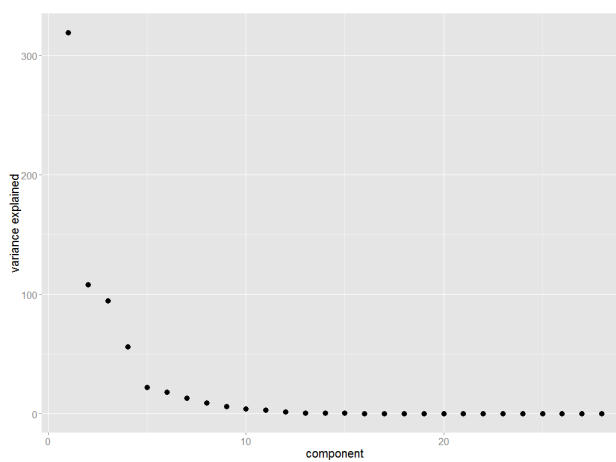

**Figure 7.34. Variance explained for each component in approach A.**

Table 7.5 shows the first four components of this model. The table includes only variables that have a high correlation with at least one of the 4 components. The full table of correlations is displayed in the Appendix (Table 10.12). The correlations with the greatest absolute value for each component are displayed in red.

Table 7.5. First four principal components of model with alpha=0.01 for the dataset in approach A.

|  | PC1 | PC2 | PC3 | PC4 |
|---|---|---|---|---|
| **AccelerationsZone3** | -0.22842 | 0.47640 | 0.02004 | -0.18828 |
| **DecelerationsZone3** | -0.22305 | 0.43085 | 0.07311 | -0.07318 |
| **DecelerationsZone4** | -0.09469 | 0.14696 | -0.00709 | -0.00245 |
| **DistancePerMin** | 0.00317 | -0.1908 | 0.43666 | -0.19408 |
| **DistanceZone5** | -0.76398 | -0.48592 | 0.22479 | -0.11721 |
| **DynamicStressLoadZone2** | -0.18703 | 0.04578 | -0.39766 | 0.37565 |
| **DynamicStressLoadZone3** | -0.16017 | 0.06806 | -0.17699 | 0.40768 |
| **ImpactsZone4** | -0.17954 | 0.08479 | -0.16280 | 0.47654 |
| **SpeedIntensityZone3** | -0.02075 | -0.13704 | -0.59876 | -0.33671 |

Figure 7.35 shows a plot of the first the component loadings plotted in pairs. The plot on the left shows the first principal component plotted against the second and the plot on the right shows the third component against the fourth.

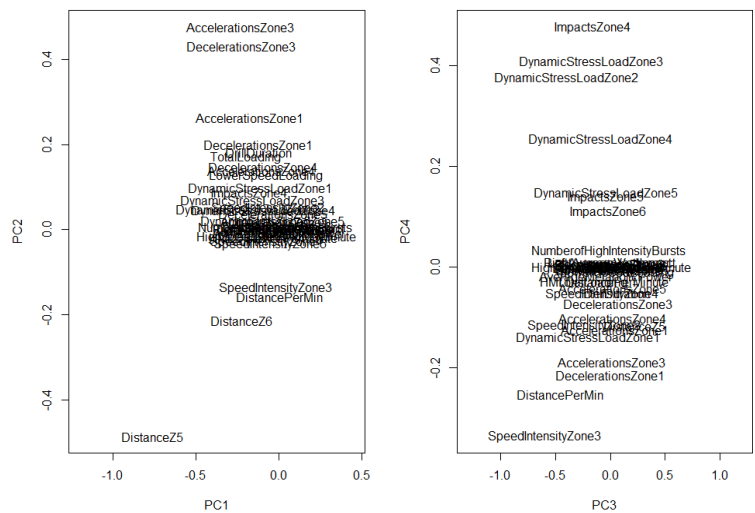

Figure 7.35. Biplot of the loadings of the first 4 principal components (approach A).

Based on the table and the graphs the following interpretation can be made. The first component is dominated by "Distance Zone 5" with which it is negatively correlated. A larger value in this component indicates that the player did not cover a large distance running in high speeds. Therefore, this component seems to be good proxy for "Absence of Running in High Speeds".

The second component has a large negative correlation with "Distance in Zone 5" and a large positive correlation with accelerations and decelerations in zone 3. A large value in this component seems to indicate the absence of running in higher speed zones and many sprints in zone 3. Therefore, this component seems to represent "Medium Speed Sprinting".

The third component is highly correlated with "Distance per Minute" and a negative correlation with "Dynamic Stress Load in Zone 3" and "Speed Intensity in Zone 3". A high value on this component seems to indicate that the player has a high average distance per minute and the absence of average level activity. Therefore, this component can be a good proxy for "High Average Speed".

The fourth component has a high correlation with variables that indicate physical stress in the middle speed zones (zones 2-4). Therefore, a good name for this component is "Medium Intensity Work".

Figure 7.36 and Figure 7.37 below shows the boxplots of the principal components versus injury. There are differences in the median and the range of the injured and non-injured groups for all components. Components 1 and 3 have higher medians for the injured athletes, whereas components 2 and 4 have lower.

Wilcoxon's signed rank test that the median of the differences between the injured and the non-injured groups is zero outputs a p-value for the null hypothesis of less than 0.01 for the first component and the third component, and less than 0.05 for the second and the fourth components. This strengthens the evidence that the components are associated with injury.

Based on the boxplot for components 1 and 3 it seems that a high distance per minute correlates with injury, but this does not hold when an athlete covers a long distance on a high speed. Perhaps the ability of an athlete to cover a large distance on high speeds is an indication that his body is not at risk of injury.

Components 2 and 4 measure activity the middle zones, in the form of accelerations/decelerations, impacts and the dynamic stress load and are negatively correlated with injury. This relationship could be due to medium stress activity having a preventative effect on injury or due to the fact that sessions of low intensity might not be stressful enough to cause injury.

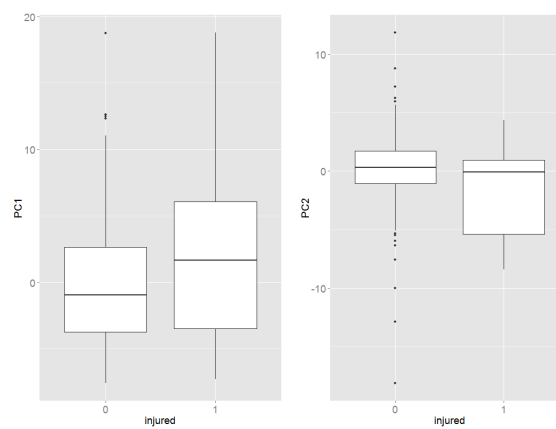

**Figure 7.36. Boxplots of the first two principal components versus the response variable.**

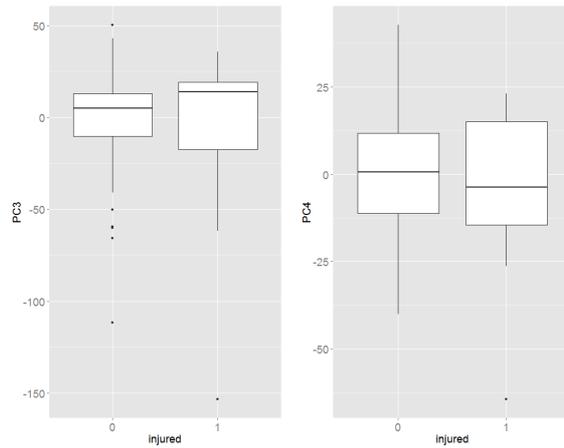

**Figure 7.37. Boxplots of principal components 3 and 4 versus the response variable.**

**Approach B**

Figure 7.38 shows the variance per component for approach B. The first two components explain the greatest percentage of the variance.

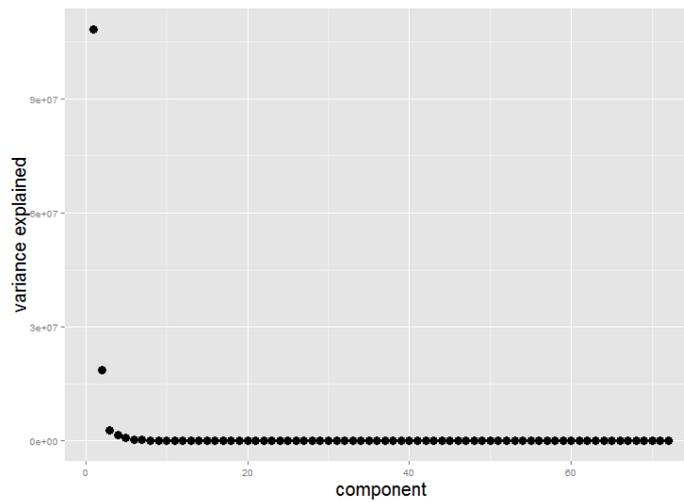

**Figure 7.38. Variance explained for each component for approach B.**

Table 7.6 shows the correlations for the first two components of this model. Like in the previous part, the full table of correlations is displayed in the Appendix (Table 10.13) and the correlations with the greatest absolute value for each component are displayed in red.

Table 7.6. First two principal components of model with alpha=0.01 for the dataset in approach B.

|  | PC1 | PC2 |
|---|---|---|
| **DynamicStressLoadZone6** | -0.7336374014 | -0.141046 |
| **ImpactsZone6** | -0.2726364018 | -0.083711 |
| **RightLateralImpact** | -0.1904632684 | -0.00039 |
| **RightMagnitudeImpact** | -0.1737929754 | 0.3375606 |
| **MetabolicDistanceZonal** | 0.0075269287 | 0.2539652 |
| **RightVerticalImpact** | 0.0037938173 | 0.3199798 |
| **LeftVerticalImpact** | 0.0442883442 | 0.3203094 |
| **LeftMagnitudeImpact** | 0.0477825359 | 0.3527447 |

Figure 7.39 below shows a biplot of the first 2 principal components.1

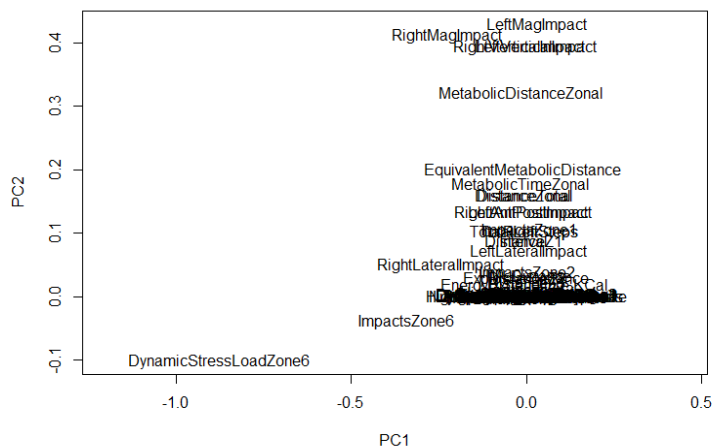

Figure 7.39. Biplot of the loadings of the first 2 principal components (approach B).

The first component correlates negatively with "Dynamic Stress Load in Zone 6" and "Impacts in Zone 6". It seems to represent the absence of high speeds in a session. Therefore, a good name for this component is "Absence of High Speed Stress".

The second component has a high correlation with the magnitude and the number of right and left vertical impacts. Therefore, this component is a good proxy for "Impacts".

Figure 7.40 shows the boxplots for the two components versus the response variable. The differences between the injured and the non-injured groups are prominent. The first component has a lower median for injured athletes, whereas the second component clearly has a larger median for injured athletes. The interpretation of the components in this case is that stress in the higher speed zones and a large number of impacts correlate with injury.

Wilcoxon's signed rank test that the median of the differences between the injured and the non-injured groups is zero outputs a p-value for the null hypothesis of less than 0.01 for both components. The test provides further evidence of the association between the extracted components and injury incidence.

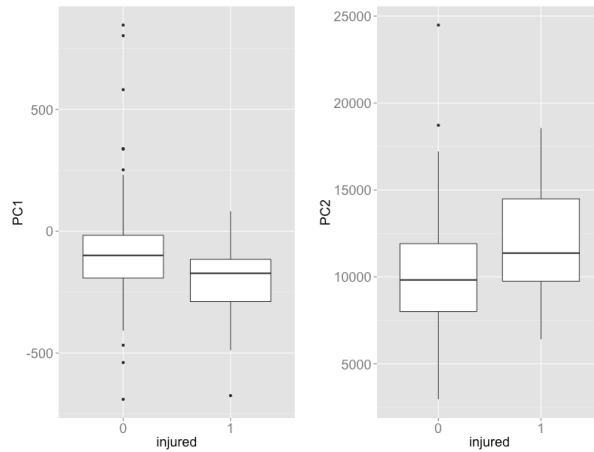

**Figure 7.40. Boxplots of the two principal components versus the response variable.**

## 7.6 Conclusion

This investigation studied the problem of predicting intrinsic injuries from GPS data gathered from training. The results demonstrated that GPS data gathered from training contain information which can be used for predicting injuries.

First, an issue with this data is that the variables are highly correlated with each other. Supervised PCA was by far the best method out of all those that were tried, probably due to the fact that it handles correlated attributes very well.

Secondly, it is possible to reduce the large number of variables to a few meaningful factors that correlate with injury. The factors provide a more parsimonious explanation of the data and can be used more easily by a practitioner.

The study also suffers from some limitations. First, the total number of fatigue-related injuries in this dataset was small, relative to the number of non-injured instances. Many of the injuries take place during matches, or are not directly related to fatigue. For these cases, GPS data are not very informative. This is the case for traumatic injuries for example, which can come as a result of a single event (e.g. a very fast sprint), and prior sensitivity, which only a physical examination might be able to uncover.

Another issue with this task is the fact that GPS data are not collected from matches, which can, nevertheless, have a huge impact on the physical condition of an athlete. This might be posing a ceiling to the performance that a predictive model can reach for this problem.

# 8 Conclusion and future work

*This thesis contributed to the field of sports analytics by studying applications of predictive modelling in football injuries. The contributions of this thesis are summarized below, along with future research directions.*

## 8.1 Contributions

The field of sports analytics has started to attract lots of attention in recent years. The increase in the volume of data available in addition to the increase in computing power and the development of powerful algorithms allows different approaches for quantitative research in sports, such as predictive modelling. This comes in contrast to the traditional way of conducting studies in sports which uses control groups and significance testing. Predictive modelling can be a convenient and powerful tool for the modern sports professional, since it allows the professional to directly predict outcomes, instead of making educated guesses.

The general contribution of this thesis to sports analytics is the examination of a particular area that is still a green field: the use of predictive modelling for football injuries. Injuries in football have a big impact both on the team, as a whole, but also on the player on many different levels. The performance of the team can suffer, but also an injury can have a huge impact on a player's career.

The thesis examined the effectiveness of various methods and algorithms for particular problems, while also outlining the importance of data collection and standardization of data collection and storage in football.

More specifically, the contributions with regards to each investigation are outlined below:

**Investigation 1: Predicting the recovery time after a football injury using the UEFA injury recordings**

This investigation performed an extensive study of the UEFA injury recordings of two professional football clubs. The UEFA injury recordings is one of the few data collection standards followed in football. The fact that this research was based on that standard makes it replicable and helps improve the dissemination of research in the field of sports analytics.

This investigation made two core contributions.

First, it studied the potential of the UEFA injury recordings as an input for predictive models that predict the recovery period after an injury. Secondly, it studied the importance of individual variables in the UEFA injury recordings with regards to that goal.

Regarding the first contribution, two points became clear. First, not all variables are equally important for predicting injuries. In fact, the removal of some variables can actually improve the performance of predictive models. Secondly, there are many issues as to how the variables are being recorded, since there can be differences as to how clubs record variables, while there can also be errors as to how they are recorded.

With regards to the second contribution, the study compared different models for predicting the recovery time after an injury. It is clear that some models, such as SVMs and Gaussian processes, work better when the noisy variables are removed, whereas random forests provide reasonably good performance even with noisy data. An important contribution of the comparison between different models is that it sets a benchmark for future work.

**Investigation 2: Predicting injuries in professional football using exposure records**

Exposure is common statistic recorded in football and is related to injury incidence. This study made the following contributions.

First, it provides a model for predicting the first injury of the season. The study focused on the first injury of the season. The model predicts the number of days a player will go uninjured from the beginning of the season.

The second contribution is that the dynamic time warping kernel used in the Gaussian process model allows the calculation of the similarity of training and match schedules, as expressed by the training and match exposure, even if those have different lengths. Therefore, the sports professional can use this model in order to compare different training schedules and understand the risk of each one with regards to their potential to cause injury.

Since exposure records are being kept by all clubs, this study provides a way for clubs to utilize already existing exposure data in order to predict injuries and sets a benchmark for future models.

**Investigation 3: Predicting intrinsic injury incidence using in-training GPS measurements**

The third investigation contributed in the understanding of GPS data gathered from training sessions and their potential for predicting injuries. This study makes the following contributions.

First, it provides a general framework for predicting injuries by using weekly aggregation for the GPS data. Secondly, it compares two different approaches to this framework: using data only from injured players and all players (injured and non-injured). Thirdly, it compares various machine learning methods as a way to build a predictive model.

The main contribution is the adjustment of supervised principal components to handle classification and its application to the current problem. The model not only performs well enough to show that the task is indeed feasible, but it also allows the derivation of components which can summarize the large number of variables (69 in this research), but do it in a way so that the components extracted are correlated to injury. This feature is particularly useful to the sports professional who is seeking an understanding of the way the training schedule can lead to injury.

Given that GPS units have been introduced to football only recently, this study provides a useful tool to use the data being collected, while also setting a benchmark for future studies.

## 8.2 Future research directions

This thesis contributed to the field of sports analytics, by investigating the use of predictive modelling for injuries. Given that there has been little work done in that area this thesis sets the standards upon which future research can expand. There are several directions that can be explored in the future. This section here will outline some general suggestions.

First of all, a wider range of models can be explored for each one of the investigations that were outlined in this thesis. Even though this thesis examined the suitability of various models, there are more methods which could be used. Obviously, a full comparison of every classification or regression algorithm was beyond the scope of this thesis. The current choice of algorithms was motivated by a number of factors, such as the suitability of an algorithm for a particular problem and its success in previous problems with similar characteristics. Machine learning, however, is a very active area of research with new algorithms being continuously published. Therefore, there is still fertile ground for research on which methods are the best for a particular problem.

Secondly, future research could also indicate additional data that could be used by the current models. Data collection in football is problematic and difficult, something which was outlined in Chapter 3. What this thesis achieved in every investigation was to illustrate that even with limited data it is possible to build predictive models for football injuries. Nevertheless, the performance of all the models could greatly benefit from additional data.

More specifically, all studies would benefit by including data from more clubs. Also, there are many different sources relevant to each individual investigation that could be used. The UEFA injury recordings that were used for predicting the recovery time could be enhanced by medical information, such as blood exams or physical tests conducted on regular intervals. Exposure records, as well as the GPS data could be enhanced by the rate of perceived exertion or more detailed information provided by the training coach.

Thirdly, the studies require replication in order to understand the degree to which the results generalize. These particular studies were conducted in collaboration with two clubs from the English Premier League. However, the question remains as to whether these studies can generalize well to other situations, since there might be intricacies, ranging from playing style to nutrition, that might be special to the Premier League teams.

One question worth investigating is whether the particular models that were chosen the best models for the particular tasks across all leagues and divisions. For example, in the first investigation an SVM with a polynomial kernel of degree 2-4 was found to be a very good model for predicting the time of recovery after a football injury based on the UEFA injury recordings. Further investigation could check whether this result holds for other leagues in Europe or not. Additional research with more clubs and leagues from different countries is required in order to reach safe conclusions regarding these questions.

# 10 Appendix

Table 10.1. Recovery (mean and standard deviation) per mechanism of injury for the WW dataset in Chapter 5.

|  | mean | sd |
|---|---|---|
| Contact - Collision | 3.20 | 2.86 |
| Contact - Fall | 15.78 | 18.29 |
| Contact - Tackle | 7.53 | 11.09 |
| Contact - With Ball | 24.60 | 21.29 |
| Goalkeeping | 1.50 | 1.41 |
| Insidious Onset | 10.93 | 27.27 |
| Non-Contact – Jumping | 12.53 | 12.02 |
| Non-Contact – Kicking | 5.75 | 4.92 |
| Non-Contact – Running | 11.13 | 20.97 |
| Non-Contact – Stretching | 18.03 | 4.69 |
| Non-Contact – Twisting | 10.66 | 4.04 |
| Other | 1.00 | NA |

Table 10.2. Recovery (mean and standard deviation) per mechanism of injury for the THFC dataset in Chapter 5.

|  | mean | sd |
|---|---|---|
| Blocked | 15.79 | 33.39 |
| Collision | 3.40 | 4.77 |
| Falling/diving | 16.83 | 36.55 |
| Heading | 14.72 | 32.27 |
| Jumping/landing | 20.2 | 28.87 |
| Kicked by other player | 2.62 | 3.46 |
| N/A | 2.63 | 7.11 |
| Other acute mechanism | 15.45 | 36.58 |
| Overuse | 3.33 | 4.99 |
| Passing/crossing | 6.96 | 10.41 |
| Running/sprinting | 35.08 | 60.93 |
| Shooting | 4.71 | 4.99 |
| Stretching | 5.01 | 3.46 |
| Tackled by other player | 8.43 | 16.73 |
| Tackling other player | 14.46 | 32.67 |
| Twisting/turning | 29.21 | 51.28 |
| Unknown mechanism | 21.89 | 40.52 |

**Table 10.3. Recovery (mean and standard deviation) per mechanism of injury for the integrated dataset in Chapter 5.**

|  | mean | sd |
|---|---|---|
| **Blocked** | 15.79 | 33.39 |
| **Collision** | 3.31 | 3.71 |
| **Fall** | 17.88 | 30.42 |
| **Goalkeeping** | 1.5 | 1.41 |
| **Heading** | 14.72 | 32.27 |
| **Insidious Onset** | 10.93 | 27.27 |
| **Jumping** | 20.27 | 27.39 |
| **Kicked by other player** | 2.62 | 3.46 |
| **Kicking** | 5.75 | 4.92 |
| **N/A** | 2.63 | 7.11 |
| **Other** | 33.43 | 37.01 |
| **Overuse** | 3.33 | 4.99 |
| **Passing/crossing** | 6.00 | 10.41 |
| **Running/sprinting** | 69.01 | 102.84 |
| **Shooting** | 4.71 | 4.99 |
| **Stretching** | 10.77 | 7.82 |
| **Tackle** | 8.03 | 14.24 |
| **Tackling other player** | 14.46 | 32.67 |
| **Twisting/turning** | 22.71 | 47.07 |
| **Unknown mechanism** | 21.89 | 40.52 |
| **With Ball** | 24.6 | 21.29 |

**Table 10.4. Recovery (mean and standard deviation) per body part for the WW dataset in Chapter 5.**

|  | mean | sd |
|---|---|---|
| **Ankle** | 9.57 | 25.11 |
| **Foot / Toe** | 7.15 | 7.07 |
| **Forearm** | 10.00 | NA |
| **Hand / Finger / Thumb** | 32.5 | 24.74 |
| **Head / Face** | 45.00 | NA |
| **Hip / Groin** | 7.82 | 8.48 |
| **Knee** | 10.08 | 13.97 |
| **Lower Back / Sacrum / Pelvis** | 7.41 | 10.50 |
| **Lower Leg /Achilles Tendon** | 15.7 | 31.93 |
| **Sternum / Ribs / Upper Back** | 4.45 | 4.94 |
| **Thigh** | 9.07 | 12.36 |

**Table 10.5. Recovery (mean and standard deviation) per body part for the THFC dataset in Chapter 5.**

|  | mean | sd |
|---|---|---|
| **Ankle Injuries** | 26.22 | 54.06 |
| **Buttock/pelvis injuries** | 14.74 | 34.03 |
| **Chest Injury** | 15.17 | 10.57 |
| **Foot Injuries** | 4.25 | 3.32 |
| **Head Injuries** | 15.52 | 35.76 |
| **Hip and Groin Injuries** | 6.41 | 8.51 |
| **Knee Injuries** | 22.60 | 39.42 |
| **Lower Leg Injuries** | 12.88 | 18.70 |
| **Lumbar Spine Injury** | 0.42 | 1.13 |
| **Neck Injuries** | 0.50 | 0.70 |
| **Paediatric Diagnoses** | 65.00 | NA |
| **Post Surgical Patient** | 101.00 | NA |
| **Shoulder Injuries** | 18.80 | 40.38 |
| **Thigh Injuries** | 9.26 | 28.91 |
| **Thoracic Spine Injury** | 3.50 | 3.10 |
| **Wrist/hand Injuries** | 18.72 | 23.05 |

**Table 10.6. Recovery (mean and standard deviation) per body part for the integrated dataset in Chapter 5.**

|  | Mean | sd |
|---|---|---|
| **Ankle** | 23.46 | 101.21 |
| **Buttock/pelvis** | 14.74 | 34.03 |
| **Foot** | 5.75 | 3.84 |
| **Forearm** | 10.00 | NA |
| **Hand/Finger/Thumb** | 32.50 | 24.74 |
| **Head/Face** | 32.56 | 37.28 |
| **Hip/Groin** | 7.21 | 8.52 |
| **Knee** | 18.31 | 33.24 |
| **Lower Leg** | 13.19 | 22.62 |
| **Lumbar Spine Injury** | 0.50 | 1.22 |
| **Neck** | 0.50 | 0.70 |
| **Shoulder** | 18.80 | 40.38 |
| **Sternum/Ribs/UpperBack** | 4.45 | 4.94 |
| **Thigh** | 9.21 | 25.29 |
| **Thoracic Spine Injury** | 3.50 | 3.10 |

**Table 10.7. Recovery (mean and standard deviation) per type of injury for the WW dataset.**

|  | mean | sd |
|---|---|---|
| **Fracture** | 8.50 | 2.12 |
| **Haematoma / Contusion / Bruise** | 7.67 | 13.79 |
| **Lesion of Meniscus or Cartilage** | 7.00 | NA |
| **Muscle Rupture / Strain / Tear / Cramps** | 11.05 | 18.49 |
| **Nerve Injury** | 1.00 | 0.00 |
| **Other** | 5.00 | 4.36 |
| **Other Bone Injury** | 36.67 | 51.63 |
| **Sprain / Ligament Injury** | 11.13 | 16.26 |
| **Tendon Injury / Rupture / Tendonosis / Bursitis** | 5.50 | 6.36 |

**Table 10.8. Recovery (mean and standard deviation) per type of injury for the THFC dataset.**

|  | mean | sd |
|---|---|---|
| **Dental injury** | 1.00 | NA |
| **Dislocation/subluxation** | 96.00 | 7.07 |
| **Fracture** | 1.75 | 3.50 |
| **Haematoma/contusion/bruise** | 1.75 | 1.50 |
| **Haematoma/contusion/bruise** | 4.00 | 3.74 |
| **Lesion of meniscus/cartilage** | 39.65 | 63.08 |
| **Muscle rupture/tear/strain** | 13.49 | 27.87 |
| **Nerve injury** | 15.06 | 34.61 |
| **Other bone injury** | 14.74 | 34.29 |
| **Other type** | 4.00 | 6.93 |
| **Overuse symptoms unspecified** | 2.33 | 3.21 |
| **Overuse/hypertonia** | 17.42 | 48.26 |
| **Sprain/ligament injury** | 12.48 | 22.35 |
| **Synovitis/effusion** | 4.40 | 6.19 |
| **Tendon rupture/tendinopathy** | 31.23 | 33.81 |

**Table 10.9. Recovery (mean and standard deviation) per type of injury for the integrated dataset.**

|  | mean | sd |
|---|---|---|
| **Dislocation/subluxation** | 91.00 | NA |
| **Fracture** | 4.80 | 4.55 |
| **Haematoma/Contusion/Bruise** | 5.35 | 9.72 |
| **Lesion of meniscus/cartilage** | 37.54 | 65.08 |
| **Muscle Rupture/Strain/Tear/Cramps** | 12.47 | 23.94 |
| **Nerve Injury** | 9.06 | 7.58 |
| **Other bone injury** | 22.00 | 41.67 |
| **Other type** | 2.83 | 4.04 |
| **Overuse symptoms unspecified** | 1.60 | 2.51 |
| **Overuse/hypertonia** | 17.42 | 48.26 |
| **Sprain/ligament injury** | 12.22 | 19.88 |
| **Synovitis/effusion** | 4.40 | 6.19 |
| **Tendon rupture/tendinopathy** | 25.14 | 32.14 |

|  | mean | sd |
|---|---|---|

Table 10.10. Results for Gaussian process model in Chapter 6.

| Ccc | Gamma | Abserror | Noise | T-a |
|---|---|---|---|---|
| 0.979769 | 1.66E-06 | 15.53392 | 2.00E-04 | 0 |
| 0.983208 | 1.71E-06 | 12.34728 | 4.00E-04 | 1 |
| 0.977935 | 2.00E-06 | 12.28539 | 6.00E-04 | 2 |
| 0.971624 | 2.00E-06 | 11.61365 | 0.001 | 3 |
| 0.967901 | 2.00E-06 | 12.88794 | 9.00E-04 | 4 |
| 0.964158 | 2.00E-06 | 13.0202 | 0.0012 | 5 |
| 0.962211 | 2.00E-06 | 13.58325 | 0.0012 | 6 |
| 0.961997 | 2.00E-06 | 14.48357 | 9.00E-04 | 7 |
| 0.961198 | 2.00E-06 | 14.76523 | 9.00E-04 | 8 |
| 0.960079 | 2.00E-06 | 14.78614 | 0.0011 | 9 |
| 0.95704 | 2.00E-06 | 15.34858 | 0.0012 | 10 |
| 0.952366 | 2.00E-06 | 15.50641 | 0.0016 | 11 |
| 0.94859 | 2.00E-06 | 16.04802 | 0.0018 | 12 |

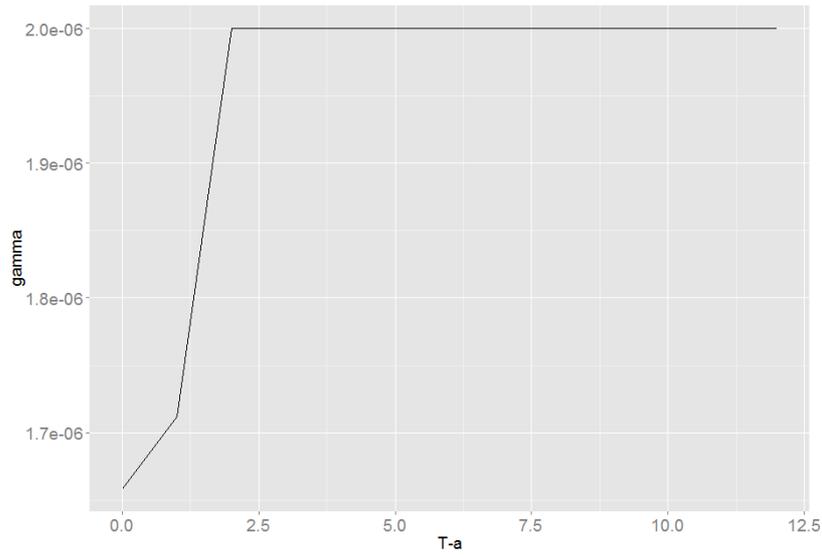

Figure 10.1. Plot of the optimal value of the noise parameter of the Gaussian Process model in Chapter 6.

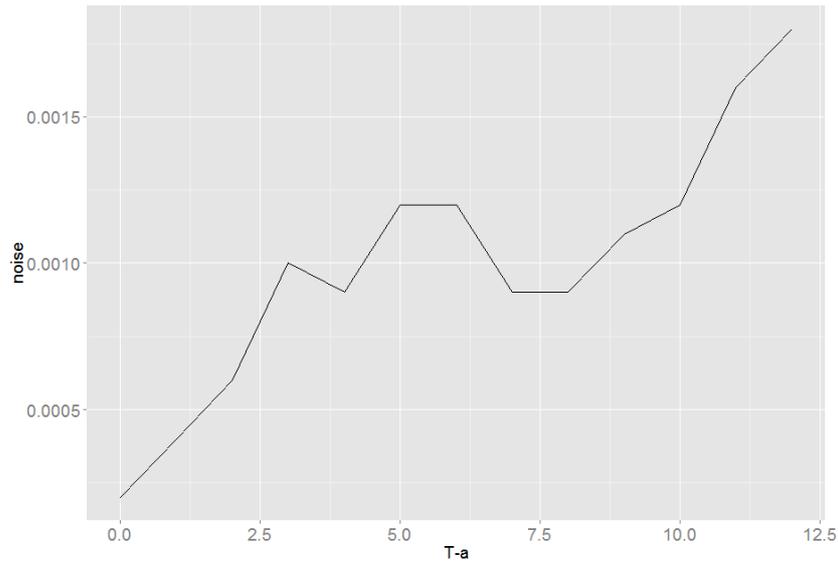

**Figure 10.2. Plot of the optimal value of the parameter gamma for the Gaussian Process model in Chapter 6.**

**Table 10.11. List of GPS variables for Chapter 7.**

| |
|---|
| AccelerationsZone1,AccelerationsZone2,AccelerationsZone3,AccelerationsZone4,AccelerationsZone5,AccelerationsZone6, AverageMetabolicPower, AverageSpeed, HighSpeedRunning, DecelerationsZone1, DecelerationsZone2,DecelerationsZone3,DecelerationsZone4,DecelerationsZone5,DecelerationsZone6,DistancePerMin,DistanceTotal, DistanceZ1, DistanceZ2, DistanceZ3, DistanceZ4, DistanceZ5, DistanceZ6, DurationofHI, DynamicStressLoadZone1, DynamicStressLoadZone2, DynamicStressLoadZone3, DynamicStressLoadZone4, DynamicStressLoadZone5, DynamicStressLoadZone6, EnergyExpenditure.KCal., EquivalentMetabolicDistance, ExplosiveDistance, HighSpeedRunning, HighSpeedRunningPerMinute, HMLDistance, HMLDistancePerMinute, HMLEfforts, ImpactsZone1, ImpactsZone2, ImpactsZone3, ImpactsZone4, ImpactsZone5, ImpactsZone6, LeftAntPostImpact, LeftAverageVertImpact, LeftLateralImpact, LeftMagnitudeImpact, LeftVerticalImpact, LowerSpeedLoading, MaxSpeed, MetabolicDistanceZonal, MetabolicTimeZonal, NumberofHighIntensityBursts, RightAverageVertImpact, RightLateralImpact, RightMagnitudeImpact, RightVerticalImpact, SpeedIntensityZone1, SpeedIntensityZone2, SpeedIntensityZone3, SpeedIntensityZone4, SpeedIntensityZone5, SpeedIntensityZone6, Sprints, StepBalance, TotalLeftSteps, TotalLoading, TotalRightSteps, DrillDuration |

**Table 10.12. Correlations of the first four components of PCA with the variables conducted for Approach A in Chapter 7. Variables with value below abs($10^{-5}$) for all components have been left out.**

|  | PC1 | PC2 | PC3 | PC4 |
|---|---|---|---|---|
| **AccelerationsZone1** | -0.17478 | 0.262532 | 0.047553 | -0.12477 |
| **AccelerationsZone3** | -0.22842 | 0.476407 | 0.020041 | -0.18829 |
| **AccelerationsZone4** | -0.10572 | 0.139499 | 0.023712 | -0.10246 |
| **AccelerationsZone5** | -0.0239 | 0.019581 | 0.026336 | -0.0424 |
| **AccelerationsZone6** | -0.00211 | -0.00055 | 0.005064 | -0.00314 |
| **AverageMetabolicPower** | -0.00264 | -0.01007 | -0.03587 | -0.02023 |
| **AverageSpeed** | 4.91E-05 | -0.00264 | -0.00763 | -0.0042 |
| **DecelerationsZone1** | -0.12199 | 0.20015 | 0.002862 | -0.21557 |
| **DecelerationsZone3** | -0.22306 | 0.430853 | 0.073116 | -0.07319 |
| **DecelerationsZone4** | -0.09469 | 0.146962 | -0.0071 | -0.00246 |
| **DecelerationsZone5** | -0.03516 | 0.037976 | -0.02209 | -0.00053 |
| **DecelerationsZone6** | -0.01121 | 0.007951 | -0.00631 | 0.005909 |
| **DistancePerMin** | 0.002838 | -0.15847 | -0.45817 | -0.25289 |
| **DistanceZ5** | -0.76398 | -0.48593 | 0.224796 | -0.11721 |
| **DistanceZ6** | -0.2269 | -0.21307 | 0.071945 | 0.002381 |
| **DynamicStressLoadZone1** | -0.11521 | 0.096693 | -0.19965 | -0.14145 |
| **DynamicStressLoadZone2** | -0.18704 | 0.04578 | -0.39766 | 0.375656 |
| **DynamicStressLoadZone3** | -0.16018 | 0.068063 | -0.17699 | 0.407681 |
| **DynamicStressLoadZone4** | -0.0948 | 0.044533 | -0.08566 | 0.252666 |
| **DynamicStressLoadZone5** | -0.04013 | 0.018942 | -0.04087 | 0.14582 |
| **HighSpeedRunningPerMinute** | -0.01591 | -0.01835 | 0.007384 | -0.0015 |
| **HMLDistancePerMinute** | -0.03392 | -0.01779 | -0.05331 | -0.02893 |
| **ImpactsZone4** | -0.17955 | 0.084792 | -0.1628 | 0.47654 |
| **ImpactsZone5** | -0.03914 | 0.018602 | -0.04001 | 0.139177 |
| **ImpactsZone6** | -0.00844 | 0.009527 | -0.01896 | 0.109988 |
| **LeftAverageVertImpact** | -0.00231 | 0.000804 | -0.00235 | 0.007278 |
| **LowerSpeedLoading** | -0.07649 | 0.126651 | 0.067832 | -0.01039 |
| **MaxSpeed** | -0.00777 | 0.003061 | -2.82E-06 | -0.00158 |
| **NumberofHighIntensityBursts** | -0.01851 | 0.004454 | -0.01033 | 0.031695 |
| **RightAverageVertImpact** | -0.00227 | 0.000246 | -0.00242 | 0.007538 |
| **SpeedIntensityZone2** | -0.06818 | 0.051936 | -0.23443 | -0.11655 |
| **SpeedIntensityZone3** | -0.02075 | -0.13705 | -0.59876 | -0.33672 |
| **SpeedIntensityZone4** | -0.07639 | -0.02801 | -0.0727 | -0.0543 |
| **SpeedIntensityZone5** | -0.05293 | -0.03391 | 0.015589 | -0.008 |
| **SpeedIntensityZone6** | -0.01749 | -0.01606 | 0.005157 | 0.00076 |
| **Sprints** | -0.01706 | -0.01064 | 0.00316 | 0.003275 |
| **StepBalance** | -3.97E-05 | 0.000102 | -3.91E-05 | 9.92E-05 |
| **TotalLoading** | -0.19934 | 0.171139 | -0.16064 | -0.03076 |
| **DrillDuration** | -0.12377 | 0.181335 | 0.062398 | -0.05195 |

**Table 10.13. Correlations of the first two components of PCA with the variables conducted for Approach B. Variables with value below abs(0.01) for all components have been left out.**

|  | PC1 | PC2 |
|---|---|---|
| **DynamicStressLoadZone6** | -0.71238 | -0.13873 |
| **ImpactsZone6** | -0.26405 | -0.08302 |
| **RightLateralImpact** | -0.18413 | -0.00140 |
| **RightMagImpact** | -0.16789 | 0.32925 |
| **DurationofHI** | -0.00595 | -0.04857 |
| **RightVerticalImpact** | 0.00481 | 0.31174 |
| **HighMetabolicPowerDistance** | 0.00846 | 0.24538 |
| **MetabolicTimeZonal** | 0.00851 | 0.11428 |
| **RightAntPostImpact** | 0.01015 | 0.07324 |
| **EquivalentMetabolicDistance** | 0.01256 | 0.13327 |
| **DistanceTotal** | 0.01489 | 0.09889 |
| **TotalRightSteps** | 0.01639 | 0.04616 |
| **HMLDistance** | 0.02046 | -0.01641 |
| **DistanceZ2** | 0.02055 | -0.01865 |
| **ExplosiveDistance** | 0.02062 | -0.02173 |
| **AccelerationsZone2** | 0.02080 | -0.04301 |
| **DecelerationsZone2** | 0.02093 | -0.04300 |
| **AccelerationsZone3** | 0.02109 | -0.04499 |
| **DecelerationsZone3** | 0.02110 | -0.04518 |
| **DecelerationsZone1** | 0.02113 | -0.04611 |
| **AccelerationsZone1** | 0.02116 | -0.04573 |
| **HighSpeedRunning** | 0.02117 | -0.04230 |
| **DrillDuration** | 0.02121 | -0.04628 |
| **AccelerationsZone4** | 0.02122 | -0.04663 |
| **DistanceZ5** | 0.02122 | -0.04346 |
| **HMLEfforts** | 0.02124 | -0.04405 |
| **DynamicStressLoadZone5** | 0.02127 | -0.04715 |
| **DistanceZ6** | 0.02127 | -0.04645 |
| **ImpactsZone5** | 0.02128 | -0.04716 |
| **DecelerationsZone4** | 0.02129 | -0.04658 |
| **SpeedIntensityZone2** | 0.02129 | -0.04610 |
| **EnergyExpenditure.KCal.** | 0.02129 | -0.03139 |
| **AccelerationsZone5** | 0.02130 | -0.04744 |
| **TotalLoading** | 0.02131 | -0.04509 |
| **DecelerationsZone5** | 0.02131 | -0.04724 |
| **LowerSpeedLoading** | 0.02132 | -0.04684 |
| **DecelerationsZone6** | 0.02132 | -0.04750 |
| **SpeedIntensityZone5** | 0.02132 | -0.04732 |
| **SpeedIntensityZone6** | 0.02133 | -0.04752 |
| **HighSpeedRunningPerMinute** | 0.02133 | -0.04755 |